\documentclass[11pt, oneside]{article} 
\pdfoutput=1

\usepackage{jheppub}
\usepackage{amsmath,amssymb,amsfonts,amsthm}
\usepackage{color}
\usepackage{booktabs}
\definecolor{darkblue}{rgb}{0.1,0.1,.7}
\usepackage[]{amsmath}
\usepackage{graphicx} % For including images
\usepackage{soul}
\usepackage[normalem]{ulem}

\usepackage[]{latexsym}
\usepackage{amscd}
\usepackage[all,cmtip]{xy}
\usepackage{mathrsfs}
\usepackage{bbold}
\usepackage[margin=10pt,font=small,labelfont=bf]{caption}
\usepackage{subcaption} % For subfigures
\usepackage{simplewick}
\usepackage{changepage}
\usepackage[]{algorithm2e}
\usepackage{booktabs,multirow}
\usepackage{float}
\usepackage{tensor}
\usepackage{cancel}
 
\usepackage{hyperref}

\usepackage{comment}
\usepackage{empheq}

\usepackage[OT2,OT1]{fontenc}
\usepackage{braket}
\usepackage{tikz}
\usepackage{pgfplots}
\pgfplotsset{compat=1.10}
\usepgfplotslibrary{fillbetween}
\usetikzlibrary{patterns}

\usepackage{placeins}

\newcommand{\CB}{\mathfrak{f}}

\newcommand{\polylog}[1]{\text{Li}_{#1}}

\theoremstyle{remark}

\makeatletter
\def\@fpheader{\ }
\makeatother

\title{Yang-Mills Flux Tube in AdS II: Effective String Theory}

\author{Barak Gabai, Victor Gorbenko, Bendeguz Offertaler}
\affiliation{Laboratory for Theoretical Fundamental Physics, Institute of Physics, École Polytechnique Fédérale de Lausanne (EPFL), CH-1015 Lausanne, Switzerland\\ }

\abstract{We continue the study of flux tubes in confining gauge theories placed in a rigid AdS background, focusing on the three-dimensional case. Our analysis is performed in the large-radius regime, where effective string theory provides a good approximation of the dynamics. Using a combination of techniques, primarily the analytic transcendentality ansatz bootstrap, we compute observables up to two-loop order in the expansion in powers of the string length over the AdS radius, which constitutes the main result of this work. Finally, we employ Padé resummations to explore the possible compatibility of our results with a smooth interpolation of observables between large-radius AdS and small-radius AdS, in which gauge theory is weakly coupled.}

\begin{document}
	
\maketitle

\section{Introduction}
In this paper we continue the study of the confining flux tube in Yang-Mills theory placed in an AdS background, which was initiated in \cite{Gabai:2025hwf}. As before, we focus on a long flux tube stretching radially across AdS space. By varying the AdS radius with respect to the confining scale, we can make one of two descriptions of the flux tube weakly coupled: the gauge theory description at small AdS radius or the Effective String Theory (EST) description at large AdS radius. The main new development in this work is a series of calculations performed on the EST side. While the EST description is very universal and different confining theories differ only by the values of their Wilson coefficients, we have in mind large $N$ pure Yang-Mills theory in 3D.

It is useful to think of the flux tube as a 1D conformal defect on the boundary of AdS. This lets us characterize the flux tube by the defect CFT (DCFT) data consisting of the dimensions of defect operators and their OPE coefficients. Our goal is to test the hypothesis that, when Neumann boundary conditions are imposed on the gauge fields on the boundary of AdS, the DCFT data interpolates smoothly between the two weakly coupled descriptions. An intuitive picture behind this conjecture is that the EST worldsheet degrees of freedom, or branons, are essentially gluons that build up the string. It is therefore plausible, at least in principle, that the gluon description takes over once the EST description becomes strongly coupled. Some of the most relevant previous works studying confining gauge theories in AdS are \cite{Callan:1989em,Aharony:2012jf,Ciccone:2024guw,DiPietro:2025ozw,Ciccone:2025dqx}. See \cite{Gabai:2025hwf} for a more detailed motivation and additional references to previous works.

While there are some similarities between the properties of the defect at asymptotically small and large AdS radius, there are also differences, as we review below. In order to test the hypothesis of smooth interpolation, we would like to calculate perturbative corrections in the two regimes and see if the properties of the defect become more similar as we go to higher orders. A more ambitious goal would be to compute observables at high enough order at weak coupling so that we can predict the EST properties. Unfortunately, calculations in the gauge theory description at small AdS radius appear to be more technically challenging at the moment. That is why in this paper we focus on making progress on calculations in the EST description at large radius. Our results are indeed consistent with the idea that the EST observables get closer and closer to those in the free gauge theory as we decrease the radius, but more computations are needed to make this statement robust.

Let us present a brief summary of the results and outline of the paper. In section \ref{sec:EST}, we describe the rules for constructing the effective action for the string in AdS. This is a review of the well known flat space construction \cite{Aharony:2013ipa,Dubovsky:2012sh} applied to the case of AdS$_3$. The main conclusion is that, up to two-loop order, which is the order that we get to in this paper, there are two independent parameters in the action: the string tension, which sets the overall coupling scale, and one free Wilson coefficient. The same conclusion holds in flat space EST in 3D. Section~\ref{sec:counting operators} studies the operator spectra in the two free theory limits and compares them. A shorter version of this discussion was presented in \cite{Gabai:2025hwf}. Here, we perform a more detailed study, including the counting of operators within each symmetry sector, which requires some new techniques \cite{Julius:2025dce}. In section \ref{sec:tree-level four point function via Witten diagrammatics}, we perform a direct computation of the leading order anomalous dimensions and free theory OPE coefficients of various operators using Witten diagrams (which were previously computed in \cite{Giombi:2017cqn}) and point-splitting regularization. We also find the expressions for several conformal primaries in terms of the basic fields. The point splitting method is quite tedious and can hardly be extended to significantly higher orders, but its advantage is that it allows us to find some CFT data for operators that consist of more than two fields. In particular, we find that at first order there is no mixing between operators with different numbers of fields, which is important for our later calculations.

Section \ref{sec:ansatz bootstrap} presents our main technical results. There, we obtain the branon four-point function at two-loop order using the so-called {\it{transcendentality ansatz}} bootstrap. This method has a long history, with its roots in \cite{Kotikov:2001sc,Kotikov:2002ab}, and has been used to bootstrap flat space amplitudes in $\mathcal{N}=4$ SYM \cite{Caron-Huot:2020bkp}, supergravity correlators in AdS$_5\times S^5$ \cite{Aharony:2016dwx,Aprile:2017bgs,Alday:2017xua,Huang:2021xws,Drummond:2022dxw}, and correlators in the Wilson-Fisher CFT \cite{Alday:2017zzv,Henriksson:2018myn}. Most relevant for us, it has also been used to study loop correlators in 1d defect CFTs \cite{Liendo:2018ukf,Ferrero:2019luz,Ferrero:2021bsb,Ferrero:2023znz,Ferrero:2023gnu}. By bootstrapping the branon four-point function to two loops, we are able to extract some of the flux tube DCFT data up to three non-trivial orders in perturbation theory. Our main results are summarized in section \ref{sec:summary of select OPE data}. We should note, however, that while we find that the transcendentality ansatz bootstrap is the most efficient way of computing the branon four-point function, its reliance on an ansatz makes it not fully rigorous. Therefore, in appendix \ref{app:dispMethod} we use 1d CFT dispersion relations \cite{Mazac:2018qmi,Bonomi:2024lky,Carmi:2024tmp} to verify the correctness of the ansatz.

An important role in our computations is played by the integrated constraints that correlation functions of branons satisfy as a consequence of the non-linearly realized conformal symmetry \cite{Gabai:2025zcs,Kong:2025sbk,Girault:2025kzt,Drukker:2025dfm}. These integrated constraints were also crucial in \cite{Gabai:2025hwf}, where they made it possible to fix a subset of anomalous dimensions in the small radius regime with very little additional input. In the EST description in the large AdS regime, the integrated constraints are not enough to fix any CFT data on their own, but they complement the ansatz bootstrap method by fixing a finite number of constants that would otherwise be undetermined. In addition, they allow us to calculate the normalization of the two-point function of the displacement operator. In agreement with our expectations based on the effective string action, our result for the branon four point function at two loops depends on a single free Wilson coefficient.

Finally, in section \ref{sec:bridging small and large AdS radius with Pade}, we collect the calculated DCFT data and analyze how well it interpolates between large and small AdS radius. We find it convenient to visualize our present results using two-sided Pad\'e approximants. While these approximants are consistent with smooth interpolation, we do not have a systematic method of judging how well they converge at intermediate values of the AdS radius, where both expansions can potentially break down. In order to get a handle on this issue, following \cite{Julius:2025dce}, we begin by analyzing a very similar problem for the BPS Wilson line in $\mathcal{N}=4$ SYM, where integrability results that are practically exact \cite{Grabner:2020nis,Cavaglia:2021bnz,Cavaglia:2022qpg} make it possible to test the convergence of analogous perturbative expansions in gauge theory and on the worldsheet. As was already emphasized in \cite{Gabai:2025hwf}, a straight $\mathcal{N}=4$ Wilson line, which becomes a radially stretched string in AdS$_5$ at strong coupling, has a lot in common with the Yang-Mills confining string in AdS. We find that the Pad\'e approximants in the two set-ups also behave quite similarly. Moreover, comparison of the $\mathcal{N}=4$ SYM Pad\'e approximants with the exact results from integrability shows that they converge to within a few percent for all values of the coupling. This makes it plausible that our calculations in pure Yang-Mills theory are close to reaching a similar precision. 

\FloatBarrier
\section{The effective string action in AdS$_3$}
\label{sec:EST}

In this section, we derive the effective action for a long string in euclidean AdS$_3$ up to the order that will be relevant for our analysis in this paper. The discussion is very similar to that of the long string in flat space $\mathbb{R}^3$ \cite{Dubovsky:2012sh,Aharony:2013ipa}.\footnote{A summary of more recent developments on flat space flux tubes can be found in \cite{Gabai:2025hwf}.} Readers interested in the final result can skip to eq.~\eqref{eq:effective string action - NG + HC}.

We want to study the excitations of a long string in AdS$_3$ with AdS$_2$ worldsheet geometry. For many purposes, it is convenient to have in mind a specific choice of coordinates that foliate AdS$_3$ by AdS$_2$ slices such that the metric takes the form
\begin{align}\label{eq:AdS_2 coordinates}
    ds^2&=G_{\mu\nu}(X)dX^\mu dX^\nu=R_{\rm AdS}^2\biggr[\cosh^2{z}\frac{dx^2+d\tau^2}{x^2}+dz^2\biggr],
\end{align}
where $x\in [0,\infty)$, $\tau,z\in \mathbb{R}$, and $X^\mu=(\tau,x,z)$. When it will be time to choose a gauge, we will work in static gauge, in which points on the worldsheet are labeled by the AdS$_2$ coordinates $\tau$ and $x$. The shape of the string in the target space is then specified by the single function $z(x,\tau)$ corresponding to transverse fluctuations of the worldsheet about the classical solution $z=0$. We call $z$ the ``branon.'' 

In analogy with the long string in flat space, the branon is governed by a two-dimensional effective action, whose expansion parameter is the string length $\ell_s$ and whose terms in the Lagrangian density are invariant under AdS$_3$ isometries and worldsheet reparameterizations. We will see that, as in flat space, the effective action for the string in AdS displays low energy universality, meaning that the first term in the action is the Nambu-Goto or area term and the first non-universal correction appears is suppressed by the relatively high power $\ell_s^4$. 

In principle, one could work directly in static gauge and write the effective action in terms of the transverse coordinate $z$, but this makes the action of the symmetries less transparent. It is simpler to write the terms using the general coordinate $X^\mu$, in which case the symmetries are manifest. We can then choose the specific coordinates in eq.~\eqref{eq:AdS_2 coordinates}, fix static gauge, and expand in powers of $z$ at the end.

We write the allowed invariant terms in the effective action by combining various target space and worldsheet building blocks. The primary target space building block is the metric, $G_{\mu\nu}$, together with geometric quantities that can be constructed from it, including the connection $\Gamma^\mu_{\nu\rho}$, the covariant derivative $\nabla_\mu$, and the Riemann tensor $R_{\mu\nu\rho\sigma}$. The primary worldsheet building blocks are the frame fields  $e_\alpha^\mu\equiv \partial_\alpha X^\mu$, and various geometric quantities that can be constructed from them, including the induced metric $g_{\alpha\beta}\equiv G_{\mu\nu} \partial_\alpha X^\mu \partial_\beta X^\nu$, the worldsheet connection $\hat{\Gamma}^\alpha_{\beta\gamma}$ (defined using $g_{\alpha\beta}$), the worldsheet covariant derivative $\hat{\nabla}_\alpha$ and Riemann tensor $\hat{R}_{\alpha\beta\gamma\delta}$ (both of which are defined using $\hat{\Gamma}^\alpha_{\beta\gamma}$), and the extrinsic curvature:
\begin{align}\label{eq:extrinsic curvature}
        K_{\alpha\beta}&\equiv n_\mu e_\alpha^\nu \nabla_\nu e_\beta^\mu.
\end{align}
Here, $n^\mu$ is a unit vector normal to the surface. The extrinsic curvature will be the main actor in what follows. Some of its useful properties include that it transforms as a target-space scalar and worldsheet two-tensor, it is symmetric $K_{\alpha\beta}=K_{\beta\alpha}$, and the mean curvature is $\text{tr}(K)\equiv K_{\alpha\beta}g^{\alpha\beta}=0$ if the surface satisfies the Nambu-Goto equation of motion \cite{Lee:2018IRM, Poisson:2009pwt}.

The terms in the effective action combine the above worldsheet quantities such that worldsheet indices are fully contracted. To the order that we are interested, the terms involve only the metric $g_{\alpha\beta}$, which is dimensionless, and the extrinsic curvature $K_{\alpha\beta}$, which has dimensions of inverse length. Thus, we can order the terms by the number of factors of $K_{\alpha\beta}$ that appear, with each power suppressed by an extra factor of $\ell_s$. We can also ignore terms that are zero on shell--- i.e., terms that contain $\text{tr}(K)$--- because these can always be removed by a suitable field redefinition \cite{Aharony:2013ipa}. Finally, each term can only have even powers of $K_{\alpha\beta}$, since it is parity odd under the change of orientation of the normal vector $n^\mu$, which we assume the action does not depend on.

Given these rules, the Nambu-Goto term is the only one allowed with no factors of the extrinsic curvature. The only allowed term with two factors of the extrinsic curvature is:
\begin{align}\label{eq:one 2K term}
    \text{tr}(K^2)\equiv g^{\alpha\delta}g^{\beta\gamma}K_{\alpha\beta}K_{\gamma\delta}.
\end{align}
(In the following paragraphs, traces and matrix multiplication of $K$ are implicitily performed by contracting worldsheet indices with the worldsheet metric.) And there is only one independent term with four factors of the extrinsic curvature:
\begin{align}\label{eq:one 4K term}
    \text{tr}(K^2)^2.
\end{align}
Note that 
$\text{tr}(K^4)=\frac{1}{2}\text{tr}(K^2)^2$ is not independent.\footnote{The trace relation is a consequence of the more basic trace relation $K^2=\frac{1}{2}\text{tr}(K^2)\mathbb{I}$. This in turn follows from the tracelessness of $K_{\alpha\beta}$ implying that $K_{\alpha\beta}\equiv \vec{K} \cdot \vec{\sigma}_{\alpha\beta}$ for some $\vec{K}$, together with the Clifford algebra property of the Pauli matrices, $\{\sigma_a,\sigma_b\}=2\delta_{ab}\mathbb{I}$.}

A further simplification is that eq.~\eqref{eq:one 2K term} is essentially topological. This follows from the Gauss equation, which in a general curved space relates the extrinsic curvature, the worldsheet Riemann tensor and the target space Riemann tensor:
\begin{equation}\label{eq:simplified Gauss-Codazzi}
\begin{aligned}
    K_{\alpha\gamma} K_{\beta\delta}-K_{\alpha\delta} K_{\gamma\beta}&=\hat{R}_{\alpha\beta\gamma\delta}-e_\alpha^\mu e_\beta^\nu e_\gamma^\rho e_\delta^\sigma R_{\mu\nu\rho\sigma}=\big(R_{\rm AdS}^{-2}+\frac{1}{2}\hat{R}\big)(g_{\alpha\gamma}g_{\beta\delta}-g_{\alpha\delta}g_{\beta\gamma})
\end{aligned}
\end{equation}
In the second equality, we used $R_{\mu\nu\rho\sigma}=-R_{\rm AdS}^{-2}(G_{\mu\rho}G_{\nu\sigma}-G_{\mu\sigma}G_{\nu\rho})$ because AdS$_3$ is maximally symmetric and $\hat{R}_{\alpha\beta\gamma\delta}=\frac{\hat{R}}{2}(g_{\alpha\gamma}g_{\beta\delta}-g_{\alpha\delta}g_{\beta\gamma})$ because the worldsheet is two dimensional. One consequence of eq.~\eqref{eq:simplified Gauss-Codazzi} is that the worldsheet Riemann tensor $\hat{R}_{\alpha\beta\gamma\delta}$ is not an independent object.\footnote{We should note that, at this order, acting with worldsheet covariant derivatives on $K_{\alpha\beta}$ also does not produce new terms in the action. This follows from the Codazzi equation, $\hat{\nabla}_\gamma K_{\alpha\beta}-\hat{\nabla}_\beta K_{\alpha\gamma}=-R_{\mu\nu\rho\sigma} n^\mu e_\alpha^\nu e_\beta^\rho e_\gamma^\sigma=0$, and the Simons identity $\hat{\nabla}_\gamma \hat{\nabla}^\gamma K_{\alpha\beta}=\hat{\nabla}_\alpha \hat{\nabla}_\beta \text{tr}(K)+\hat{R}K_{\alpha\beta}-\frac{\hat{R}}{2}g_{\alpha\beta}\text{tr}(K)$, both of which take the stated form for any surface in a maximally symmetric background. For a minimal surface, these imply $\hat{\nabla}^\alpha K_{\alpha\beta}=0$ and $\hat{\nabla}_\gamma \hat{\nabla}^\gamma K_{\alpha\beta}=\hat{R}K_{\alpha\beta}$, which let us exchange covariant derivatives for $K$s. Any term in the action with two copies of $K_{\alpha\beta}$ and two covariant derivatives can be reduced via integration by parts to one of these two forms.} Furthermore, contracting both sides of eq.~\eqref{eq:simplified Gauss-Codazzi} with $g^{\alpha\delta}g^{\beta\gamma}$ yields:
\begin{align}\label{eq:simplifying identity for curvature squared term}
    \text{tr}(K^2)=\text{tr}(K)^2-\hat{R}-\frac{2}{R_{\rm AdS}^2}.
\end{align}
The first term on the RHS is zero on shell, the second is topological due to the Gauss-Bonnet theorem, and the third is proportional to the Nambu-Goto term. Thus, the only effect of adding the term quadratic in the curvature to the action would be to modify the string tension by a term of order $\ell_s^2/R_{\rm AdS}^2$, and it therefore does not constitute an independent term in the effective action. More generally, for a string in a non-maximally symmetric spacetime, the curvature-squared term  is not topological or proportional to the Nambu-Goto term, which means that non-universal effects become important earlier in perturbation theory.

We conclude that the effective string action up to second order in $\ell_s^2$ has only two independent terms, given by:
\begin{align}\label{eq:effective string action - NG + HC}
    S&=\int d^2\sigma \sqrt{g}\left[\ell_s^{-2}+\kappa \ell_s^2(G_{\mu\nu} g^{\alpha\gamma}g^{\beta\delta}K^\mu_{\alpha\beta}K^\nu_{\gamma\delta})^2+O(\ell_s^4)\right],
\end{align}
where $\kappa$ is a Wilson coefficient. This is the covariantized form of the effective action for a string in three-dimensional flat space.

\FloatBarrier

\section{Counting operators at small and large AdS radius}\label{sec:counting operators}

In this section, we introduce the operators that can be placed on the AdS flux tube in both the gauge theory and effective string descriptions, and study their properties in the two free theory limits. In particular, we compare the spectrum of operators on the Wilson line in Yang-Mills in the limit $g_{\rm YM}R_{\rm AdS}\to 0$ to the spectrum of operators on the confining string in the limit $\lambda\equiv  \ell_s^{-2}R_{\rm AdS}^2\to \infty$. The conjecture of smooth interpolation between the two descriptions leads to what we call the ``square-root formula'' relating an operator's dimension in the EST description to its dimension in the gauge theory description. These ideas were already introduced in \cite{Gabai:2025hwf}, but here we give a more detailed discussion; the main novelty in this section is to determine the spectrum of the operators within each global symmetry sector, which we will show all follow the square-root formula separately.

The symmetries preserved by the flux tube consist of the 1d conformal group and two discrete symmetries: a reflection transverse to the line, and a reflection along the line accompanied by charge conjugation. The resulting symmetry group is
\begin{equation}
    SL(2,\mathbb{R})\times \mathbb{Z}_2^\perp\times \mathbb{Z}_2^\parallel.
\end{equation}
We refer to the discrete symmetries as $\textbf{R}$ and $\textbf{CT}$, respectively. We can thus work in a basis of primary operators labeled by their scaling dimension $\Delta$ and two parities:
\begin{equation}
    x_\perp, x_\parallel = \pm 1 \, .
\end{equation}

The basic strategy we use to count primaries takes advantage of the simplicity of conformal families in one dimension. Given any primary operator $O$ which is not the identity,  its conformal family is
\begin{align}
    \underset{\Delta}{O}\rightarrow \underset{\Delta+1}{\partial_\tau O}\rightarrow \underset{\Delta+2}{\partial_\tau^2O}\rightarrow \underset{\Delta+3}{\partial_\tau^3O}\rightarrow \ldots
\end{align} 
If $\mathcal{N}_O(\Delta)$ denotes the number of operators with dimension $\Delta$, it follows that the number of primaries with dimension $\Delta$ is simply:
\begin{align}\label{eq:N_P in terms of N_O}
    \mathcal{N}_P(\Delta)=\left\{\begin{array}{ll}\mathcal{N}_O(\Delta)-\mathcal{N}_O(\Delta-1) & \Delta >1 \\ \mathcal{N}_O(\Delta) & \Delta\leq 1\end{array}\right..
\end{align}
We also want to count operators within each global symmetry sector. Let $\mathcal{N}^{x_\perp x_\parallel}_O(\Delta)$ and $\mathcal{N}^{x_\perp x_\parallel}_P(\Delta)$ denote the number of operators and primaries with dimension $\Delta$ and parities $x_\perp,x_\parallel$. Then, because $\partial_\tau$ has parity $+-$,

\begin{align}\label{eq:number of primaries related to number of operators in different reps}
    \mathcal{N}_{P}^{x_\perp x_\parallel}(\Delta)&=\left\{\begin{array}{ll}\mathcal{N}_O^{x_\perp x_\parallel}(\Delta)-\mathcal{N}_{O}^{x_\perp(-x_\parallel)}(\Delta-1)& \Delta>1 \\ \mathcal{N}_O^{x_\perp x_\parallel}(\Delta) & \Delta\leq 1\end{array}\right..
\end{align}

We now turn to the counting of operators in each symmetry sector in the free limits corresponding to both the effective string and Wilson line descriptions of the flux tube. Explicit expressions for light primary operators up to dimension 12 in the free effective string description are given in Table~\ref{tab:table of GFF primaries} in Appendix~\ref{app:unmixing at tree level}.

\subsection{EST in the free limit} 

In the limit $\lambda^2 \equiv R_{\rm AdS}^2\ell_s^{-2}\to \infty$, the only relevant term in eq.~\eqref{eq:effective string action - NG + HC} is the piece of the Nambu-Goto action quadratic in $z$ (see eq.~\eqref{eq:effective action expanded} for more details):
\begin{align}
    S[z]&=\frac{R^2}{\ell_s^2}\int \frac{d^2\sigma}{x^2} \left[1+\frac{1}{2}R^2 g^{\alpha\beta} \partial_\alpha z \partial_\beta z +z^2+\ldots \right]
\end{align}
This is the action of a free scalar in AdS$_2$ with dimension $\Delta_z=2$. Its boundary correlators are those of a generalized free field, meaning that the two-point function takes the form
\begin{align}
    \braket{z(\tau_1)z(\tau_2)}&\propto \frac{1}{\tau_{12}^4},
\end{align}
and higher-point functions obey Wick's theorem. Furthermore, all other operators in the theory are given by taking products and derivatives with respect to $\tau$ of $z$.\footnote{Whereas the AdS branon was denoted $X$ in \cite{Gabai:2025hwf} to match the conventional notation for the flat space branon \cite{Polchinski:1991ax,Dubovsky:2012sh,Aharony:2013ipa}, in the present work we denote it by the AdS coordinate $z$ instead.}

In particular, a basis of operators (both primary and secondary) is given by arbitrary products of $z$ with arbitrary numbers of derivatives:
\begin{align}\label{eq:GFF generic operator}
     O=\underbrace{\partial^{n_1}z \partial^{n_2}z \ldots \partial^{n_k}z}_{k\text{ copies of }z},\qquad \Delta_O=2k+\sum_{i=1}^k n_k.
\end{align}
Because of bose symmetry, different choices of $(n_1,n_2,\ldots,n_k)$ that are related by a permutation are equivalent. This means we may restrict to $n_1\leq n_2\leq \ldots \leq n_k$, which implies that the operators are in one-to-one correspondence with partitions of integers.

We will first count the number of operators of a given dimension using a brute-force approach. Given our choice of basis in eq.~\eqref{eq:GFF generic operator},
\begin{equation}
\begin{aligned}
     p_{\leq k}(\ell)&\equiv \text{number of operators with $k$ copies of $z$ and $\ell$ derivatives},\\&=\text{number of partitions of $\ell$ into $k$ or fewer positive numbers}.
\end{aligned}
\end{equation}
The number of operators of dimension $\Delta$ is then:
\begin{align}\label{eq:ertyuytres}
    \mathcal{N}^{\rm EST}_O(\Delta)&=\delta_{\Delta,0}+\sum_{k=1}^{\lfloor \frac{\Delta}{2}\rfloor}p_{\leq k}(\Delta-2k).
\end{align}
For example, combining this result with eq.~\eqref{eq:N_P in terms of N_O}, it follows that the numbers of operators and primaries with dimensions up to $\Delta=18$ are
\begin{center}
\begin{tabular}{l|llllllllllllllllllll}
 $\Delta$ & 0 & 1 & 2 & 3 & 4 & 5 & 6 & 7 & 8 & 9 & 10 & 11 & 12 & 13 & 14 & 15 & 16 & 17 & 18  \\ \hline
$\mathcal{N}^{\rm EST}_O(\Delta)$ & 1& 0& 1& 1& 2& 2& 4& 4& 7& 8& 12& 14& 21& 24& 34& 41& 55& 66& 88 \\ 
$\mathcal{N}^{\rm EST}_P(\Delta)$ & 1 & 0 & 1 & 0 & 1 &0   &2 & 0 & 3 & 1 & 4 & 2 & 7 & 3 & 10 & 7 & 14 & 11 & 22 
\end{tabular}
\end{center}

\paragraph{Asymptotic spectrum.} We can also study the spectrum in terms of the character for the free boson. First, we define
\begin{align}\label{eq:free boson operator partition function}
    \chi^{\rm EST}_O(q)&\equiv \sum_O q^{\Delta_O}=\sum_{\Delta=0}^\infty q^\Delta \mathcal{N}^{\rm EST}_O(\Delta). 
\end{align}
The first sum is over operators $O$ with dimensions $\Delta_O$, while the second is over dimensions $\Delta$ with degeneracy $\mathcal{N}^{\rm EST}_O(\Delta)$. Instead of labeling a given operator by an integer partition, as in eq.~\eqref{eq:GFF generic operator}, we can also label it by the sequence of non-negative integers $m_0,m_1,m_2,\ldots$, such that $m_k$ is the number of factors of $\partial^k z$ appearing. The dimension is then $\Delta_O=\sum_k (2+k)m_k$ and the character is
\begin{align}
    \chi_O^{\rm EST}(q)&=\sum_{m_0,m_1,\ldots,=0}^\infty q^{\sum_k(2+k)m_k}=\frac{1-q}{(q)_\infty}.
\end{align}
We expressed the result in terms of the Euler function, $(q)_\infty\equiv \prod_{i=1}^\infty (1-q^i)$. Furthermore, from eq.~\eqref{eq:N_P in terms of N_O}, it follows that the character for primary states is related to $\chi_O^{\rm EST}(q)$ by
\begin{align}\label{eq:free boson primary partition function}
    \chi_P^{\rm EST}(q)\equiv\sum_\Delta q^\Delta \mathcal{N}^{\rm EST}_P(\Delta)=q+(1-q)\chi_O^{\rm EST}(q)=q+\frac{(1-q)^2}{(q)_\infty}.
\end{align}
Expanding $\chi_{O/P}^{\rm EST}$ in powers of $q$ and identifying $\mathcal{N}_{O/P}^{\rm EST}(\Delta)$ as the coefficient of $q^\Delta$, we can thus reproduce the results of our brute-force analysis summarized in the table above.

The character in eq.~\eqref{eq:free boson primary partition function} also makes it possible to read off the spectrum of primaries at large $\Delta$. In the limit $q\to 1^-$, the sum in eq.~\eqref{eq:free boson operator partition function} is dominated by large values of $\Delta$, where the spectrum becomes effectively continuous. This suggests that we write
\begin{align}\label{eq:primary partition function continuous approximation}
    \chi_P^{\rm EST}(q)\overset{q\to 1^-}\approx \int_0^\infty d\Delta \;\mathcal{N}_P^{\rm EST}(\Delta)e^{-\Delta \log(1/q)},
\end{align}
where we now treat $\mathcal{N}_P^{\rm EST}(\Delta)$ as a smooth function. Meanwhile, the Euler function is related to the Dedekind eta function by $\eta(q) = q^{1/24}(q)_\infty$. From the modularity property of $\eta(q)$, one finds the following asymptotic behavior of the character of primaries: 
\begin{align}\label{eq:asymptotic form of primary partition function}
    \chi_P^{\rm EST}(q)\overset{q\to 1^-}{\sim}\frac{\log^{\frac{5}{2}}(1/q)}{\sqrt{2\pi}}e^{\frac{\pi^2}{6\log(1/q)}}. 
\end{align}
If we now make an ansatz $\mathcal{N}_P^{\rm EST}(\Delta)\overset{\Delta\to \infty}{\sim} B\Delta^\beta e^{A\Delta^\alpha}$ for the asymptotic approximation to the number of primaries with dimension $\Delta$, we can fix $A,B,\alpha,\beta$ by approximating the integral in eq.~\eqref{eq:primary partition function continuous approximation} by its saddle point and ensuring that it reproduce the exponential term in eq.~\eqref{eq:asymptotic form of primary partition function}. The result is:
\begin{align}\label{eq:EST asymptotic DOS}
    \mathcal{N}_P^{\rm EST}(\Delta)\overset{\Delta\to \infty}{\sim} \frac{\pi^2}{8\cdot 3^{\frac{3}{2}}\Delta^2} e^{\pi\sqrt{\frac{2\Delta}{3}}}. 
\end{align}
The growth is consistent with the Cardy formula. We will compare this with the asymptotic spectrum on the free Wilson line. 

\paragraph{Symmetry resolved operator spectrum.} We can use the method of eq.~\eqref{eq:GFF generic operator}-\eqref{eq:ertyuytres} to also count the operators within each global symmetry sector. The two reflections send $\textbf{R}: z\to -z$ and $\textbf{CT}:\tau\to -\tau$, which means that $z$ has parity $-+$ and $\partial_\tau$ has parity $+-$. 

Consequently, the number of operators with dimension $\Delta$ with parity $x_\perp x_\parallel$ is the number of distinct ways of distributing $\ell$ derivatives on $k$ copies of $z$ such that $2k+\ell=\Delta$, and such that $k$ is even if $x_\perp=1$ and odd if $x_\perp=-1$, and $\ell$ is even if $x_\parallel=1$ and odd if $x_\parallel=-1$. This leads to:
\begin{align}
    \mathcal{N}^{++,\text{EST}}_{O}(\Delta)&=\left\{\begin{array}{cl} 0 &\Delta \;\text{odd}\\ \sum_{k=2,4,6,\ldots}^{\lfloor\Delta/2\rfloor}p_{\leq k}(\Delta-2k)&\Delta\;\text{even}\end{array}\right.,\label{eq:counting formula 1}\\
    \mathcal{N}^{+-,\text{EST}}_{O}(\Delta)&=\left\{\begin{array}{cl}\sum_{k=2,4,6,\ldots}^{\lfloor \Delta/2\rfloor} p_{\leq k}(\Delta-2k) &\Delta\;\text{odd}\\ 0 &\Delta\;\text{even}\end{array}\right.\\
    \mathcal{N}^{-+,\text{EST}}_{O}(\Delta)&=\left\{\begin{array}{cl} 0 & \Delta \;\text{odd}\\ \sum_{k=1,3,5,\ldots}^{\lfloor\Delta/2\rfloor}p_{\leq k}(\Delta-2k)& \Delta\;\text{even}\end{array}\right.\\
    \mathcal{N}^{--,\text{EST}}_{O}(\Delta)&=\left\{\begin{array}{cl}\sum_{k=1,3,5,\ldots}^{\lfloor \Delta/2\rfloor} p_{\leq k}(\Delta-2k) & \Delta\;\text{odd}\\ 0 & \Delta\;\text{even}\end{array}\right.\label{eq:counting formula 4}
\end{align}
Combined with eq.~\eqref{eq:number of primaries related to number of operators in different reps}, this lets us determine the number of primaries the four global symmetry sectors up to a fairly large value of $\Delta$. The results up to $\Delta=20$ are summarized in Table~\ref{tab:strong coupling}  shown in Figure~\ref{table:counts of primaries in different representations}.

\begin{figure}[ht]
    \centering
    % First table
    \begin{minipage}{0.45\textwidth}
        \centering
            \begin{tabular}{c|cccc}
            $\Delta$ & $++$ & $+-$ & $-+$ & $--$ \\ \hline
            0 & 1 &  &  &  \\
            1 &  &  &  &  \\
            2 &  &  & 1 &  \\
            3 &  &  &  &  \\
            4 & 1 &  &  &  \\
            5 &  &  &  &  \\
            6 & 1 &  & 1 &  \\
            7 &  &  &  &  \\
            8 & 2 &  & 1 &  \\
            9 &  &  &  & 1 \\
            10 & 2 &  & 2 &  \\
            11 &  & 1 &  & 1 \\
            12 & 4 &  & 3 &  \\
            13 &  & 1 &  & 2 \\
            14 & 5 &  & 5 &  \\
            15 &  & 3 &  & 4 \\
            16 & 8 &  & 6 &  \\
            17 &  & 5 &  & 6 \\
            18 & 11 &  & 11 &  \\
            19 &  & 8 &  & 9 \\
            20 & 17 &  & 15 & 
            \end{tabular}
        \captionof{table}{Effective string}
        \label{tab:strong coupling}
    \end{minipage}
    \hfill
    % Second table
    \begin{minipage}{0.45\textwidth}
        \centering
            \begin{tabular}{c|cccc}
                $\Delta$ & $++$ & $+-$ & $-+$ & $--$ \\ \hline
                0 & 1 &  &  &  \\
                1 &  &  &  &  \\
                2 &  &  & 1 &  \\
                3 & 1 &  &  &  \\
                4 & 1 &  & 1 &  \\
                5 & 2 &  & 1 & 1 \\
                6 & 3 & 1 & 3 & 1 \\
                7 & 6 & 2 & 4 & 4 \\
                8 & 10 & 6 & 10 & 6 \\
                9 & 20 & 12 & 16 & 16 \\
                10 & 36 & 28 & 36 & 28
            \end{tabular}
        \captionof{table}{Wilson line}
        \label{tab:weak coupling}
    \end{minipage}
    \caption{Number of primaries transforming under the four different representations of $\mathbb{R}:z\to -z$ and $\mathbb{CT}:\tau\to -\tau$, for each dimension $\Delta$, in both the effective string (left) and Wilson line (right) descriptions of the confining flux tube.}
    \label{table:counts of primaries in different representations}
\end{figure}

\paragraph{Symmetry resolved counting using characters.}

We can also count the operators with each parity more elegantly using the method of characters.\footnote{We thank Julius Julius for discussions about symmetry-resolved counting of superconformal primaries on the half-BPS Wilson line in $\mathcal{N}=4$ SYM using characters. See \cite{Julius:2025dce}.} Let's introduce a 
character that counts states with dimension $\Delta$, \textbf{R} parity $x_\perp$ and \textbf{CT} parity $x_\parallel$, weighted by $q^\Delta g^{\frac{1}{2}(1-x_\perp)} h^{\frac{1}{2}(1-x_\parallel)}$, where $q\in [0,\infty)$ and $g,h=\pm 1$. In other words,
\begin{align}
    \chi_O(q,g,h)\equiv \sum_{\Delta=0}^\infty \sum_{x_\perp,x_\parallel=\pm1} \mathcal{N}_O^{x_\perp x_\parallel}(\Delta)q^\Delta g^{\frac{1-x_\perp}{2}} h^{\frac{1-x_\parallel}{2}}.
\end{align}
We can recover the operator counts using the orthogonality of characters:
\begin{align}\label{eq:projection formula}
    \mathcal{N}_O^{x_\perp x_\parallel}(\Delta)&=\oint \frac{dq}{2\pi i}q^{-\Delta-1} \frac{1}{4}\sum_{g,h=\pm 1} g^{\frac{1-x_\perp}{2}}h^{\frac{1-x_\parallel}{2}}\chi_O(q,g,h).
\end{align}
The symmetry-resolved single-particle character for $z$ and its descendants is:
\begin{align}
    \chi_{1}^{\rm EST}(q,g,h)=\sum_{n=0}^\infty q^{2+n} gh^n =\frac{g q^2}{1-hq}.
\end{align}
The character for multiparticle states is then given by the plethystic exponential (because the fields are commuting):\footnote{Note that $\prod_{i=1}^\infty (1+q^i)=(q^2)_\infty/(q)_\infty$.}
\begin{align}
    \chi_O^{\rm EST}(q,g,h)&=\text{exp}\left(\sum_{k=1}^\infty \frac{1}{k} \chi_1(q^k,g^k,h^k)\right)=\frac{1-hq}{(hq)_\infty}\delta_{g,1}+\frac{1+hq}{(q^2)_\infty/(hq)_\infty}\delta_{g,-1}.
\end{align}
Finally, it follows from eq.~\eqref{eq:number of primaries related to number of operators in different reps} that the symmetry-resolved character for primaries is related to the symmetry-resolved character for all operators by 
\begin{align}\label{eq:symmetry resolved characters primary vs all operators}
    \chi_P^{\rm EST}(q,g,h)&=h q+(1-hq)\chi_O^{\rm EST}(q,g,h).
\end{align}
Projecting out the numbers of primaries with different parities using eq.~\eqref{eq:projection formula}, we get a closed form result that agrees with the brute force counting formulas given in eq.~\eqref{eq:counting formula 1}-\eqref{eq:counting formula 4} up to any dimension we checked ($\Delta\sim 100$). 

Furthermore, we can again read off the asymptotic behavior as $q\to 1^-$ of the characters of primaries with a given parity, $\chi_P^{x_\perp x_\parallel}(q)\equiv \frac{1}{4}\sum_{g,h=\pm 1} g^{\frac{1-x_\perp}{2}}h^{\frac{1-x_\parallel}{2}}\chi_P(q,g,h)$. We find:
\begin{align}
    \chi_P^{x_\perp x_\parallel,\rm EST}(q)\overset{q\to 1^-}{\sim}\frac{1}{4}\chi_P^{\rm EST}(q),
\end{align}
independent of the parity. This implies that the number of primaries within each symmetry sector are asymptotically the same:
\begin{align}\label{eq:symmetry resolve num primaries asymptotic EST}
    \mathcal{N}_P^{x_\perp x_\parallel,\text{EST}}(\Delta)\overset{\Delta\to \infty}{\sim}\frac{1}{4}\mathcal{N}_P^{\rm EST}(\Delta).
\end{align}

\subsection{YM Wilson line in the free limit}

We will now count the primaries in the YM Wilson line in the free limit $g_{\rm YM}R_{\rm AdS}\to 0$. We begin by reviewing the counting of the operators at each dimension $\Delta$, using a slightly different approach from the one in \cite{Gabai:2025hwf}, and then proceed to count the operators in each global symmetry sector.

{Whereas for the effective string description of the flux tube it is convenient to work with the coordinates in eq.~\eqref{eq:AdS_2 coordinates}, which make the AdS$_2$ geometry of the classical string worldsheet manifest, for the gauge theory description of the operators on the Wilson line it is preferable to work with coordinates that make the metric on the conformal boundary explicit. Therefore, in the rest of this subsection, we will write the AdS$_3$ metric in Poincar\'e coordinates: 
\begin{equation}
ds^2=\frac{d\tau^2+dy^2+dw^2}{w^2},
\end{equation}
where $w\geq 0$ is the bulk coordinate with the boundary at $w=0$, $\tau\in \mathbb{R}$ is the boundary coordinate along the Wilson line, and $y\in \mathbb{R}$ is the boundary coordinate perpendicular to the line.}\footnote{We hope denoting the Poincar\'e coordinate along the Wilson line by $\tau$ will not cause confusion. Although it is technically different from the $\tau$ in eq.~\eqref{eq:AdS_2 coordinates}, for the present discussion they are functionally the same.}

The basic gauge-invariant object on the Wilson line is the field-strength $F_{\mu\nu}$. In AdS$_3$ in Poincar\'e coordinates, it has three independent components: $F_{\tau y}$, $F_{\tau w}$ and $F_{yw}$. After imposing the Neumann boundary condition $F_{\tau w}=F_{yw}=0$, as well as the Bianchi identity $\partial_{[\mu} F_{\nu \rho]}=0$ and the Maxwell equation $\partial_\mu F_{\mu\nu}=0$, one can show that a linearly independent basis for operators containing a single $F$ is given by taking any number of $\partial_\tau$ and $\partial_\perp\equiv \partial_y$ derivatives acting on $F_{\tau y}$:
\begin{align}
    \partial_\perp^{m}\partial_\tau^{n} F.
\end{align}
Here, we write $F\equiv F_{\tau y}$ for short, which we recall is the displacement operator on the Wilson line, and has protected dimension $\Delta=2$. 

The most general ``multiparticle'' operator then takes the schematic form:
\begin{align}
    \partial_\perp^{m_1} \partial_\tau ^{n_1}F\partial_\perp^{m_2} \partial_\tau ^{n_2}F\ldots \partial_\perp^{m_k} \partial_\tau ^{n_k}F.
\end{align}
Note that, unlike in the discussion of composite operators made out of the branon in EST, the order of different terms matters because the $F$'s are $N\times N$ matrices and we take $N\to \infty$. 

The number of operators with $k$ copies of $F$, $n_\tau$ $\partial_\tau$ derivatives and $n_\perp$ $\partial_\perp$ derivatives is
\begin{align}
    N^{\text{WL}}(k,n_\tau,n_\perp)=\binom{k-1+n_\tau}{n_\tau}\binom{k-1+n_\perp}{n_\perp}.
\end{align}
The first factor counts the number of ways of distributing the $n_\tau$ $\partial_\tau$ derivatives, and the second factor counts the number of ways of distributing the $n_\perp$ $\partial_\perp$ derivatives. Then the total number of operators with $k$ fields and $n$ total derivatives (of either type) is:
\begin{align}
    N^{\text{WL}}(k,n)=\sum_{n_\tau=0}^n N^{\text{WL}}(k,n_\tau,n-n_\tau)=\binom{2k-1+n}{n}.
\end{align}
The total number of operators with a given dimension $\Delta$ is given by the sum over all allowed values of $k$, and the result simplifies to: 
\begin{align}\label{eq:number of operators gauge theory YM3}
    \mathcal{N}_O^{\text{WL}}(\Delta)=\sum_{k=0}^{\lfloor \Delta/2\rfloor}N^{\text{WL}}(k,\Delta-2k)=\left\{\begin{array}{cl} 1 & \Delta = 0\\ 0 & \Delta = 1\\ 2^{\Delta-2} & \Delta = 2,3,\ldots \end{array}\right..
\end{align}
The total number of primaries follows from the total number of operators via eq.~\eqref{eq:N_P in terms of N_O}, with the result:
\begin{align}\label{eq:number of primaries gauge theory YM3}
    \mathcal{N}_P^{\rm \text{WL}}(\Delta)&=\left\{\begin{array}{cl} 1 & \Delta = 0 \\ 0 & \Delta =1 \\ 1 & \Delta = 2\\ 2^{\Delta-3} & \Delta = 3,4,\ldots\end{array}\right..
\end{align}
We note that the density of primaries grows exponentially. These results were previously given in \cite{Gabai:2025hwf}. 

\paragraph{Symmetry resolved counts.} We will now count the operators with each of the four possible parities under $\mathbb{Z}_2^\perp\times \mathbb{Z}_2^\parallel$. These parities act as follows on the field strength $F$:
\begin{align}
    \mathbf{R}:F(\tau,y)\to -F(\tau,-y),\\
    \mathbf{CT}:F(\tau,y)\to F(-\tau,y)^T.\label{eq:action of P parallel}
\end{align}
Thus, the parity under \textbf{R} of a composite operator with $k_F$ copies of $F$ and $n_\perp$ copies of $\partial_\perp$ is $(-1)^{k_F+n_\perp}$. On the other hand, the transpose in eq.~\eqref{eq:action of P parallel} implies that \textbf{CT} reverses the order of operators appearing in a composite operator, which makes the determination of the parity of composite operators under \textbf{CT} slightly more complicated than just counting the number of $\partial_\tau$ derivatives.

More precisely, to discuss parity of composite operators under \textbf{CT}, it is useful to introduce the notions of the reverse of a composite operator, and of operators that are palindromes. The \textit{reverse} of an operator is the corresponding operator with the same single trace operators multiplied in the reverse order. For example, given the composite operator $O=\partial_\tau F \partial_\perp F F \partial_\tau^2 \partial_\perp F$, its reverse is $RO=\partial_\tau^2 \partial_\perp F F \partial_\perp F \partial_\tau F$. A \textit{palindrome} is then a composite operator that is equal to its reverse. Some examples include $F$, $\partial_\tau^3 \partial_\perp^7 F$, $FF$, $\partial_\tau F\partial_\tau F$, $FFF$, $\partial_\perp^3 F\partial_\tau^2 F \partial_\perp^3 F$. These notions are useful because palindromes have definite parity under \textbf{CT}, given by $(-1)^{n_\tau}$, where $n_\tau$ is the total number of $\partial_\tau$ derivatives appearing in the operator. By contrast, an operator $O$ that is not a palindrome does not have a definite parity under \textbf{CT}, but the linear combinations $O\pm RO$ of $O$ do have definite parities, given by $\pm (-1)^{n_\tau}$.

Therefore, to count the number of operators with a given dimension $\Delta$ and given paritities $x_\perp$ and $x_\parallel$ under \textbf{R} and \textbf{CT}, we should first count the number of palindromes with these quantum numbers. There are two ways to construct a palindrome: we can either glue an operator to its reverse (which always yields an operator with even dimension and parities $x_\perp=x_\parallel=+1$) or we can sandwich a single-$F$ operator between another operator and its reverse. Thus, we first note the number of operators with given quantum numbers with only one field strength insertion:
\begin{equation}
    \mathcal{N}^{x_\perp, x_\parallel}_{1}(\Delta) = \left\{\begin{array}{ll} \frac{\Delta + (x_\parallel -x_\perp - 2)/2}{2}  & \quad \Delta\geq 2\;\text{and}\;\Delta + (x_\parallel -x_\perp - 2)/2\;\text{is even}\\ 0 & \quad\text{otherwise}\end{array}\right.\ .
\end{equation}
Then, accounting for the two ways of constructing palindromes, the total number of palindromes with given quantum numbers is:
\begin{equation}
    \mathcal{N}^{x_\perp,x_\parallel}_{\rm palin}(\Delta) = \sum_{j=0}^{\lfloor\Delta/2\rfloor-1} \mathcal{N}_O^{\text{WL}}(j) \mathcal{N}^{x_\perp,x_\parallel}_{1}(\Delta-2j)\;+\;\left\{\begin{array}{ll} \mathcal{N}_O^{\text{WL}}(\Delta/2)\delta_{x_\perp,+1}\delta_{x_\parallel,+1}  & \quad\Delta\;\text{even}\\ 0 & \quad\text{otherwise}\end{array}\right..
\end{equation} 
The last ingredient we need is the naive symmetry-resolved counting of operators, which is the counting that \textit{mislabels} \textbf{CT} parity as simply the number of $\partial_\tau$ derivatives. This is easily obtained with either of the methods we used to count operators in the effective string description.\footnote{For example, the naive symmetry-resolved character for operators with a single $F$ field and any number of $\partial_\perp$ and $\partial_\tau$ derivatives is $\chi_{1,\text{naive}}^{\rm WL}(q,g,h)=\sum_{m,n=0}^\infty g^{m+1} h^n q^{m+n+2}=\frac{gq^2}{(1-gq)(1-hq)}$. The character for operators with $N$ fields is $(\chi_{1,\text{naive}}^{\rm WL})^N$ because the fields are non-commuting, and the character for operators with any number of fields is $\chi_{O,\text{naive}}^{\rm WL}(q,g,h)=\frac{1}{1-\chi_{1,\text{naive}}^{\rm WL}(q,g,h)}$. Finally, extracting the operator counts within each symmetry sector using the orthogonality of characters, we get eq.~\eqref{eq:naive symmetry resolved counts WL}.} The result can be written relatively compactly as:\footnote{The cosine in eq.~\eqref{eq:naive symmetry resolved counts WL} plays the role of a discrete, periodic ``switch.'' Namely, for odd $\Delta$, it obeys $\cos(\frac{\pi}{4}(\Delta-3))=0,1,0,-1$ if $\frac{\Delta-1}{2}=0,1,2,3$ modulo $4$, while for even $\Delta$, it obeys $\sqrt{2}\cos(\frac{\pi}{4}(\Delta-3))=-1,1,1,-1$ if $\frac{\Delta}{2}=0,1,2,3$ modulo $4$.}
\begin{equation}\label{eq:naive symmetry resolved counts WL}
    \mathcal{N}_{\rm naive}^{x_\perp,x_\parallel}(\Delta) =
    {
    \renewcommand{\arraystretch}{1.1}
    \left\{\begin{array}{ll} \,\delta_{x_\perp,+1}\delta_{x_\parallel,+1} & \quad\Delta=0\\ 0 & \quad \Delta =1\\ \delta_{x_\perp,-1}\delta_{x_\parallel,+1} & \quad \Delta=2\\2^{\Delta-4} + x_\parallel x_\perp \ 2^{\frac{\Delta-5}{2}} \ \cos(\frac{\pi}{4}(\Delta-3))& \quad\Delta\geq 3\; \text{and}\;\Delta\;\text{odd}\\2^{\Delta-4} + x_\parallel 2^{\frac{\Delta}{2}-3} - x_\parallel x_\perp \ 2^{\frac{\Delta-5}{2}}\ \cos(\frac{\pi}{4}(\Delta-3)) & \quad  \Delta \geq 4\;\text{and}\;\Delta\;\text{even}\end{array}\right. 
    }.
\end{equation}
Finally, this naive counting is related to the correct counting by the formula,
\begin{equation}
    \mathcal{N}_O^{x_\perp,x_\parallel,\text{WL}}(\Delta) = \mathcal{N}_{\rm palin}^{x_\perp, x_\parallel}(\Delta) + \sum_{\pm} \frac{\mathcal{N}_{\rm naive}^{x_\perp,\pm x_\parallel}(\Delta) - \mathcal{N}_{\rm palin}^{x_\perp,\pm x_\parallel}(\Delta)}{2} \ .
\end{equation}

Evaluating this for the different choices of parity, and using eq.~\eqref{eq:number of primaries related to number of operators in different reps} to relate the counts of operators to the counts of primaries, we arrive at the following expressions for the symmetry-resolved counts of primaries:
\begin{align}
    \mathcal{N}_P^{++,\text{WL}}(\Delta)=\left\{\begin{array}{cc}1 & \Delta=0,3,\\ 0 & \Delta=1,2,\\ 2^{\frac{\Delta}{2}-3}+2^{\Delta-5}& \Delta\text{ even and }\Delta\geq 4\\ 2^{\frac{\Delta}{2}-\frac{5}{2}}+2^{\Delta-5}& \Delta\text{ odd and }\Delta\geq 5\end{array}\right.\label{eq:rty5est 1}
\end{align}
\begin{align}
    \mathcal{N}_P^{+-,\text{WL}}(\Delta)=\left\{\begin{array}{cc} 0 & 0\leq \Delta\leq 5,\\ -2^{\frac{\Delta}{2}-3}+2^{\Delta-5}& \Delta\text{ even and }\Delta\geq 6\\ -2^{\frac{\Delta}{2}-\frac{5}{2}}+2^{\Delta-5}& \Delta\text{ odd and }\Delta\geq 7\end{array}\right.\label{eq:rty5est 2}
\end{align}
\begin{align}
    \mathcal{N}_P^{-+,\text{WL}}(\Delta)=\left\{\begin{array}{cc} 0 & \Delta=0,1,3,\\ 1 & \Delta=2\\ 2^{\frac{\Delta}{2}-3}+2^{\Delta-5}& \Delta\text{ even and }\Delta\geq 4\\ 2^{\Delta-5}& \Delta\text{ odd and }\Delta\geq 5\end{array}\right.\label{eq:rty5est 3}
\end{align}
\begin{align}
    \mathcal{N}_P^{--,\text{WL}}(\Delta)=\left\{\begin{array}{cc} 0 & 0\leq \Delta \leq 3 \\ -2^{\frac{\Delta}{2}-3}+2^{\Delta-5}& \Delta\text{ even and }\Delta\geq 4\\ 2^{\Delta-5}& \Delta\text{ odd and }\Delta\geq 5\end{array}\right.\label{eq:rty5est 4}
\end{align}
The results up to $\Delta=10$ are summarized in Table~\ref{tab:weak coupling} in Figure~\ref{table:counts of primaries in different representations}. In addition, we see that, as in the effective string description, the numbers of primaries within each symmetry sector are asymptotically the same (with exponentially small corrections): 
\begin{align}\label{eq:symmetry resolve num primaries asymptotic WL}
    \mathcal{N}_P^{x_\perp x_\parallel,\text{WL}}(\Delta)\overset{\Delta\to \infty}{\sim}\frac{1}{4}\mathcal{N}_P^{\rm WL}(\Delta).
\end{align}

\subsection{Additional comments on the square-root formula}

The simplest possibility consistent with the conjecture that there is no phase transition in Yang-Mills in AdS with Neumann boundary conditions as one dials the AdS radius from small to large is that the anomalous dimensions of operators in the defect CFT interpolate ``as smoothly as possible.'' This would mean that the operator with the $n$th lowest dimension in the confining string description is identified with the operator with the $n$th lowest dimension in the Wilson line description, and that its dimension varies smoothly, and perhaps even monotonically, as the combination $g_{\rm YM}R_{\rm AdS}$, or equivalently, $\lambda^2\equiv R_{\rm AdS}^2l_{s}^{-2}$, is varied. It was argued in \cite{Gabai:2025hwf} that this would lead to a ``square-root formula'' relating the dimensions of a given operator at large AdS radius and small AdS radius. We now make a few additional comments in support of these ideas using our results for the symmetry-resolved spectrum of primaries on the line defect in both weakly coupled limits.

To be more precise, we expect that the $n$th lowest operator at weak coupling is mapped to the $n$th lowest operator at strong coupling only within each symmetry sector, assuming there are no additional hidden symmetries, due to the general absence of level crossing of states with the same quantum numbers (see e.g., \cite{Korchemsky:2015cyx}). We may call this the no-crossing rule. (The fact that this does not uniquely identify the effective string operator with the corresponding Wilson line operators due to significant degeneracies on both sides of the dictionary, which increase at higher dimensions, does not affect our argument.) However, as we probe the spectrum at asymptotically large dimensions, because the number of primaries within each symmetry sector have the same growth, both in the effective string and the Wilson line limits (see eqs.~\eqref{eq:symmetry resolve num primaries asymptotic EST} and eq.~\eqref{eq:symmetry resolve num primaries asymptotic WL}), we can lump all the operators together and apply the no-crossing rule without worrying about symmetry.

Let us label dimensions of the $n$th operator in the spectrum in the EST and WL descriptions as $\Delta_n^{\rm EST}$ and $\Delta_n^{\rm WL}$, and introduce a function $f$ that satisfies $\Delta_n^{\rm WL}=f(\Delta_n^{\rm EST})$. Then the densities of states are related by
\begin{align}\label{eq:matching densities of states}
    n&=\int_0^{\Delta_n^{\rm EST}}d\Delta \mathcal{N}_P^{\rm EST}(\Delta)=\int_0^{f(\Delta_n^{\rm EST})}d\Delta \mathcal{N}_P^{\rm WL}(\Delta).
\end{align}
Asymptotically, the densities $\mathcal{N}_P^{\rm EST}(\Delta)$ and $\mathcal{N}_P^{\rm WL}(\Delta)$ and the function $f$ are approximately smooth (and independent of parities). We can thus differentiate eq.~\eqref{eq:matching densities of states} with respect to $\Delta_n^{\rm EST}$, which gives $\mathcal{N}_P^{\rm EST}(\Delta_n^{\rm EST})=f'(\Delta_n^{\rm EST})\mathcal{N}_P^{\rm WL}(f(\Delta_n^{\rm EST}))$. Finally, given the asymptotic behaviors of the density of states in eqs.~\eqref{eq:EST asymptotic DOS} and~\eqref{eq:number of primaries gauge theory YM3}, and matching the exponents, we deduce that the mapping between dimensions in the EST and WL descriptions is asymptotically given by
\begin{align}
    \Delta_n^{\rm WL}=f(\Delta_n^{\rm EST})\overset{n\to \infty}{\sim }\frac{\pi}{\log{2}}\sqrt{\frac{2}{3}}\sqrt{\Delta_n^{\rm EST}}. 
\end{align}
This square-root behavior was introduced in \cite{Gabai:2025hwf} to describe the interpolation of all operators. Our present analysis confirms that the same square-root behavior holds asymptotically within each symmetry sector.

\begin{figure}[ht]
    \centering
    \includegraphics[width=0.9\textwidth]{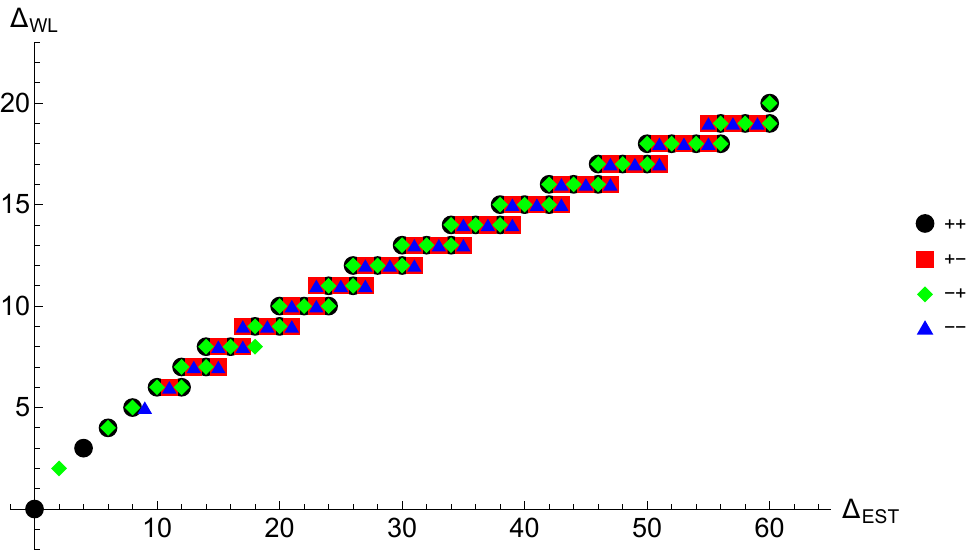}
    \caption{For each operator on the flux tube in AdS, we plot its dimensions, $(\Delta^{\rm EST},\Delta^{\rm WL})$, in the free limits of both the EST and Wilson line descriptions. These pairs as determined within each global symmetry sector using the no-crossing rule. The map between $\Delta^{\rm EST}$ and $\Delta^{\rm WL}$ is remarkably smooth and uniform across the four symmetry sectors for all dimensions.
    }
    \label{fig:Delta EST to Delta WL and square root fit}
\end{figure}

Besides asymptotics, it is also interesting to consider the behavior of the mapping between operators at small and large AdS radius when the dimensions are relatively small. In Figure~\ref{fig:Delta EST to Delta WL and square root fit}, we plot the pairings $(\Delta_{\rm EST},\Delta_{\rm WL})$ using the no-level-crossing rule within each global symmetry sector for all operators with $\Delta_{\rm EST}\leq 60$. The plot makes it clear that the map from $\Delta^{\rm EST}$ to $\Delta^{\rm WL}$ is remarkably smooth as a function of $\Delta^{\rm EST}$ and has very little spread between different symmetry sectors. It is interesting that these properties, which we understand analytically in the limit of large dimensions, hold very well for all dimensions all the way down to $\Delta=0$. 

\FloatBarrier

\section{The branon four-point function via Witten diagrams}\label{sec:tree-level four point function via Witten diagrammatics}

The main observable that we will focus on for the remainder of the paper is the branon four-point function. As discussed at length in \cite{Gabai:2025hwf}, the branon is the effective string representation of the displacement operator on the Wilson line in Yang-Mills in AdS. The displacement operator plays a fundamental role in the line defect CFT because it nonlinearly realizes the AdS isometries broken by the defect. Among other things, this means that the displacement has protected dimension $\Delta_z=2$ as well as a natural scheme-independent normalization, and that correlators in which the displacement is one of the external operators satisfy certain integral identities. We are mainly interested in the branon four-point function because it efficiently packages various OPE data, which we will extract in section~\ref{sec:ansatz bootstrap} and discuss in section~\ref{sec:bridging small and large AdS radius with Pade}.

In this section, we determine the free and tree-level contributions to the branon four-point function using  Witten diagrams, with the effective string action providing the propagators and interaction vertices. The computation is the same as the one presented in \cite{Giombi:2017cqn} for the displacement four-point function of the fundamental string in AdS$_5\times S^5$ (the analysis is the same after turning off the fluctuations in $S^5$ and along two of the AdS directions), and the result was also previously discussed in the context of the Yang-Mills flux tube in \cite{Gabai:2025hwf}. We review the analysis here for a few reasons: to set our notation, for the completeness of the present narrative, and to serve as a check of our analytic bootstrap result for the tree-level four point function in section~\ref{sec:bootstrap tree-level}. In addition, we generalize the Witten diagram computation beyond the Nambu-Goto quartic interaction, considering also contact diagrams that arise from zero- and eight-derivative quartic interactions. These three contact diagrams span the space of tree-level solutions to crossing (and other constraints) that appear in the two-loop analytic bootstrap in section~\ref{sec:ansatz bootstrap}.

Finally, we close this section by using the result for the branon four-point function to study the properties of the primaries in the theory at tree-level. (This analysis differs from the one for the fundamental string in AdS$_5\times S^5$ due to the different particle content.) In particular, by explicitly constructing the composite primaries in the EST and computing their two point functions in point-splitting regularization, we demonstrate the absence of tree-level mixing between operators with different particle number. We also arrive at a simple formula for the tree-level anomalous dimensions of all composites with dimension less than or equal to $12$, which we conjecture holds for all operators in the theory. These results are summarized in section~\ref{sec:comments on mixing at tree-level}, and the details are provided in Appendix~\ref{app:unmixing at tree level}.

\subsection{Branon four-point function}

The branon boundary correlators can be computed perturbatively in inverse powers of $1/\lambda^2$, where $\lambda\equiv R_{\rm AdS}\ell_s^{-1}$. In principle, this can be done by computing the bulk correlator of $n$ branon excitations on the string worldsheet via the path integral, and then pushing the bulk points to the boundary with a suitable rescaling by the AdS bulk coordinate:
\begin{align}
    \braket{z(\tau_1)\ldots z(\tau_n)}&=Z^n\lim_{x_1,\ldots,x_n\to 0}\prod_i x_i^{-\Delta_z}\int \mathcal{D}ze^{-S[z]}z(\tau_1,x_1)\ldots z(\tau_n,x_n).
\end{align}
Here, we work with the AdS$_2$ slicing coordinates of AdS$_3$ in eq.~\eqref{eq:AdS_2 coordinates}, and $S[z]$ is the effective action in eq.~\eqref{eq:effective string action - NG + HC} in static gauge. The multiplicative factor $Z$ is fixed by the condition,\footnote{{We normalize the branon canonically for simplicity. Another common choice is $\mathbb{D}\equiv\sqrt{\Lambda} z$, with $\Lambda$ fixed so that $\mathbb{D}$ generates displacements of the Wilson line via $\braket{\delta W}=\int d\tau \braket{W\mathbb{D}(\tau)}$. At the end of our analysis, we evaluate $\Lambda$ in eq.~\eqref{eq:displacement two-pt function}.}}
\begin{equation}\label{eq:z 2pt normalization}
    \braket{z(\tau)z(0)} = \frac{1}{\tau^4}\ .
\end{equation}

The procedure for evaluating tree-level Witten diagrams contributing to the string two- and four-point functions is by now well understood \cite{Giombi:2017cqn}, but results for higher-point tree-level diagrams and loop diagrams are more cumbersome (see \cite{Bliard:2022xsm,Giombi:2023zte} and \cite{Giombi:2017hpr,Carmi:2019ocp,Carmi:2021dsn,Carmi:2024tzj} for some progress). Therefore, in this section we restrict our analysis to tree level, postponing until section~\ref{sec:ansatz bootstrap} the analysis of higher orders via an analytic bootstrap approach.

Because the Wilson line preserves the $SL(2,\mathbb{R})$ isometry of the $\mathrm{AdS}_2$ worldsheet, the two- and three-point functions of conformal primaries take the standard CFT form,
\begin{align}
    \braket{O_i(\tau_1)O_j(\tau_2)}&=\frac{\delta_{ij}}{\tau_{12}^{2\Delta_i}},\\
    \braket{O_i(\tau_1)O_j(\tau_2)O_k(\tau_3)}&=\frac{c_{O_iO_jO_k}}{|\tau_{12}|^{\Delta_{ij|k}}|\tau_{13}|^{\Delta_{ik|j}}|\tau_{23}|^{\Delta_{jk|i}}},
\end{align}
where $\Delta_{ij|k}\equiv \Delta_i+\Delta_j-\Delta_k$. Furthermore, the branon four point function is fixed by conformal symmetry to take the form of a kinematical prefactor times a function of a single cross ratio:
\begin{align}\label{eq:normalized branon four point function}
   \braket{z(\tau_1)z(\tau_2)z(\tau_3)z(\tau_4)}&=\frac{1}{\tau_{12}^4\tau_{34}^4}G(\chi),
\end{align}
where we define the cross-ratio 
\begin{align}
    \chi\equiv \frac{\tau_{12}\tau_{34}}{\tau_{13}\tau_{24}}.
\end{align}

Our goal is to evaluate the reduced four-point function $G(\chi)$ as a perturbative series:
\begin{align}
\label{eq:pertExpG}
    G(\chi)&=1+G^{(0)}(\chi)+\varsigma G^{(1)}(\chi)+\varsigma^2 G^{(2)}(\chi)+\varsigma^3 G^{(3)}(\chi)+O(\varsigma^4). 
\end{align}
Here, $\varsigma$ is our scheme independent\footnote{We work with $\varsigma$ rather than $1/\lambda^2$ because the latter is scheme-dependent beyond tree-level. In principle, any monotonic physical observable may be used as a scheme-independent coupling. The evidence collected at small and large $\lambda$ \cite{Gabai:2025hwf} indicates that $\Delta_{z^2}$ is indeed monotonic. Another possible choice of coupling would be the normalization of the displacement operator $\Lambda$, but this is less convenient for our purposes.
} expansion parameter, which we define to be
\begin{equation}
\label{eq:defVarsig}
    \varsigma \equiv 4 - \Delta_{z^2}\ ,
\end{equation}
where $z^2$ is the lowest dimension two-particle operator. At large $\lambda$, 
\begin{align}\label{eq:varsigma in terms of 1/lambdaSq}
    \varsigma=\frac{2}{\pi\lambda^2}+O(1/\lambda^4).
\end{align}
The proportionality constant $2/\pi$ will be fixed in eq.~\eqref{eq:varsigma in terms of 1/lambdaSq 2} by extracting the leading anomalous dimension of $z^2$ from the $4$-point function. 

\subsection{Effective string action expanded}

To read off the properties of the branon, we expand the effective action in eq.~\eqref{eq:effective string action - NG + HC} in $z$:
\begin{equation}\label{eq:effective action expanded}
\begin{aligned}
    S[z]&=\int d^2\sigma\sqrt{h}\bigg[\ell_s^{-2}\left(1+\frac{1}{2}h^{\alpha\beta}\partial_\alpha z \partial_\beta z+\frac{z^2}{R^2}-\frac{1}{8}(h^{\alpha\beta}\partial_\alpha z \partial_\beta z)^2+\frac{1}{3}\frac{z^4}{R^4}+O(z^6)\right)\\&+\kappa\ell_s^2\left((\nabla_\alpha\nabla_\beta z \nabla^\alpha\nabla^\beta z)^2-4R^{-4}(h^{\alpha\beta}\partial_\alpha z \partial_\beta z)^2+\frac{20}{3}R^{-8}z^4+O(z^6)\right)+\ldots\bigg]
\end{aligned}
\end{equation}
Here, $h_{\alpha\beta}\equiv \frac{1}{x^2}\delta_{\alpha\beta}$ is the AdS$_2$ metric. It is convenient to rescale $(\tau,x)\to R(\tau,x)$, $z\to Rz/\lambda$. Then the action becomes
\begin{align}\label{eq:effective string action expanded}
    S[z]&=\int d^2\sigma\sqrt{h}\bigg[\lambda^2+\frac{1}{2}h^{\alpha\beta}\partial_\alpha z \partial_\beta z+z^2-\frac{1}{8\lambda^2}(h^{\alpha\beta}\partial_\alpha z \partial_\beta z)^2+\frac{1}{3\lambda^2}z^4\\&+\frac{\kappa}{\lambda^6}\left((\nabla_\alpha\nabla_\beta z \nabla^\alpha\nabla^\beta z)^2-4(h^{\alpha\beta}\partial_\alpha z \partial_\beta z)^2+\frac{20}{3}z^4+O(z^6)\right)+O(z^6\lambda^{-4},z^4\lambda^{-8})\bigg]\nonumber
\end{align}
This makes it clear that $z$ is a scalar of mass $m_z^2=2$ on AdS$_2$ with a tower of interactions suppressed by powers of $1/\lambda^2$. The non-universal higher-curvature term contributes to the four-point function at order $1/\lambda^6$, which is the same order as diagrams containing three Nambu-Goto quartic vertices and two loops.

The main ingredients for computing Witten diagrams are the bulk-to-bulk and bulk-to-boundary propagators. We write the bulk-to-bulk propagator as $G^{\rm BB}_{\Delta}(u)$, where $u=\frac{x_{12}^2+\tau_{12}^2}{2x_1x_2}$ is the squared chordal distance between the bulk points in embedding space. It obeys:
\begin{align}
    \left[\square-m_z^2\right]G^{\rm BB}_\Delta(u)=-\frac{1}{\sqrt{h}}\delta(\tau_1-\tau_2)\delta(x_1-x_2),
\end{align}
where $m^2=\Delta(\Delta-1)$. Its explicit solution is:
\begin{align}
    G^{\rm BB}_{\Delta}(u)&=\frac{\mathcal{C}_{\Delta}}{2^\Delta}\frac{1}{(1+u)^\Delta}{_2F_1}(\frac{\Delta}{2},\frac{\Delta}{2}+\frac{1}{2},\Delta+\frac{1}{2},\frac{1}{(1+u)^2}),
\end{align}
where 
\begin{align}
    \mathcal{C}_\Delta&=\frac{\Gamma(\Delta)}{2\sqrt{\pi}\Gamma(\Delta+\frac{1}{2})}.
\end{align}
Pushing one of the bulk points to the boundary defines the boundary-to-bulk propagator:
\begin{align}
    K_\Delta(x_1,\tau_1,\tau_2)&=\lim_{x_2\to 0}{x_2}^{-\Delta}G_\Delta(u)=\mathcal{C}_\Delta \tilde{K}_\Delta(x_1,\tau_1,\tau_2), 
\end{align}
where
\begin{align}
    \tilde{K}_\Delta(x_1,\tau_1,\tau_2)\equiv \left(\frac{x_1}{x_1^2+(\tau_1-\tau_2)^2}\right)^\Delta.
\end{align}
Pushing the second point to the boundary defines the boundary-to-boundary propagator. 

For the leading contribution to the branon two-point function, setting $\Delta=2$, we have
\begin{align}\label{eq:ertyy54e}
    \braket{z(\tau_1)z(\tau_2)}=Z^2\lim_{x_1,x_2\to 0}x_1^{-2}x_2^{-2} [G^{\rm BB}_2(u)+O(1/\lambda^2)]=\frac{Z^2\mathcal{C}_2}{\tau_{12}^4}\left(1+O(1/\lambda^2))\right).
\end{align}
Fixing the convention in eq.~\eqref{eq:z 2pt normalization} gives 
\begin{align}
    Z=1/\sqrt{\mathcal{C}_2}+O(1/\lambda^2).
\end{align}

\subsection{Free correlator}

In the free theory (at zeroth order in $\varsigma$), the four-point function consists of the sum of three ways of contracting the four operators with boundary-to-boundary propagators, giving:
\begin{align}
    \braket{z(\tau_1)z(\tau_2)z(\tau_3)z(\tau_4)}\bigg\rvert_{\varsigma^0}=\frac{1}{\tau_{12}^4\tau_{34}^4}+\frac{1}{\tau_{13}^4\tau_{24}^4}+\frac{1}{\tau_{14}^4\tau_{23}^4}.
\end{align}
The reduced four-point function at zeroth order is therefore:
\begin{align}\label{eq:free correlator}
    1+G^{(0)}(\chi)&=1+\chi^4+\frac{\chi^4}{(1-\chi)^4}.
\end{align}

\subsection{Tree-level correlator: four-point contact diagrams}

Next, we compute the order $\varsigma$ (or $1/\lambda^2$) correction. The nontrivial contribution arises from four-point contact diagrams generated by the quartic interactions in the first line of eq.~\eqref{eq:effective string action expanded}. In principle, bubble corrections to the boundary-to-boundary propagator are also present; however, because we expect that the branon dimension is protected \cite{Gabai:2025hwf}, the effect of the bubble diagrams is simply absorbed into the normalization $Z$.

Only one linear combination of the no-derivative and four-derivative quartic interactions appears at order $1/\lambda^2$ in the effective action \eqref{eq:effective string action expanded}. However, it will be useful for us -- in particular, for the analytic bootstrap analysis in Section~\ref{sec:ansatz bootstrap} -- to consider the following three types of quartic vertices, with different numbers of vertices, separately:\footnote{The vertices in \eqref{eq:three interaction vertices} are the only quartic vertices that appear in \eqref{eq:effective string action expanded} up to order $1/\lambda^6$, and there are no other independent vertices with eight or fewer derivatives (up to equations of motion and integration by parts). We will essentially prove this in section~\ref{sec:bootstrap tree-level} when we show that there are only three independent solutions to the tree-level analytic bootstrap if we specify a certain bounded Regge behavior that corresponds to bounding the number of derivatives in the interaction vertex.}
\begin{align}
    \mathcal{V}_{0-\text{der}}&=z^4,&\mathcal{V}_{4-\text{der}}&=(h^{\alpha\beta}\partial_\alpha z \partial_\beta z)^2,&\mathcal{V}_{8-\text{der}}&=(h^{\alpha\gamma}h^{\beta\delta}\nabla_\alpha\nabla_\beta z \nabla_\gamma \nabla_\delta z)^2.\label{eq:three interaction vertices}
\end{align}
We then define the corresponding contact diagrams:
\begin{align}
    X_{n-\text{der}}(\vec{\tau})&=\braket{z(\tau_1)z(\tau_2)z(\tau_3)z(\tau_4)\mathcal{V}_{n-\text{der}}}_{\rm connected}
\end{align}
Here, $n=0,4,8$ and we use the shorthand $\vec{\tau}\equiv (\tau_1,\tau_2,\tau_3,\tau_4)$. The explicit expressions for these contact diagrams are:
\begin{align}
    X_{0-\text{der}}(\vec{\tau})&=24\mathcal{C}_2^4 \int \frac{d\tau dx}{x^2} \tilde{K}_2(x,\tau,\tau_1)\tilde{K}_2(x,\tau,\tau_2)\tilde{K}_2(x,\tau,\tau_3)\tilde{K}_2(x,\tau,\tau_4),\label{eq:0 deriv contact diagram}\\
    X_{4-\text{der}}(\vec{\tau})&=X_{4-\text{der}}^{\rm S}(\tau_1,\tau_2;\tau_3,\tau_4)+\tau_2\leftrightarrow \tau_3+\tau_2\leftrightarrow \tau_4,\label{eq:4 deriv contact diagram}\\
    X_{8-\text{der}}(\vec{\tau})&=X_{8-\text{der}}^{\rm S}(\tau_1,\tau_2;\tau_3,\tau_4)+\tau_2\leftrightarrow \tau_3+\tau_2\leftrightarrow \tau_4.\label{eq:8 deriv contact diagram}
\end{align}
The four-derivative and eight-derivative contact diagrams split into different channels, with the $s$-channels given by:
\begin{align}
    X_{4-\text{der}}^{\rm S}(\tau_1,\tau_2;\tau_3,\tau_4)&=8\mathcal{C}_2^4 \int \frac{d\tau dx}{x^2} [h^{\alpha\beta}\partial_\alpha \tilde{K}_2(x,\tau,\tau_1)\partial_\beta \tilde{K}_2(x,\tau,\tau_2)]\nonumber\\&\qquad\qquad\qquad\times[h^{\gamma\delta}\partial_\gamma \tilde{K}_2(x,\tau,\tau_3)\partial_\delta \tilde{K}_2(x,\tau,\tau_4)],\label{eq:4 deriv contact diagram S channel}\\
    X_{8-\text{der}}^{\rm S}(\tau_1,\tau_2;\tau_3,\tau_4)&=8\mathcal{C}_2^4\int \frac{d\tau dx}{x^2}[h^{\alpha \gamma}h^{\beta \delta}\nabla_\alpha \nabla_\beta \tilde{K}_2(x,\tau,\tau_1) \nabla_\gamma \nabla_\delta \tilde{K}_2(x,\tau,\tau_2)]\nonumber\\&\qquad\qquad\qquad\times[h^{\kappa \rho}h^{\lambda \sigma}\nabla_\kappa \nabla_\lambda \tilde{K}_2(x,\tau,\tau_3)\nabla_\rho \nabla_\sigma \tilde{K}_2(x,\tau,\tau_4)].\label{eq:8 deriv contact diagram S channel}
\end{align}
Next, we define the following linear combinations of the three contact diagrams:
\begin{align}
    \frac{1}{\tau_{12}^4\tau_{34}^4}\xi_{z^4}(\chi) &\equiv -\frac{27\pi^3}{20} X_{0-\text{der}}(\vec{\tau}),\label{eq:z-to-the-fourth contact diagram def}\\
    \frac{1}{\tau_{12}^4\tau_{34}^4}\xi_{\rm NG}(\chi) &\equiv -\frac{9\pi^3}{8}\left[\frac{1}{8}X_{4-\text{der}}(\vec{\tau})-\frac{1}{3}X_{0-\text{der}}(\vec{\tau})\right],\label{eq:nambu-goto contact diagram def}\\
    \frac{1}{\tau_{12}^4\tau_{34}^4}\xi_{\rm HC}(\chi) &\equiv \frac{\pi^3}{160}\left[-X_{8-\text{der}}(\vec{x})+4X_{4-\text{der}}(\vec{x})-\frac{20}{3}X_{0-\text{der}}(\vec{x})\right].\label{eq:higher curvature contact diagram def}
\end{align}
Here, `NG' stands for Nambu-Goto and `HC' stands for higher-curvature, and $\xi_{\bullet}(\chi)$ is conformally invariant. Up to overall factors, these correspond to the hypothetical contributions to the four-point function that would arise from a $z^4$ interaction vertex, from the pair of vertices in the first line of eq.~\eqref{eq:effective action expanded} originating from the Nambu-Goto action, and from the three vertices in the second line of eq.~\eqref{eq:effective action expanded} originating from the higher curvature term in the action. The numerical factors in the definitions of the $\xi_\bullet$ were chosen, with the advantage of hindsight, so that in the OPE limit $\chi\to0$ we have $\xi_\bullet(\chi)\big\rvert_{\log|\chi|}=2\chi^4+O(\chi^5)$. We will see momentarily that this is a convenient normalization.

The contact diagrams can be computed using the techniques of \cite{Giombi:2017cqn}; we provide the details in Appendix~\ref{app:Witten diagrams and D functions}. The result for the $z^4$ contact
diagram is:
\begin{equation}\label{eq:z-to-the-fourth contact diagram final expression}
\begin{aligned}
    \xi_{z^4}(\chi)&=\frac{4 \left(\chi
   ^2-\chi +1\right) \chi ^2}{5 (\chi -1)^2}-\frac{2 \left(2 \chi
   ^2-5 \chi +5\right) \chi ^4 \log | \chi |}{5 (\chi -1)^3}\\&\qquad\qquad+\frac{2}{5} \left(2 \chi ^2+\chi +2\right) \chi  \log | 1-\chi |,
\end{aligned}
\end{equation}
the result for the Nambu-Goto contact diagram is:
\begin{equation}\label{eq:xi NG}
\begin{aligned}
    \xi_{\rm NG}(\chi)&=\frac{12 \chi ^8-48 \chi ^7+73 \chi ^6-51 \chi ^5+40 \chi ^4-51 \chi
   ^3+73 \chi ^2-48 \chi +12}{6 (\chi -1)^4}\\&\qquad +\frac{\left(-2 \chi
   ^6+11 \chi ^5-25 \chi ^4+30 \chi ^3-20 \chi ^2+6 \chi -2\right) \chi ^4 \log | \chi |
   )}{(\chi -1)^5}\\&\qquad +\left(2 \chi ^5-\chi ^4+\frac{2}{\chi }-1\right) \log | 1-\chi |,
\end{aligned}
\end{equation}
and the result for the higher curvature diagram is:
\begin{equation}\label{eq:higher curvature contact diagram final expression}
\begin{aligned}
    \xi_{\rm HC}(\chi)&=\frac{1}{3 (\chi -1)^6 \chi ^2}\bigg[60 \chi ^{14}-420 \chi
   ^{13}+1271 \chi ^{12}-2166 \chi ^{11}+2265 \chi ^{10}\\&\qquad -1480 \chi ^9+607 \chi ^8 -214
   \chi ^7+607 \chi ^6\\&\qquad-1480 \chi ^5+2265 \chi ^4-2166 \chi ^3+1271 \chi ^2-420 \chi
   +60\bigg]\\&\qquad +\frac{\chi^4}{(1-\chi)^7}\bigg[20 \chi ^{10}-170 \chi ^9+642
   \chi ^8-1415 \chi ^7+2009 \chi ^6-1911 \chi ^5\\&\qquad +1225 \chi ^4-516 \chi ^3+144 \chi ^2-10
   \chi +2 \bigg]\log | \chi |\\&\qquad +\frac{\left(60 \chi ^{10}-90 \chi ^9+36 \chi ^8-3 \chi ^7-3 \chi ^3+36 \chi ^2-90 \chi
   +60\right) \log | 1-\chi |}{3 \chi ^3}.
\end{aligned}
\end{equation}

Given all of our preparatory work, we can now simply write down the tree-level contribution to the branon four-point function:
\begin{align}\label{eq:45ty54}
    \varsigma G^{(1)}(\chi)&=\frac{Z^4\tau_{12}^4\tau_{34}^4}{\lambda^2}\left[\frac{1}{8}X_{4-\text{der}}(\vec{\tau})-\frac{1}{3}X_{0-\text{der}}(\vec{\tau})\right]=-\frac{2}{\pi\lambda^2}\xi_{\rm NG}(\chi).
\end{align}

The final step is to relate $\varsigma$ to $1/\lambda^2$ by computing the anomalous dimension of $z^2$. Although we will extract OPE data from the four-point function more systematically later, here we adopt the minimal approach sufficient for this purpose. Let us write $\Delta_{z^2}=4+\varsigma \gamma_{z^2}^{(1)}+O(\varsigma^2)$ and $a_{zzz^2}\equiv c_{zzz^2}^2=a_{zzz^2}^{(0)}+O(\varsigma)$ for the dimension and squared OPE coefficient of $z^2$, respectively. Our choice of coupling in eq.~\eqref{eq:defVarsig} is equivalent to setting $\gamma_{z^2}^{(1)}=-1$ and higher anomalous dimensions to zero. Meanwhile, the leading order squared OPE coefficient is $a_{zzz^2}^{(0)}=2$. Finally, the anomalous dimension of $z^2$ appears in the tree-level four-point function as the coefficient of the leading term in the small $\chi$ expansion of the prefactor of $\log|\chi|$: 
\begin{align}\label{eq:varsigma in terms of 1/lambdaSq 2}
    \varsigma G^{(1)}\rvert_{\log|\chi|}=\varsigma a_{zzz^2}^{(0)}\gamma_{z^2}^{(1)}\chi^4+O(\chi^5)=-2\varsigma\chi^4+O(\chi^5).
\end{align}
Using eq.~\eqref{eq:45ty54} and recalling $\xi_{\rm NG}(\chi)\rvert_{\log|\chi|}=2\chi^4+O(\chi^5)$ leads to $\varsigma=\frac{2}{\pi \lambda^2}+O(1/\lambda^4)$, as anticipated in eq.~\eqref{eq:varsigma in terms of 1/lambdaSq}.

Our final result for the tree-level correlator is therefore
\begin{align}\label{eq:tree level four point function}
    G^{(1)}(\chi)=-\xi_{\rm NG}(\chi).
\end{align}

\subsection{Comments on mixing at tree-level}\label{sec:comments on mixing at tree-level}

The conformal block expansion is often the most systematic and efficient tool for extracting OPE data from the four-point function and the one that we will use in section~\ref{sec:ansatz bootstrap}. However, a priori, one might think that a naive implementation is complicated by operator and OPE data mixing. In Appendix~\ref{app:unmixing at tree level}, we treat the mixing problem in detail (see also \cite{Ferrero:2023gnu} for an extensive discussion of mixing in the context of the supersymmetric Wilson line in $\mathcal{N}=4$ SYM), but the final conclusion is that we mostly do not need to worry about mixing in the present analysis. We summarize the main points below, and refer the reader to the Appendix for details.

First, we should briefly clarify what we mean by the related concepts of ``operator mixing'' and ``OPE data mixing.'' ``Operator mixing'' is the phenomenon in perturbation theory where, when there are degeneracies in the free theory, the good basis of primaries at a given order involves linear combinations of the degenerate primaries in the free theory. This is the CFT analog of the usual state mixing that occurs in degenerate perturbation theory in quantum mechanics. Meanwhile, ``OPE data mixing'' refers to the phenomenon where, if one expands a four-point function first as a sum over conformal blocks, and then expands the conformal blocks and their coefficients in a perturbative expansion, then the OPE coefficients and anomalous dimensions of operators that are degenerate in the free theory always appear together in the expansion. This means that one cannot in general determine the individual OPE data of such operators from a single four point function.

In Appendix~\ref{app:unmixing at tree level}, we show that:
\begin{itemize}
    \item In the free theory, primaries are composite operators constructed out of the branon field $z$. We may choose a basis such that a given primary can be written as a linear combination of terms with $n$ derivatives acting on $k$ copies of $z$, which we denote as $[z^k]_n^0$. We call this a $k$-particle operator; it has dimension $2k+n$. There is one two-particle primary $[z^2]_n^0$ for each non-negative even $n$, while multi-particle primaries $[z^k]_{n,A}^0$ with $k\geq 3$ are in general degenerate and come with an extra label $A$. 
    \item The tree-level four point function lets us compute the tree-level anomalous dimensions and identify the good basis of primaries in the free limit. We find that the good basis only mixes free theory primaries that have the same number of particles. In particular, we can unambiguously identify the two-particle primaries in the interacting theory, $[z^2]_n$, as those which satisfy $[z^2]_n=[z^2]_n^0+O(\varsigma)$. All other primaries reduce to multiparticle primaries, with no overlap with the two-particle primaries, in the limit $\varsigma\to 0$. These properties hold for any weakly coupled scalar in AdS${_2}$ with  quartic interaction.
    \item One important consequence of the previous bullet is that only the two-particle primaries have a non-zero three-point function with two copies of $z$ in the limit $\varsigma\to 0$. Namely, the two-particle and multi-particle operators have the following squared OPE coefficients:
    \begin{align}\label{eq:squared OPE coefficient two- and multi-particle}
        c_{zz[z^2]_n}^2=&O(\varsigma^0),&c_{zz[z^k]_{n,A}}^2&=O(\varsigma^2),\quad k\geq 3.
    \end{align}
    This result is important because it means that we can extract the OPE data of the two-particle operators from the four-point function to relatively high order in perturbation theory by proceeding naively with the conformal block expansion.
    \item Finally, in the appendix we explicitly compute the anomalous dimensions of all primaries with dimension less than or equal to $12$. We find that the degeneracies in the free theory between primaries with different numbers of particles are lifted, while primaries with the same number of particles remain degenerate. Moreover, the tree-level anomalous dimension $\gamma^{(1)}$ of a generic operator with $k$ particles and free theory dimension $\Delta^{(0)}$ takes the simple form:
    \begin{align}\label{eq:general formula for tree-level anomalous dimensions summary}
        \gamma^{(1)}(\Delta^{(0)},k)=-\frac{c_{\Delta^{(0)}}}{8}+\frac{k}{4},
    \end{align}
    where $c_\Delta\equiv \Delta(\Delta-1)$ is the conformal Casimir. We conjecture that this formula holds for all operators in the theory. This result, and the tree-level degeneracy of operators with the same number of particles, appears to be special to the Nambu-Goto action.

    We also computed the free and tree-level OPE coefficients of some primaries; see eqs.~\eqref{eq:OPE coefficients zeroth order} and \eqref{eq:OPE coefficients direct method}.
\end{itemize}
These observations will be useful for the next section.

\FloatBarrier

\section{Four-point function via the ansatz bootstrap}\label{sec:ansatz bootstrap}

In this section, we resume the computation of the branon four point function. We will first reproduce the tree-level result from section~\ref{sec:tree-level four point function via Witten diagrammatics} and then extend it up to two-loops using the transcendentality ansatz bootstrap. The basic idea of this bootstrap method is to write an ansatz for the correlator using generalized logarithms whose maximum transcendentality is set by the perturbative order, and then impose crossing symmetry and other constraints. This method has been used to compute, to multiple loops, flat space amplitudes in $\mathcal{N}=4$ SYM (see the review in \cite{Caron-Huot:2020bkp}), supergravity correlators in AdS$_5\times S^5$ (see, e.g., \cite{Aharony:2016dwx,Aprile:2017bgs,Alday:2017xua,Huang:2021xws,Drummond:2022dxw}), and correlators in the Wilson-Fisher CFT \cite{Alday:2017zzv,Henriksson:2018myn}. It was developed in the context of general 1d CFTs corresponding to effective field theories in AdS$_2$ in \cite{Ferrero:2019luz} and very successfully applied to the half-BPS Wilson line in $\mathcal{N}=4$ SYM in the supergravity regime in \cite{Liendo:2018ukf,Ferrero:2021bsb,Ferrero:2023znz,Ferrero:2023gnu}. (For related work on the analytic bootstrap of line defects in 2d and 3d superconformal theories, see \cite{Bianchi:2020hsz,Pozzi:2024xnu,Bliard:2024bcz}.) Our computation of the two-loop branon four-point function for the effective string in AdS$_3$ is most similar to \cite{Ferrero:2021bsb,Ferrero:2023znz,Ferrero:2023gnu}. Our main results are summarized in section \ref{sec:summary of select OPE data}.

It should be noted that the ansatz-based approach we adopt in this section is not fully rigorous because it could, in principle, miss solutions to the analytic bootstrap constraints presented below. To ensure that this does not occur, in Appendix~\ref{app:dispMethod} we verify our results for the four-point function using an alternative analytic bootstrap approach based on the Lorentzian inversion formula for 1d CFTs \cite{Mazac:2018qmi,Bonomi:2024lky,Carmi:2024tmp}. This approach is technically more involved than the ansatz bootstrap and we were able to implement it analytically only up to one loop. However, we have used it to also numerically verify our two-loops results.

\subsection{The ansatz bootstrap}
\label{sec:ansatzBootEnum}
We begin by reviewing the transcendentality bootstrap method.
The conformal block expansion of $G(\chi)$, expanded perturbatively in $\varsigma$, implies that $G^{(L)}(\chi)$ (the $L$-th order term in the perturbative expansion in \eqref{eq:pertExpG}) can be expressed as a sum of functions multiplying powers of $\log(\chi)$, with the highest power being $L$. Namely:
\begin{align}
    G^{(1)}(\chi)&=G^{(1)}_{\log^0}(\chi)+G^{(1)}_{\log^1}(\chi)\log{\chi},\label{eq:G(1) expanded in log terms}\\
    G^{(2)}(\chi)&=G^{(2)}_{\log^0}(\chi)+G^{(2)}_{\log^1}(\chi)\log{\chi}+G^{(2)}_{\log^2}(\chi)\log^2{\chi},\label{eq:G(2) expanded in log terms}\\
    G^{(3)}(\chi)&=G^{(3)}_{\log^0}(\chi)+G^{(3)}_{\log^1}(\chi)\log{\chi}+G^{(3)}_{\log^2}(\chi)\log^2{\chi}+G^{(3)}_{\log^3}(\chi)\log^3{\chi}, \label{eq:G(3) expanded in log terms}
\end{align}
and so on. Each $G^{(L)}_{\log^k}(\chi)$ is analytic around $\chi=0$, and can {only} have poles or branch points at $\chi=1,\infty$. The ansatz bootstrap approach at order $L$ consists of the following steps:
\begin{enumerate}
    \item Write an ansatz for $G^{(L)}_{\log^k}(\chi)$, for each $k=1,\ldots,L$, 
    as a linear combination of generalized logarithms with coefficients that are rational functions of $\chi$ with poles at most at $\chi=0,1,\infty$.
    \item Fix the ``highest log'' terms $G^{(L)}_{\log^k}$ for $k=2,\ldots,L$ using the conformal block expansion and lower order OPE data.
    \item Impose the ``cyclic'' and ``braiding'' permutation symmetries of the correlator.\footnote{These follow from $s$--$t$ and $s$--$u$ crossing, respectively. While $s$--$u$ crossing is not a generic symmetry of one-dimensional correlators, we prove that an appropriate version holds in this case.}
    \item Impose correct behavior for $G(\chi)$ in the OPE limit $\chi\to 0$.
    \item Fix remaining undetermined coefficients by requiring that:
    \begin{enumerate}
        \item $\gamma_{z^2}^{(L)}=-\delta_{L1}$, which defines our coupling $\varsigma$,
        \item the branon four-point function satisfies certain universal integral constraints due to non-linearly realized isometries, and
        \item the flat-space limit of the AdS correlator reproduces the branon S-matrix.
    \end{enumerate}
    \item Having completely fixed the correlator, read off the anomalous dimensions and OPE coefficients at order $L$ from the functions $G^{(L)}_{\log^1}$ and $G^{(L)}_{\log^0}$.
    \item Repeat the process at order $L+1$.
\end{enumerate}

Ultimately, there are various reasons why this procedure eventually breaks down. First, the number of undetermined coefficients in step $5$ increases at higher orders, and there are eventually not enough physical constraints to fix all of them. Second, it is possible that at some order the space of functions chosen for step 1 (which we specify below) is no longer sufficient. What obstructs us in practice is a third complication: starting at a certain order, multiple operators contribute to most OPE channels, and it becomes impossible to disentangle their conformal data from the displacement four-point function alone. We commented on this phenomenon, which we call OPE data mixing, at the end of the previous section and discuss it in more detail in Appendix~\ref{app:unmixing at tree level}. When OPE data mixing occurs, steps~2 and~6 cannot be implemented without substantial additional effort. For this reason, in the present work we will be satisfied with bootstrapping the correlator up to two-loops.

We now discuss each of these steps in some more detail.

\paragraph{The ansatz.} The transcendentality ansatz is an assumption about the possible functions that can appear in the various prefactors, $G^{(L)}_{\log^n}(\chi)$ in \eqref{eq:G(1) expanded in log terms}-\eqref{eq:G(3) expanded in log terms}. Concretely, we will assume -- in analogy with the analysis of the correlator on the half-BPS Wilson line in $\mathcal{N}=4$ SYM \cite{Ferrero:2021bsb,Ferrero:2023gnu,Ferrero:2023znz} -- that the allowed functions are products of rational functions and the following (poly)logarithms:
\begin{align}
    \text{Transcendentality}\; 0&:\quad 1,\\
    \text{Transcendentality}\; 1&:\quad\log(1-\chi),\\
    \text{Transcendentality}\; 2&:\quad\log^2(1-\chi),\quad\polylog{2}(\chi),\\
    \text{Transcendentality}\; 3&:\quad\log^3(1-\chi),\quad \log(1-\chi)\polylog{2}(\chi),\quad \polylog{3}(\chi),\quad S_{1,2}(\chi).
\end{align}
Here, $\polylog{n}(\chi)$ is the standard polylogarithm, and $S_{1,2}(\chi)$ is a Nielsen polylogarithm.

{A further important assumption of the transcendentality ansatz is that the total transcendentality of $G^{(L)}(\chi)$ does not exceed $L$.  In particular, the coefficient $G^{(L)}_{\log^k}(\chi)$ may contain functions of transcendentality up to $L-k$.}

This ansatz was motivated in \cite{Ferrero:2019luz,Ferrero:2023gnu}. We will simply take it as an assumption of our bootstrap analysis, justifying it only a posteriori by comparison with the approach based on the Lorentzian inversion formula that is reviewed in appendix~\ref{app:dispMethod}.

\paragraph{Conformal block expansion in perturbation theory.} Much of the heavy lifting in the ansatz bootstrap is done by the conformal block expansion of the reduced correlator, which takes the form:
\begin{align}\label{eq:5tyhr4w}
    G(\chi)=\sum_O a_O \CB_{\Delta_O}(\chi),
\end{align}
where $O$ is the exchanged primary with dimension $\Delta_O$ and squared OPE coefficient $a_O\equiv c_{zzO}^2$. Because the external operators are identical, the operators that appear in the conformal block expansion have $++$ parities, which includes the tower of two-particle operators $[z^2]_{2n}$, $n=0,1,2,\ldots$, as well as multi-particle operators. The conformal block for a 1d CFT is:
\begin{align}
\label{eq:ConfBlocks}
    \CB_{\Delta}(\chi)\equiv \chi^\Delta F_{\Delta}(\chi),\qquad F_\Delta(\chi)\equiv {_2F_1}(\Delta,\Delta,2\Delta,\chi).
\end{align}
It will also be convenient to introduce the following notation for the $n$th derivative of the conformal block with respect to the dimension:
\begin{align}
    \CB^{(n)}_{\Delta}\equiv \chi^\Delta \partial_\Delta^n(\chi^{-\Delta} \CB_{\Delta})=\chi^\Delta \partial_\Delta^n F_\Delta(\chi).
\end{align}
Next, we expand the conformal dimensions and squared OPE coefficients in powers of $\varsigma$:
\begin{align}
    \Delta_O&=\Delta_O^{(0)}+\varsigma \gamma_O^{(1)}+\varsigma^2 \gamma_O^{(2)}+\varsigma^3 \gamma_O^{(3)}+\ldots,\label{eq:conformal dimensions perturbative expansion}\\
    a_O&=a_O^{(0)}+\varsigma a_O^{(1)}+\varsigma^2 a_O^{(2)}+\varsigma^3 a_O^{(3)}+\ldots.\label{eq:OPE coefficients perturbative expansion}
\end{align}
Because they play an especially important role in what follows, we also introduce abbreviated notation for the two-particle operators:
\begin{align}
    \hat{\Delta}_n^{(0)}\equiv \Delta^{(0)}_{[z^2]_{2n}}\equiv 4+2n, \quad \hat{\gamma}^{(i)}_n \equiv \gamma^{(i)}_{[z^2]_{2n}},\quad \hat{a}^{(i)}_n\equiv a_{[z^2]_{2n}}^{(i)},
\end{align}
where now $n=0,1,2,\ldots$.

The OPE data of operators that are degenerate in the GFF limit appear together in the perturbative conformal block expansion. Therefore, it is convenient to introduce notation for sums of OPE data over all operators that have the same dimension {as the $n$th two-particle operator}
in the GFF limit:
\begin{align}\label{eq:notation for mixed OPE data}
    \braket{a^{(i)}\gamma^{(j_1)}\gamma^{(j_2)}\ldots }_n\equiv \sum_{O:\Delta_O^{(0)}=\hat{\Delta}_n^{(0)}} a_O^{(i)}\gamma^{(j_1)}_O\gamma_O^{(j_2)}\ldots.
\end{align}

We now substitute eqs.~\eqref{eq:conformal dimensions perturbative expansion}-\eqref{eq:OPE coefficients perturbative expansion} into eq.~\eqref{eq:5tyhr4w}, expand up to order $O(\varsigma^3)$, and collect terms. This lets us express the functions $G^{(L)}_{\log^k}(\chi)$ in eqs.~\eqref{eq:G(1) expanded in log terms}-\eqref{eq:G(3) expanded in log terms} as sums over exchanged operators, with the mixed OPE data in eq.~\eqref{eq:notation for mixed OPE data} appearing as the expansion coefficients. Up to two-loops, we have:
\begin{align}
    G^{(0)}(\chi)&=\sum_n \langle a^{(0)}\rangle_n\CB_{\Delta_n^{(0)}}(\chi),\label{eq:CB expansion 1}\\
    G^{(1)}_{\log^0}(\chi)&=\sum_n [\langle a^{(1)}\rangle_n\CB_{\Delta_n^{(0)}}(\chi)+\langle a^{(0)}\gamma^{(1)}\rangle_n \CB_{\Delta_n^{(0)}}^{(1)}(\chi)],\label{eq:CB expansion 2}\\
    G^{(1)}_{\log^1}(\chi)&=\sum_n \langle a^{(0)}\gamma^{(1)}\rangle_n \CB_{\Delta_n^{(0)}}(\chi),\label{eq:CB expansion 3}\\
    G^{(2)}_{\log^0}(\chi)&=\sum_n \big[\braket{a^{(2)}}_n\CB_{\Delta_n^{(0)}}(\chi)+(\langle a^{(1)}\gamma^{(1)}\rangle _n+\langle a^{(0)}\gamma^{(2)}\rangle_n)\CB_{\Delta_n^{(0)}}^{(1)}(\chi)\label{eq:CB expansion 4}\\&\qquad\qquad +\frac{1}{2}\langle a^{(0)}(\gamma^{(1)})^2\rangle_n\CB_{\Delta_n^{(0)}}^{(2)}(\chi)\big],\nonumber\\
    G^{(2)}_{\log^1}(\chi)&=\sum_n \big[(\langle a^{(1)}\gamma^{(1)}\rangle_n+\langle a^{(0)}\gamma^{(2)}\rangle_n)\CB_{\Delta_n^{(0)}}(\chi)+\langle a^{(0)}(\gamma^{(1)})^2\rangle_n\CB^{(1)}_{\Delta_n^{(0)}}(\chi)\big],\label{eq:CB expansion 5}\\
    G^{(2)}_{\log^2}(\chi)&=\sum_n \frac{1}{2}\langle a^{(0)}(\gamma^{(1)})^2\rangle_n \CB_{\Delta_n^{(0)}}(\chi),\label{eq:CB expansion 6}\\
    G^{(3)}_{\log^0}(\chi)&=\sum_n \big[\braket{a^{(3)}}_n\CB_{\Delta_n^{(0)}}(\chi)+(\braket{a^{(2)}\gamma^{(1)}}_n+\langle a^{(1)}\gamma^{(2)}\rangle_n+\langle a^{(0)}\gamma^{(3)}\rangle_n)\CB_{\Delta_n^{(0)}}^{(1)}(\chi)\label{eq:CB expansion 7}\\&\qquad \qquad +\big(\frac{1}{2}\langle a^{(1)}(\gamma^{(1)})^2\rangle_n+\langle a^{(0)}\gamma^{(1)}\gamma^{(2)}\rangle_n\big)\CB_{\Delta_n^{(0)}}^{(2)}(\chi)+\frac{1}{3!}\langle a^{(0)} (\gamma^{(1)})^3\rangle_n \CB_{\Delta_n^{(0)}}^{(3)}(\chi)\big],\nonumber\\
    G^{(3)}_{\log^1}(\chi)&=\sum_n \big[(\braket{a^{(2)}\gamma^{(1)}}_n+\langle a^{(1)}\gamma^{(2)}\rangle_n+\langle a^{(0)}\gamma^{(3)}\rangle_n)\CB_{\Delta_n^{(0)}}(\chi)\label{eq:CB expansion 8}\\&\qquad\qquad +(\langle a^{(1)}(\gamma^{(1)})^2\rangle_n+2\langle a^{(0)}\gamma^{(1)}\gamma^{(2)}\rangle_n)\CB_{\Delta_n^{(0)}}^{(1)}(\chi)+
    \frac{1}{2}\langle a^{(0)}(\gamma^{(1)})^3\rangle_n\CB_{\Delta_n^{(0)}}^{(2)}(\chi)\big],\nonumber\\
    G^{(3)}_{\log^2}(\chi)&=\sum_n \big[\big(\frac{1}{2}\langle a^{(1)}(\gamma^{(1)})^2\rangle_n+\langle a^{(0)}\gamma^{(1)}\gamma^{(2)}\rangle_n\big)\CB_{\Delta_n^{(0)}}(\chi)+\frac{1}{2}\langle a^{(0)} (\gamma^{(1)})^3\rangle_n \CB_{\Delta_n^{(0)}}^{(1)}(\chi)\big],\label{eq:CB expansion 9}\\
    G^{(3)}_{\log^3}(\chi)&=\sum_n \frac{1}{3!}\langle a^{(0)} (\gamma^{(1)})^3\rangle_n \CB_{\Delta_n^{(0)}}(\chi).\label{eq:CB expansion 10}
\end{align}

These series expansions will play a key role in what follows because they let us read off OPE data from different parts of the correlator, and vice versa, but the OPE data mixing (which we discuss in more detail in appendix \ref{app:unmixing at tree level}) makes this procedure complicated. The problem is that the sum over degenerate operators of a product of OPE quantities is not equal to the product of the sums. This complication is significantly mitigated by the observation that $a_O^{(0)}$ and $a^{(1)}_O$ are non-zero only for the two-particle operators $O=[z^2]_{2n}$ (see~\eqref{eq:squared OPE coefficient two- and multi-particle} or ~\eqref{eq:OPE coefficients direct method}). This implies that any mixed product of OPE data with either coefficient as a factor will simplify to the corresponding product for a single two-particle primary:
\begin{align}
    \braket{a^{(0)}\gamma^{(j_1)}\gamma^{(j_2)}\ldots }_n&=\hat{a}_n^{(0)}\hat{\gamma}_n^{(j_1)}\hat{\gamma}_n^{(j_2)}\ldots ,&\braket{a^{(1)}\gamma^{(j_1)}\gamma^{(j_2)}\ldots }_n&=\hat{a}_n^{(1)}\hat{\gamma}_n^{(j_1)}\hat{\gamma}_n^{(j_2)}\ldots .
\end{align}
Thus, the only coefficients in the conformal block expansion up to two loops that have mixing are $\braket{a^{(2)}}_n$ in $G^{(2)}_{\log^0}$, $\braket{a^{(2)}\gamma^{(1)}}_n$ in $G^{(3)}_{\log^1}$, and $\braket{a^{(3)}}_n$ and $\braket{a^{(2)}\gamma^{(1)}}_n$ in $G^{(3)}_{\log^0}$. In particular, the conformal block expansions  of the highest-log terms at both one loop and two loop involve only the OPE data of the two-particle operators, and we will therefore be able to analytically bootstrap the full correlator. On the other hand, once we have the correlator up to two loops, we will only be able to extract the following unmixed OPE data of the two-particle operators
\begin{align}\label{eq:accessible unmixed OPE data}
    \hat{a}^{(0)}_n,\quad\hat{a}^{(1)}_n,\quad \hat{\gamma}^{(1)}_n,\quad \hat{\gamma}^{(2)}_n,
\end{align}
as well as the mixed OPE data
\begin{align}\label{eq:accessible mixed OPE data}
    \braket{a^{(2)}}_n, \quad \braket{a^{(3)}}_n,\quad \braket{a^{(2)}\gamma^{(1)}}_n+\underbrace{\braket{a^{(0)}\gamma^{(3)}}_n}_{=\hat{a}^{(0)}_n \hat{\gamma}_n^{(3)}}.
\end{align}
This is equivalent to saying that we can determine the sum of squared OPE coefficients and the squared-OPE-coefficient-weighted average of the conformal dimensions for all the operators that are classically degenerate,
\begin{align}
    \braket{a}_n\qquad \text{and}\qquad \frac{\braket{a\Delta}_n}{\braket{a}_n},
\end{align}
up to $O(\varsigma^3)$. 

\paragraph{Cyclic and braiding symmetry.} Next, we comment on the analytic structure of the four-point function and derive the two crossing relations that we will need, which are given in eq.~\eqref{eq:cyclic symmetry} and eq.~\eqref{eq:braiding symmetry}. 

Following \cite{Mazac:2018qmi}, we begin by decomposing the reduced four-point function $G(\chi)$ into three pieces, based on the value of $\chi$:
\begin{align}\label{eq:G decomposed into three}
    G(\chi)&=\left\{\begin{array}{cc} G^{\chi <0}(\chi) & \chi<0, \\ G^{0<\chi<1}(\chi) & 0<\chi<1, \\ G^{\chi>1}(\chi) & \chi>1. \end{array}\right.
\end{align}
{Fixing a conformal frame in which the first, third, and fourth operators are located at $\tau=0,1,\infty$, the cross-ratio reduces to $\chi=\tau_2$, and the three cases correspond to the position of the second operator relative to the other three. In one dimension, unlike in higher dimensions, there is no continuous path that moves the second operator between any two of the three intervals without it becoming coincident with another operator. Consequently, although the three functions $G^{\chi<0}(\chi)$, $G^{0<\chi<1}(\chi)$, and $G^{\chi>1}(\chi)$ can each be analytically continued to the complex plane and are expected to be holomorphic away from possible singularities at $\chi=0,1,\infty$, they do not generically coincide after analytic continuation.}

However, because the four external operators are identical, we can derive crossing relations on $G$, some of which put constraints on the values of $G$ on a single interval and some of which relate the values of $G$ on different intervals. Recalling the definition of $G$, \begin{align}\label{eq:5yhgfd4}
    \braket{z(\tau_1)z(\tau_2)z(\tau_3)z(\tau_4)}&=\frac{1}{\tau_{12}^4\tau_{34}^4}G(\chi).
\end{align}
and using the permutation symmetry
\begin{align}\label{eq:ty765ew}
    \braket{z(\tau_1)z(\tau_2)z(\tau_3)z(\tau_4)}=\braket{z(\tau_{\pi(1)})z(\tau_{\pi(2)})z(\tau_{\pi(3)})z(\tau_{\pi(4)})}
\end{align}
for any $\pi\in S_4$, we can derive various crossing relations. For example, the permutations $\pi:\tau_1\leftrightarrow \tau_2$ and $\pi:\tau_2\leftrightarrow \tau_3$ imply $G(\chi)=G\big(\frac{\chi}{\chi-1}\big)$ and $G(\chi)=\chi^4 G(1/\chi)$. Restricting to $\chi<0$ for the first relation and to $\chi>1$ for the second relation, and using the decomposition in eq.~\eqref{eq:G decomposed into three}, yields:
\begin{align}\label{eq:G on different intervals from unit interval}
    G^{\chi<0}(\chi)&\overset{\chi<0}{=}G^{0<\chi<1}\left(\frac{\chi}{\chi-1}\right),&G^{\chi>1}(\chi)&\overset{\chi>1}{=}\chi^4 G^{0<\chi<1}\left(\frac{1}{\chi}\right).
\end{align}
Once we start implementing the analytic bootstrap, we will restrict to $0<\chi<1$. The two relations above are useful because they allow us to reconstruct the full correlator on the real line from the correlator on the interval $\chi\in(0,1)$, but they do not constrain $G^{0<\chi<1}(\chi)$ itself. 

To get a crossing relation that constrains $G^{0<\chi<1}(\chi)$, let the permutation in eq.~\eqref{eq:ty765ew} be given by $\pi:\tau_{i}\to \tau_{i+1\;(\text{mod}\;4)}$. This yields $(1-\chi)^4 G(\chi)= \chi^4 G(1-\chi)$. When restricted to $\chi\in(0,1)$, this puts a constraint on $G^{0<\chi<1}$:
\begin{align}\label{eq:cyclic symmetry}
    (1-\chi)^4 G^{0<\chi<1}(\chi)&= \chi^4 G^{0<\chi<1}(1-\chi),\qquad (\text{cyclic symmetry}).
\end{align}
 This is the first of two crossing relations that are needed for the ansatz bootstrap and, following \cite{Ferrero:2023gnu}, we call it the ``cyclic symmetry,'' after the permutation $\pi$ of external points from which it arises.

The second crossing relation we need arises from the following useful rule for relating $G^{\chi<0}(\chi)$ and $G^{0<\chi<1}(\chi)$ in perturbation theory \cite{Ferrero:2023gnu} (see also \cite{Giombi:2017cqn,Bianchi:2020hsz}):
\begin{align}\label{eq:perturbative analytic continuation}
    G^{\chi<0}(\chi)&\overset{\chi<0}{=} G^{0<\chi<1}(\chi)\bigg\rvert_{\log{\chi}\to \log{|\chi|}}.
\end{align}
On the right-hand side we first introduce absolute values inside the logarithms and then analytically continue to negative values of $\chi$. This rule can be derived from the conformal block expansion \cite{Ferrero:2023gnu,Bianchi:2020hsz}. Concretely, for $0<\chi<1$, the conformal block expansion of the four point function is
\begin{align}\label{eq:4rthy6}
    G^{0<\chi<1}(\chi)=\sum_O a_O \chi^{\Delta_O} F_{\Delta_O}(\chi),
\end{align}
where $a_O=c_{zzO}^2$. Meanwhile, for $\chi<0$, the conformal block expansion is
\begin{align}\label{eq:456yhbvcdrt}
    G^{\chi<0}(\chi)&=G^{0<\chi<1}\left(\frac{\chi}{\chi-1}\right)=\sum_O a_O (-\chi)^{\Delta_O} F_{\Delta_O}(\chi).
\end{align}
The first equality comes from eq.~\eqref{eq:G on different intervals from unit interval}, while the second comes from combining eq.~\eqref{eq:4rthy6} with the property of the hypergeometric function that ${_2F_1}(a,b,c,\frac{\chi}{\chi-1})=(1-\chi)^{-a}{_2F_1}(a,c-b,c,\chi)$. The last step is to expand eq.~\eqref{eq:4rthy6} and eq.~\eqref{eq:456yhbvcdrt} as series in $\varsigma$, using eq.~\eqref{eq:conformal dimensions perturbative expansion} and eq.~\eqref{eq:OPE coefficients perturbative expansion}. An important simplification is that all operators that have non-zero three-point function with two $z$'s have an even dimension in the free limit. This is because the operators must be even under both \textbf{R} and \textbf{CT}, which means the number of fields and derivatives of the operator in the free limit must both be even (see Table~\ref{tab:strong coupling}). Therefore:
\begin{align}
    \chi^{\Delta_O}&=\chi^{\Delta_O^{(0)}}e^{(\varsigma \gamma_O^{(1)}+\varsigma^2 \gamma_O^{(2)}+\ldots)\log{\chi}}, &(-\chi)^{\Delta_O}&=\chi^{\Delta_O^{(0)}}e^{(\varsigma \gamma_O^{(1)}+\varsigma^2 \gamma_O^{(2)}+\ldots)\log(-\chi)}.
\end{align}
Combining this with eq.~\eqref{eq:4rthy6} and eq.~\eqref{eq:456yhbvcdrt} yields eq.~\eqref{eq:perturbative analytic continuation}. 

Finally, combining eq.~\eqref{eq:G on different intervals from unit interval} with eq.~\eqref{eq:perturbative analytic continuation} gives us our second constraint on $G^{0<\chi<1}$:
\begin{align}\label{eq:braiding symmetry}
    G^{0<\chi<1}\left(\frac{\chi}{\chi-1}\right)&\overset{\chi<0}{=}G^{0<\chi<1}(\chi)\bigg\rvert_{\log{\chi}\to \log{|\chi|}},\qquad (\text{braiding symmetry}).
\end{align}
Again following \cite{Ferrero:2023gnu}, we call this ``braiding symmetry,'' after the permutation $\pi$ of external points from which it ultimately arises. We must emphasize that this property holds order by order in perturbation theory in the effective string limit, but does not hold non-perturbatively, nor in perturbation theory in the gauge theory limit of the flux tube in AdS, which was the focus of \cite{Gabai:2025hwf}.

\paragraph{OPE and flat space limits.} Two useful constraints on the correlator come from requiring that it have the correct behavior in the OPE and flat space (i.e., Regge) limits. The first limit is simple, while the second requires some explanation.

The lowest dimension operator that appears in the OPE of $z$ with $z$ is $z^2$, which in the free limit has dimension $4$. Thus all the functions multiplying the log terms in eq.~\eqref{eq:G(1) expanded in log terms}-\eqref{eq:G(3) expanded in log terms} should have series expansions around $\chi=0$ that obey
\begin{align}\label{eq:OPE constraint}
    G^{(L)}_{\log^n}(\chi)\overset{\chi\to 0}{=}O(\chi^4).
\end{align}

Meanwhile, the Regge limit of a four-point function in a one-dimensional CFT (which is equivalent to the late-time out-of-time-order correlator \cite{Maldacena:2015waa}) is defined by analytically continuing the cross-ratio from the interval $(0,1)$ to $i\infty$ in the complex plane. This limit is related to the flat-space $2\to2$ branon scattering phase through the relation \cite{Shenker:2014cwa,deBoer:2017xdk,Lam:2018pvp}:\footnote{See section 5.4 of \cite{Giombi:2023zte} for a derivation of eq.~\eqref{eq:flat-space limit of correlator}. Note that ~\eqref{eq:flat-space limit of correlator} relates the flat-space limit of AdS$_2$ correlators of operators with external dimensions that \textit{remain fixed} as $R_{\rm AdS}\to \infty$ to the scattering of \textit{massless} particles in $\mathbb{R}^{1,1}$.},\footnote{Some readers may be more familiar with the {relation between the flat-space phase shift and the} asymptotic behavior of anomalous dimensions of two-particle operators \cite{Paulos:2016fap},
\begin{align}\label{eq:flat-space limit of anomalous dimension}
2\delta(s)
=
-\pi
\lim_{\substack{n,R\to \infty\\ s \equiv \frac{4n^2}{R^2}\ \text{fixed}}}
\big[
\Delta_{[z^2]_n}
-
(2\Delta_z+n)
\big].
\end{align}
This formula assumes that at large enough dimensions, but not too large as to invalidate the perturbative expansion, the exchanged operator content is the same as in a generlized free field theory, up to small corrections to the conformal data. It can be shown to be the case in the theory at hand. By expanding the flat-space phase shift in powers of the string length as in eq.~\eqref{eq:ujmhgfs}, we translate eq.~\eqref{eq:flat-space limit of anomalous dimension} into a constraint on the large-$n$ behavior of the anomalous dimensions:
\begin{align}\label{eq:flat space limit of anomalous dimensions}
    \gamma^{(L)}_{[z^2]_n}\overset{n\to \infty}{\sim}={
\renewcommand{\arraystretch}{1.2}\left\{\begin{array}{cc} -\frac{1}{2}n^2& L=1 \\ O(n^3) & L=2 \\ -\frac{\pi^2\gamma_{\rm SM}}{96}n^6 & L=3\end{array}\right.}.
\end{align}
From the explicit expressions for the correlator that we derive below, we have confirmed that eq.~\eqref{eq:flat space limit of anomalous dimensions} and~\eqref{eq:flat space limit of correlator} are equivalent up to two-loops. We prefer to use~\eqref{eq:flat space limit of correlator} in the main text because it works directly with the correlator.
\label{fn:flat space phase-shift anomalous dimension formula}}
\begin{align}\label{eq:flat-space limit of correlator}
    \lim_{\substack{\ell_s/R\to 0,t\to \infty\\ \ell_s^2t/R^2:\text{ fixed}}}\left[\chi^{-4}G(\chi)\biggr\rvert_{\chi=\frac{1}{2}+i t}- \int_0^\infty dp dq\frac{(pq)^{3}}{\Gamma(4)^2} e^{-p-q}e^{2i\delta(s)}\biggr\rvert_{s=ip q t/R^2}\right]=0.
\end{align}
We will therefore use the terms Regge limit and flat-space limit interchangeably.

Through unitarity of the S-matrix, the relation in \eqref{eq:flat-space limit of correlator} is consistent with the expectation of Regge boundedness for the non-perturbative correlator \cite{Mazac:2018qmi},
\begin{align}\label{eq:Regge constraint 1}
    \chi^{-4}G(\chi)\bigr\rvert_{\chi=\frac{1}{2}+it}\overset{t\to \infty}{=}\text{bounded}.
\end{align}
In our case, we are computing a perturbative correlator order by order in $\ell_s^2/R^2$, and individual orders violate the bound. Their growth as $t\to \infty$ is dictated by eq.~\eqref{eq:flat-space limit of correlator} and the known form of the flat space scattering phase \cite{Chen:2018keo,EliasMiro:2019kyf,Guerrieri:2024ckc}\footnote{{The parameter $\gamma_{\rm SM}$ obeys the bound $\gamma_{\rm SM}\geq -1$ from the S-matrix bootstrap \cite{EliasMiro:2019kyf}. From lattice calculations of the spectrum of the flux tube in 3d $SU(N)$ Yang-Mills, it is estimated to take the values $\gamma_{\rm SM}\approx-0.3$ for $N=2$ \cite{Caristo:2021tbk}, $\gamma_{\rm SM}\approx 0$ for $N=3$ \cite{Guerrieri:2024ckc}, $\gamma_{\rm SM}\approx 0.2$ for $N=6$ \cite{Dubovsky:2014fma,Chen:2018keo}, and $\gamma_{\rm SM}\approx 0.3$ for $N=\infty$ \cite{Guerrieri:2024ckc}.}}
\begin{align}\label{eq:ujmhgfs}
    2\delta(s)&=\frac{1}{4}\ell_s^2 s+\frac{\gamma_{\rm SM}}{768}\ell_s^6s^3+O(\ell_s^8 s^4).
\end{align}
Explicitly, substituting eq.~\eqref{eq:ujmhgfs} into eq.~\eqref{eq:flat-space limit of correlator} and expanding both sides in $\ell_s^2t/R^2$ (and remembering that $\varsigma=\frac{2}{\pi\lambda^2}+O(1/\lambda^4)$ and $\lambda^2\equiv \frac{R^2}{\ell_s^2}$), we get,\footnote{In eq.~\eqref{eq:flat space limit of correlator} and eq.~\eqref{eq:flat space limit of anomalous dimensions}, $f(x)\overset{x\to x_0}\sim g(x)$ means $\lim_{x\to x_0}f(x)/g(x)=1$.}
\begin{align}\label{eq:flat space limit of correlator}
    \chi^{-4}G^{(L)}(\chi)\bigg\rvert_{\chi=\frac{1}{2}+i t}\overset{t\to \infty}{\sim}{
\renewcommand{\arraystretch}{1.2}\left\{\begin{array}{cc} 1 & L = 0\\ -2\pi t & L=1 \\ \frac{25\pi^2}{8} t^2& L=2\\ -\frac{75\pi^3}{16}\left(1-\frac{\gamma_{\rm SM}}{2}\right) t^3& L=3\end{array}\right.}.
\end{align}
This is consistent with the observed behavior of the four-point function on the supersymmetric Wilson line \cite{Ferrero:2023gnu},
\begin{align}\label{eq:Regge constraint 2}
    \chi^{-4}G^{(L)}(\chi)\bigr\rvert_{\chi=\frac{1}{2}+it}\overset{t\to \infty}{\sim}O(t^L).
\end{align}

\paragraph{Definition of the coupling.}
Another constraint on the correlator arises from the perturbative definition of the coupling $\varsigma$ given in eq.~\eqref{eq:defVarsig},
\begin{align}\label{eq:anomalous dimension of z2 and definition of coupling}
    \gamma_{[z^2]_0}^{(1)}&=-1, &\gamma_{[z^2]_0}^{(L)}&=0\text{ for }L\geq 2.
\end{align}

\paragraph{Integral constraints.} The final constraint arises from a set of integral identities satisfied by correlators involving the displacement operator
\cite{Gabai:2025zcs,Kong:2025sbk,Girault:2025kzt,Drukker:2025dfm} (see also \cite{Gabai:2022vri,Cavaglia:2022qpg,Drukker:2022pxk,Gabai:2023lax}). In particular, the single independent constraint relevant for our analysis is \cite{Gabai:2025hwf}
\begin{align}\label{eq:integral constraint}
0
&=
\int_0^\infty \! dt \, \frac{1+2t+3t^2}{t^4}
\left[
G\!\left(\frac{t}{1+t}\right)
-
G^{(0)}\!\left(\frac{t}{1+t}\right)
\right].
\end{align}
This identity holds order by order in perturbation theory.

We note for later use that integrating the three contact diagrams in eq.~\eqref{eq:z-to-the-fourth contact diagram final expression}-\eqref{eq:tree level four point function} gives
\begin{align}
    \int_0^\infty dt \frac{1+2t+3t^2}{t^4}\xi_{z^4}\left(\frac{t}{1+t}\right)&=-\frac{4\pi^2}{5},\\
    \int_0^\infty dt \frac{1+2t+3t^2}{t^4}\xi_{\rm NG}\left(\frac{t}{1+t}\right)&=\int_0^\infty dt \frac{1+2t+3t^2}{t^4}\xi_{\rm HC}\left(\frac{t}{1+t}\right)=0.
\end{align}
The Nambu-Goto and higher-curvature contact diagrams individually satisfy the integral constraint. This follows from the fact that they originate from terms in the effective action of a string in AdS$_3$ and therefore respect the full AdS$_3$ isometry group, rather than only the AdS$_2$ isometries of the worldsheet. This property does not hold for the $z^4$ contact diagram.

We are now finally ready to start bootstrapping the correlator starting at tree-level.

\subsection{Tree level}\label{sec:bootstrap tree-level}
At tree-level, our ansatz for the correlator is
\begin{align}\label{eq:tree level ansatz}
    G^{(1)}(\chi)&=r_1(\chi)+r_2(\chi)\log(1-\chi)+r_3(\chi)\log(\chi). 
\end{align}
Here, the $r_i(\chi)$ are rational functions and the expression is valid for $0<\chi<1$. 

Cyclic symmetry, given in eq.~\eqref{eq:cyclic symmetry}, implies
\begin{align}\label{eq:tree level crossing relations}
    (1-\chi)^4r_1(\chi)&=\chi^4 r_1(1-\chi),&(1-\chi)^4r_2(\chi)&=\chi^4 r_3(1-\chi).
\end{align}
Braiding symmetry, given in eq.~\eqref{eq:braiding symmetry}, implies 
\begin{align}\label{eq:tree level bose symmetry relations}
    r_1(\chi)&=r_1\left(\frac{\chi}{\chi-1}\right),& r_3(\chi)&=r_3\left(\frac{\chi}{\chi-1}\right), & r_3(\chi)+\frac{\chi^4}{(1-\chi)^4}r_3(1-\chi)+\chi^4 r_3\left(\frac{1}{1-\chi}\right)&=0. 
\end{align}
Accordingly, we look for rational solutions to Eqs.~\eqref{eq:tree level crossing relations} and \eqref{eq:tree level bose symmetry relations}, with poles only at $\chi=0$ and $\chi=1$. Following \cite{Ghosh:2019rcj,Ferrero:2023gnu}, we consider functions of the form $p(\chi)/[\chi^k(1-\chi)^k]$, where $k\geq 0$ and $p(\chi)$ is a polynomial of degree at most $d$. We then gradually increase $k$ and $d$. Following this procedure, we find a single new solution for $r_1$ at each even value of $k\geq 2$. One convenient choice of basis is
\begin{align}\label{eq:theta n}
\theta_n(\chi)
=
\frac{(1-\chi+\chi^2)^{3n+1}}{\chi^{2n-2}(1-\chi)^{2n+2}},
\qquad
n=0,1,2,3,\ldots
\end{align}
For $r_3$, we find a new solution at each value of $k\geq 2$, with a convenient choice of basis given by
\begin{align}\label{eq:eta n}
    \eta_n(\chi)=\frac{(\chi-1)^{3n+2}+1-(-1)^n 2\chi^{3n+2}}{\chi^{n-2}(1-\chi)^{n+2}},\qquad n=0,1,2,3,\ldots
\end{align} 
Thus, we can write the ansatze for $r_1$, $r_2$ and $r_3$ as
\begin{align}\label{eq:tree-level r1,r2, r3}
    r_1(\chi)&=\sum_{n=0}^{n_{\rm max}}c_{1,n} \theta_n(\chi),&r_3(\chi)&=\sum_{n=0}^{n_{\rm max}}c_{3,n} \eta_n(\chi),& r_2(\chi)&=\frac{\chi^4}{(1-\chi)^4}r_3(1-\chi).
\end{align}
Imposing that the OPE limit of eq.~\eqref{eq:tree level ansatz} satisfies eq.~\eqref{eq:OPE constraint}, and that its flat-space limit obeys eq.~\eqref{eq:flat space limit of correlator}, leaves only two linearly independent solutions. Up to an overall normalization, the first solution, with $n_{\rm max}=1$, is given by
\begin{align}\label{eq:tree-level bootstrap solution 1}
    c_{1,n}=\frac{4}{5}\delta_{n,0},\qquad  c_{3,n}=-\frac{2}{3}\delta_{n,0}+\frac{4}{15}\delta_{n,1}
\end{align}
while the second, with $n_{\rm max}=3$, is given by
\begin{align}\label{eq:tree-level bootstrap solution 2}
    c_{1,n}=-\frac{47}{6}\delta_{n,0}+2\delta_{n,1}, \qquad c_{3,n}=-\frac{2}{3}\delta_{n,0}+\frac{14}{3}\delta_{n,1}-\frac{11}{3}\delta_{n,2}+\frac{2}{3}\delta_{n,3}.
\end{align}
Using eqs.~\eqref{eq:tree-level r1,r2, r3} and \eqref{eq:tree level ansatz}, each of these solutions can be translated into a reduced correlator. In this way, they reproduce the quartic contact diagram \eqref{eq:z-to-the-fourth contact diagram final expression} and the Nambu-Goto contact diagram \eqref{eq:tree level four point function}, respectively. This is consistent with our expectations, since contact Witten diagrams automatically satisfy crossing symmetry and have the correct OPE limit (At this order, the higher-curvature contact diagram~\eqref{eq:z-to-the-fourth contact diagram final expression} is ruled out by the flat space limit condition). To recap, at this intermediate stage the tree-level correlator is fixed to take the form
\begin{align}
    G^{(1)}(\chi)=A_1\xi_{z^4}(\chi)+A_2\xi_{\rm NG}(\chi).
\end{align}
Finally, imposing the integral constraint~\eqref{eq:integral constraint} sets $A_1=0$, and eliminating the coupling redefinition freedom according to~\eqref{eq:anomalous dimension of z2 and definition of coupling} fixes $A_2=-1$. The final result for the tree-level correlator is in agreement with the result from the Witten diagram computation~\eqref{eq:tree level four point function}. 

\paragraph{Space of solutions to tree-level crossing.} We will now make a few additional comments on the space of solutions to the tree-level bootstrap problem when we relax the behavior of the correlator in the {flat-space} limit. This larger set of solutions will reappear as ambiguities in the one-loop and two-loop correlators that need to be fixed by the integral constraint and flat space limit.

In addition to the solutions with $n_{\rm max}=1$ and $n_{\rm max}=3$ in eq.~\eqref{eq:tree-level bootstrap solution 1}-\eqref{eq:tree-level bootstrap solution 2}, we find a solution to the cylic and braiding relations having the correct OPE limit with $n_{\rm max}=5$, given by:
\begin{equation}\label{eq:certy54ewsd}
\begin{aligned}
    c_{1,n}&=\frac{376}{3}\delta_{n,0}-\frac{409}{3}\delta_{n,1}+20\delta_{n,2},\\c_{3,n}&=-\frac{2}{3}\delta_{n,0}+\frac{124}{3}\delta_{n,1}-157\delta_{n,2}+\frac{472}{3}\delta_{n,3}-\frac{170}{3}\delta_{n,4}+\frac{20}{3}\delta_{n,5}.
\end{aligned}
\end{equation}
Substituting this into eq.~\eqref{eq:tree-level r1,r2, r3} precisely reproduces the higher-curvature contact diagram in eq.~\eqref{eq:higher curvature contact diagram final expression}.

We can similarly proceed to higher values of $n_{\rm max}$, and will comment on their properties without giving the explicit solutions here. Let us denote the tree-level correlator corresponding to these solutions as:
\begin{align}
    \xi_{n_{\rm max}}(\chi)=r_1(\chi)+r_2(\chi)\log(1-\chi)+r_3(\chi)\log(\chi),
\end{align}
where $n_{\rm max}=1,3,5,7,\ldots$, and $r_1,r_2,r_3$ are determined by eq.~\eqref{eq:tree-level r1,r2, r3}. We have seen that $\xi_1=\xi_{\rm z^4}$, $\xi_3=\xi_{\rm NG}$, and $\xi_5=\xi_{\rm HC}$. More generally, we expect that $\xi_{n_{\rm max}}$ should equal a (sum of) four-point contact diagram(s) with interaction vertices that have a maximum number of derivatives determined by $n_{\rm max}$. The same behavior holds in QFT in AdS in higher dimensions \cite{Heemskerk:2009pn,Fitzpatrick:2010zm}. From our three explicit examples, we observe $n_{\rm derivs.}=2(n_{\rm max}-1)$.

In terms of the Regge behavior of these solutions to tree-level crossing and the OPE constraint, we observe: 
\begin{align}
    \chi^{-4}\xi_n(\chi)\big\rvert_{\chi=\frac{1}{2}+it}&\overset{s\to \infty}{=} \# t^{\ell_n},\qquad \text{where }\ell_n=n-2.
\end{align}
Here, $\#$ denotes a non-zero prefactor. We also note the corresponding behavior of the anomalous dimensions, when computed from the relation
\begin{align}
    \xi_n(\chi)\big\rvert_{\log{\chi}}&=\sum_{m=0}^\infty \hat{a}_m^{(0)} \hat{\gamma}_m^{\xi_n}\CB_{\hat{\Delta}_m^{(0)}}(\chi).
\end{align}
Here, $\hat{\gamma}_m^{\xi_n}$ is the contribution of a possible term proportional to $\xi_n$ to the two-particle operator $[z^2]_{2m}$ (possibly mixed with multi-particle operators). We find from the explicit examples we construct that asymptotically at large $m$, 
\begin{align}
    \gamma_m^{\xi_n}\overset{m\to \infty}{\sim}\# m^{p_n}, \qquad \text{where }p_n=2\ell_n.
\end{align}
A table summarizing the Regge exponent $\ell_n$ and anomalous dimension exponent $p_n$ corresponding to the contact diagram with $2n-2$ derivatives is given by:
\begin{center}
\begin{tabular}{c|cccccc}
$n$ & $1$ & $3$ & $5$ & $7$ & $9$ & $11$ \\ \hline
$\ell_n$ & $-1$ & $1$ & $3$ & $5$ & $7$ & $9$ \\
$p_n$ & $-2$ & $2$ & $6$ & $10$ & $14$ & $18$
\end{tabular}
\end{center}
We see that the space of solutions to tree-level crossing and the OPE constraint that is also consistent with the Regge constraints on the correlator and anomalous dimension in eqs.~\eqref{eq:flat space limit of correlator} and \eqref{eq:flat space limit of anomalous dimensions} is two-dimensional at tree-level and one-loop, and three-dimensional at two-loops.

\paragraph{Extracting OPE data.} Given the four-point function at free and tree-level, we can read off the OPE coefficieints in eq.~\eqref{eq:OPE coefficients perturbative expansion} and the anomalous dimensions in \eqref{eq:conformal dimensions perturbative expansion} up to order $\varsigma$ via the conformal block expansion. (In what follows, we use various techniques from \cite{Heemskerk:2009pn,Giombi:2017cqn}.)

At order $\varsigma^0$, equating the free correlator $G^{(0)}(\chi)$ (given explicitly in eq.~\eqref{eq:free correlator}) with the conformal block expansion in eq.~\eqref{eq:CB expansion 1} lets us read off the zeroth order OPE coefficients, $\hat{a}_n^{(0)}$. This can be done by expanding the correlator $G^{(0)}(\chi)$ and the conformal blocks $\CB_\Delta(\chi)$ in powers of $\chi$, and then setting the two series in eq.~\eqref{eq:CB expansion 1} equal term-by-term. The lower powers of $\chi$ fix the OPE coefficients of the lower dimension operators. Alternatively, one can use the following orthogonality property of the conformal blocks,
\begin{align}\label{eq:conformal block orthogonality}
    \oint \frac{d\chi}{2\pi i}\frac{1}{\chi^2}\CB_{4+2n}(\chi)\CB_{-3-2n'}(\chi)=\delta_{nn'},
\end{align}
which gives
\begin{align}
    \hat{a}_n^{(0)}&=\int \frac{d\chi}{2\pi i}\frac{1}{\chi^2}\CB_{-3-2n}G^{(0)}(\chi).
\end{align}
Either way, one finds
\begin{align}
    \hat{a}_n^{(0)}=\frac{\Gamma(2n+4)^2\Gamma(2n+7)}{18\Gamma(2n+1)\Gamma(4n+7)}.
\end{align}
This matches the result given in eq.~\eqref{eq:OPE coefficients zeroth order} from a direct construction of two-particle primaries in the GFF and an explicit computation of their OPE coefficients with two copies of $z$ via Wick contractions. We emphasize that, in accordance with the discussion in Appendix~\ref{app:unmixing at tree level}, these are the squared OPE coefficients of the two-particle operators. In particular, they are not affected by mixing with multi-particle operators, whose squared OPE coefficients start at order $\varsigma^2$.

At order $\varsigma^1$, equating $G^{(1)}(\chi)$ (given explicitly in eq.~\eqref{eq:tree level four point function} and eq.~\eqref{eq:xi NG}) with the conformal block expansions in eq.~\eqref{eq:CB expansion 2} and~\eqref{eq:CB expansion 3} lets us read off $\hat{\gamma}_n^{(1)}$ and $\hat{a}_n^{(1)}$. First, matching the $\log{\chi}$ term with eq.~\eqref{eq:CB expansion 3} gives
\begin{align}
    \hat{\gamma}_n^{(1)}&=\frac{1}{\hat{a}_n^{(0)}}\int \frac{d\chi}{2\pi i}\frac{1}{\chi^2}\CB_{-3-2n}G_{\log^1}^{(1)}(\chi).
\end{align}
This lets us efficiently determine $\gamma_n^{(1)}$ for many values of $n$, and then deduce the result:
\begin{align}
    \hat{\gamma}_n^{(1)}=-\frac{2n^2+7n+4}{4}.
\end{align}
This result matches the general formula in eq.~\eqref{eq:general formula for tree-level anomalous dimensions summary}, if we set $\hat{\Delta}^{(0)}_n=4+2n$ for the classical dimension and $k=2$ for the number of fields. Note also that the anomalous dimensions have the large $n$ behavior given in \eqref{eq:flat space limit of anomalous dimensions}, as predicted by the phase shift formula.

Next, we match the no-$\log(\chi)$ piece of $G^{(1)}(\chi)$ with eq.~\eqref{eq:CB expansion 2}. Taking derivatives of hypergeometric functions with respect to their parameters can be cumbersome. One way to proceed is to expand the conformal blocks in powers of $\chi$, the coefficients of which are rational functions of the exchanged dimensions $\Delta$, and then match both sides of eq.~\eqref{eq:CB expansion 2} term by term. This works well for low orders in the expansion. For higher orders, it is more efficient to use the following relation between the conformal blocks and their first derivatives with respect to $\Delta$, (which is valid for $\Delta \in \mathbb{Z}^+$ and $n,n'\in \mathbb{N}$):
\begin{align}\label{eq:conformal block derivative orthogonality}
    \oint \frac{d\chi}{2\pi i}\frac{1}{\chi^2}\CB^{(1)}_{\Delta+n}(\chi)\CB_{1-\Delta-n'}(\chi)= \left\{\begin{array}{cc}0& n'\leq n\\ A_{\Delta+n,n'-n} & n'>n\end{array}\right.
\end{align}
where
\begin{align}
    A_{n,m}=\left(-\frac{1}{4}\right)^{m-1}\frac{\Gamma(n+m)}{m(2n+m-1)(n-1)!(\frac{1}{2}+n)_{m-1}}.
\end{align}
Applying this result to eq.~\eqref{eq:CB expansion 2} yields
\begin{align}
    \hat{a}_{n'}^{(1)}+\sum_{n=0}^{n'-1}\hat{a}_n^{(0)}\hat{\gamma}_n^{(1)} A_{4+2n,2n'-2n}=\oint \frac{d\chi}{2\pi i}\frac{1}{\chi^2}\CB_{-3-2n'}(\chi)G^{(1)}_{\log^0}(\chi). 
\end{align}
This lets us efficiently determine $\hat{a}_n^{(1)}$ for many values of $n$ and deduce the general relation:
\begin{align}
    \hat{a}_n^{(1)}=\frac{1}{2}\partial_n(\hat{a}_n^{(0)}\hat{\gamma}_n^{(1)}). 
\end{align}

\subsection{One loop}

We begin by making the following ansatz for the four-point function at one-loop:
\begin{align}
    G^{(2)}(\chi)&=G^{(2)}_{\log^0}(\chi)+G^{(2)}_{\log^1}(\chi)\log{\chi}+G^{(2)}_{\log^2}(\chi)\log^2{\chi},
\end{align}
where
\begin{align}
    G^{(2)}_{\log^0}(\chi)&=r_1(\chi)+r_2(\chi)\log(1-\chi)+r_3(\chi)\log^2(1-\chi)+r_4(\chi)\text{Li}_2(\chi),\\
    G^{(2)}_{\log^1}(\chi)&=r_5(\chi)+r_6(\chi)\log(1-\chi),\\
    G^{(2)}_{\log^2}(\chi)&=r_7(\chi).
\end{align}
We first fix the highest log term using the conformal block expansion and our previous results for the OPE data and anomalous dimensions of the two-particle operators. Concretely, from eq.~\eqref{eq:CB expansion 6}, we have
\begin{align}
    r_7(\chi)&=\sum_n \hat{a}_n^{(0)}\frac{1}{2}(\hat{\gamma}^{(1)}_n)^2 \CB_{\Delta_n^{(0)}}(\chi).
\end{align}
To evaluate this sum, we first recall that the conformal blocks obey the Casimir equation,
\begin{align}
    \hat{\mathcal{C}}\CB_{\Delta}(\chi)=c_\Delta \CB_\Delta(\chi),
\end{align}
where $\hat{\mathcal{C}}=\chi^2(1-\chi)\partial_\chi^2-\chi^2\partial_\chi$ is the conformal Casimir operator and $c_\Delta=\Delta(\Delta-1)$. Furthermore, we have the free theory result:
\begin{align}
    \sum_{n=0}^\infty \hat{a}_n^{(0)} \CB_{\Delta_n^{(0)}}(\chi)=\chi^4+\frac{\chi^4}{(1-\chi)^4}.
\end{align}
Finally, we have $\hat{\gamma}_n^{(1)}=-\frac{1}{8}(c_{\Delta_n^{(0)}}-4)$. Therefore,
\begin{equation}
\begin{aligned}
    r_7(\chi)&=\frac{1}{128}(\hat{\mathcal{C}}-4)(\hat{\mathcal{C}}-4) G^{(0)}(\chi)\\&=\frac{\chi^4(25 \chi^8-174 \chi^7+523 \chi^6-884 \chi^5+915 \chi^4-590 \chi^3+234 \chi^2-32 \chi+8)}{32(1-\chi)^6}.
\end{aligned}
\end{equation}
We now impose the cylic and braiding symmetries. For this purpose, we will require the following two identities of the dilogarithm:
\begin{align}
    \text{Li}_2(1-\chi)&=-\text{Li}_2(\chi)-\log{\chi}\log(1-\chi)+\zeta(2),\label{eq:dilog identity 1}\\
    \text{Li}_2(\frac{\chi}{\chi-1})&=-\text{Li}_2(\chi)-\frac{1}{2}\log^2(1-\chi)\label{eq:dilog identity 2}
\end{align}
where $\zeta(2)=\frac{\pi^2}{6}$. Imposing the cyclic symmetry yields the conditions:
\begin{equation}
\label{eq:cyc1loop}
\begin{aligned}
    (1-\chi)^4r_1(\chi)&=\chi^4[r_1(1-\chi)+\zeta(2)r_4(1-\chi)],\\
    (1-\chi)^4 r_2(\chi)&=\chi^4 r_5(1-\chi),\\
    (1-\chi)^4 r_3(\chi)&=\chi^4 r_7(1-\chi),\\
    (1-\chi)^4 r_4(\chi)&=-\chi^4 r_4(1-\chi),
    \\(1-\chi)^4 r_6(\chi)&=\chi^4 [r_6(1-\chi)-r_4(1-\chi)],
\end{aligned}
\end{equation}
while imposing braiding symmetry yields:
\begin{equation}
\begin{aligned}
    r_1(\chi)&=r_1(\frac{\chi}{\chi-1}),\\
    r_2(\chi)&=-r_2(\frac{\chi}{\chi-1})-r_5(\frac{\chi}{\chi-1}),\\
    r_3(\chi)&=r_3(\frac{\chi}{\chi-1})-\frac{1}{2}r_4(\frac{\chi}{\chi-1})+r_6(\frac{\chi}{\chi-1})+r_7(\frac{\chi}{\chi-1}),\\
    r_4(\chi)&=-r_4(\frac{\chi}{\chi-1}),\\
    r_5(\chi)&=r_5(\frac{\chi}{\chi-1}),\\
    r_6(\chi)&=-r_6(\frac{\chi}{\chi-1})-2r_7(\frac{\chi}{\chi-1}),\\
    r_7(\chi)&=r_7(\frac{\chi}{\chi-1}).
\end{aligned}
\end{equation}
Let us focus on the equations constraining the rational functions multiplying highest log terms, which are $r_3$, $r_4$, $r_6$ and $r_7$. 
We have already determined $r_7(\chi)$. It turns out that the subset of equations involving only $r_1$ and $r_4$, together with the requirement that singularities occur only at operator collisions, forces $r_4(\chi)=0$.\footnote{In practice, we established this fact experimentally. The same phenomena was observed in \cite{Ferrero:2023gnu}. Providing a rigorous proof would be of interest.}
In that case,
\begin{align}
    r_3(\chi)&=\frac{\chi^4}{(1-\chi)^4}r_7(1-\chi),&
    r_6(\chi)&=-r_7(\chi)-\frac{\chi^4}{(1-\chi)^4}r_7(1-\chi)+\chi^4 r_7\left(\frac{1}{1-\chi}\right).
\end{align}
Next, we consider the equations constraining the rational functions multiplying the lower log terms, $r_1$, $r_2$, and $r_5$. These equations are identical to the cyclic and braiding symmetry conditions encountered at tree level. Consequently, we can write the general solution as
\begin{align}
    r_1(\chi)&=\sum_{n= 0}c_{1,n}\theta_n(\chi),& r_5(\chi)&=\sum_{n=0}c_{5,n}\eta_n(\chi),& r_2(\chi)&=\frac{\chi^4}{(1-\chi)^4}r_5(1-\chi), 
\end{align}
where the functions $\theta_n$ and $\eta_n$ are given in eq.~\eqref{eq:theta n} and~\eqref{eq:eta n}. 

Imposing that $G^{(2)}(\chi)$ exhibits the correct behavior in the OPE and flat-space limits, as specified in Eqs.~\eqref{eq:OPE constraint} and \eqref{eq:Regge constraint 2}, yields a two-parameter family of solutions:
\begin{align}
    c_{1,n}&=\frac{1315}{384}\delta_{n,0}-\frac{25}{32}\delta_{n,1}+\frac{4A_1}{5}\delta_{n,0}+A_2\big[-\frac{47}{6}\delta_{n,0}+2\delta_{n,1}\big]\\c_{5,n}&=-\frac{95}{384}\delta_{n,0}+\frac{1}{32}\delta_{n,2}+A_1\big[-\frac{2}{3}\delta_{n,0}+\frac{4}{15}\delta_{n,1}\big]\label{eq:tyedsrt54}\\&\qquad\qquad\qquad\qquad\qquad+A_2\big[-\frac{2}{3}\delta_{n,0}+\frac{14}{3}\delta_{n,1}-\frac{11}{3}\delta_{n,2}+\frac{2}{3}\delta_{n,3}\big]\nonumber
\end{align}
We see that the general solution can be written as the sum of a particular solution, which is required to cancel the contributions of the higher logarithmic terms in the OPE limit, and two linearly independent homogeneous solutions, which already appeared at tree level in eqs.~\eqref{eq:tree-level bootstrap solution 1} and \eqref{eq:tree-level bootstrap solution 2}. 
Finally, we fix the coefficients of the two homogeneous solutions $A_1$ and $A_2$ by imposing the integral identity \eqref{eq:integral constraint} and by requiring that $\hat{\gamma}_n^{(2)}=0$. The resulting one-loop correlator is:
\begin{equation}\label{eq:G2 final answer}
\begin{aligned}
    &G^{(2)}(\chi)=\\&\frac{\chi^4 \left(25 \chi^8-174 \chi^7+523 \chi^6-884 \chi^5+915 \chi^4-590 \chi^3+234 \chi^2-32 \chi+8\right)}{8(\chi-1)^6}\log^2(\chi)\\&+\frac{25 \left(\chi^8+1\right)-26 \left(\chi^6+1\right) \chi+5\left(\chi^4+1\right) \chi^2}{8\chi^2}\log^2(1-\chi)\\&+\big(50 \chi^{12}-300 \chi^{11}+759 \chi^{10}-1045 \chi^9+840 \chi^8-390 \chi^7+97 \chi^6-15 \chi^5+20 \chi^4\\&\qquad\qquad -30 \chi^3+25 \chi^2-11 \chi+2\big)\frac{\log(\chi)\log(1-\chi)}{8(1-\chi)^5 \chi}\\&+\big(2032 \chi^{10}-11116 \chi^9+25140 \chi^8-30055 \chi^7+20055 \chi^6-6181 \chi^5+2667 \chi^4\\&\qquad\qquad -1240 \chi^3+1210 \chi^2-600 \chi+120\big)\frac{\log(\chi)}{480 (1 - \chi)^5}\\&+\big(2032 (\chi^{10} + 1) - 9204 \chi (\chi^8 + 1) + 16536 \chi^2 (\chi^6 + 1) - 
 14729 \chi^3 (\chi^4 + 1)\\&\qquad\qquad + 6566 \chi^4 (\chi^2 + 1) - 2282 \chi^5\big)\frac{\log(1-\chi)}{480\chi(\chi-1)^4}\\&+\frac{(1-\chi+\chi^2)[1596(1+\chi^6)-4788 \chi(1+\chi^4)+3385 \chi^2(1+\chi^2)+1210 \chi^3]}{1440(1-\chi)^4}.
\end{aligned}
\end{equation}

\paragraph{Extracting OPE data.} First, applying the orthogonality relations in eq.~\eqref{eq:conformal block orthogonality} and eq.~\eqref{eq:conformal block derivative orthogonality} to eq.~\eqref{eq:CB expansion 5} yields 
\begin{align}
    \hat{a}_{n}^{(0)}\hat{\gamma}_{n}^{(2)}+\hat{a}_{n}^{(1)}\hat{\gamma}_{n}^{(1)}+\sum_{m=0}^{n-1}\hat{a}_m^{(0)}(\hat{\gamma}_m^{(1)})^2 A_{4+2m,2n-2m}=\oint \frac{d\chi}{2\pi i}\frac{1}{\chi^2}G^{(2)}_{\log^1}(\chi)\mathfrak{f}_{-3-2n}(\chi).
\end{align}
With this we can efficiently determine $\hat{\gamma}_n^{(2)}$ up to large values of $n$, which lets us deduce the general formula:
\begin{align}
   \hat{\gamma}_n^{(2)}=\frac{1}{2}\hat{\gamma}_n^{(1)}\partial_n \hat{\gamma}_n^{(1)}+\frac{39\hat{\gamma}_n^{(1)}}{80}+\frac{\hat{\gamma}_n^{(1)}}{4}H_{2n+3}+\frac{1}{16}-\frac{3}{16}\frac{1-\hat{\gamma}_n^{(1)}}{(1-2\hat{\gamma}_n^{(1)})(1-4\hat{\gamma}_n^{(1)})(1+4\hat{\gamma}_n^{(1)})}.
\end{align}
Here, $H_n\equiv \sum_{k=1}^n \frac{1}{k}$ are the Harmonic numbers.

Similarly, to determine the OPE coefficients, it is useful to first note the identity
\begin{align}
    \oint \frac{d\chi}{2\pi i}\frac{1}{\chi^2}\mathfrak{f}^{(2)}_{\Delta+n}(\chi)\mathfrak{f}_{1-\Delta-n'}(\chi)&=\left\{\begin{array}{cc} 0 & n'\leq n+1\\ B_{\Delta+n,n'+\Delta-2} & n'>n+1\end{array}\right.,
\end{align}
Here, $B_{m,n}$ is defined by the recursion relations
\begin{align}\label{eq:conformal block second derivative orthogonality}
    0&=\frac{(n+2)^2(n+2-m)^2(n+1+m)(n+2+m)(3n^2+18n+m^2+27)}{(2n+3)(2n+5)(n+3-m)}B_{m,n}\nonumber\\&+2(n+2+m)[3n^4+30n^3+(111-2m^2)n^2+(180-10m^2)n-m^4-11m^2+108]B_{m,n+1}\nonumber\\&+4(n+4-m)(n+3+m)^2(3n^2+12n+12+m^2)B_{m,n+2}
\end{align}
and the initial condition
\begin{align}
    B_{m,n}=\left\{\begin{array}{cc}0 & n<m\\ \frac{1}{4}-\frac{1}{4(2m+1)^3}& n=m\end{array}\right.
\end{align}
Applying the orthogonality relations in eq.~\eqref{eq:conformal block orthogonality}, eq.~\eqref{eq:conformal block derivative orthogonality} and eq.~\eqref{eq:conformal block second derivative orthogonality} to eq.~\eqref{eq:CB expansion 4} yields the linear relation
\begin{align}
    &\langle a^{(2)}\rangle_{n}+\sum_{m=0}^{n-1} (\hat{a}_m^{(1)}\hat{\gamma}_m^{(1)}+\hat{a}_m^{(0)}\hat{\gamma}_m^{(2)})A_{4+2n,2m-2n}+\frac{1}{2}\sum_{m=0}^{n-1}\hat{a}_m^{(0)}(\hat{\gamma}_m^{(1)})^2B_{4+2m,2+2n}\nonumber\\&=\oint \frac{d\chi}{2\pi i} \frac{1}{\chi^2}G^{(2)}_{\log^0}(\chi)\mathfrak{f}_{-3-2n}(\chi).
\end{align}
With this we can efficiently determine $\braket{a^{(2)}}_n$ for many values of $n$ and deduce the general expression: 
\begin{align}
    \langle a^{(2)}\rangle_n &=\frac{1}{2}\partial_n(\hat{a}_n^{(0)}\hat{\gamma}_n^{(2)}+\hat{a}_n^{(1)}\hat{\gamma}_n^{(1)})-\frac{1}{8}\partial_n^2(\hat{a}_n^{(0)}(\hat{\gamma}_n^{(1)})^2)\\&+\hat{a}_n^{(0)}\bigg[\frac{(2n^2+7n+4)(2n+5)(n+1)}{8}\big[S_{-2}(4+2n)+\frac{1}{2}\zeta(2)\big]+\frac{1}{960} \bigg(-30 n (2 n+5)\nonumber\\&+\frac{30 (2 n+11)}{(2 n+3)^2}+\frac{5 (5 n+8)}{(n+1)^2}-\frac{24 (11 n+3)}{(2 n+1)^2}+\frac{3 (22
   n+71)}{(n+3)^2}-\frac{105}{n+2}-\frac{10 (10 n+19)}{(2 n+5)^2}-15\bigg)\bigg],\nonumber
\end{align}
where
\begin{align}
    S_{-2}(\Delta)\equiv \sum_{k=1}^\Delta \frac{(-1)^k}{k^2}.
\end{align}
Note that $\braket{a^{(2)}}_n$ is the first OPE datum we have encountered in our bootstrap analysis that is expected to be mixed. For example, $\braket{a^{(2)}}_4=a_{[z^2]_4}^{(2)}+a_{z^4}^{(2)}$.

\subsection{Two loops}
At two loops, we make the following ansatz for the four point function:
\begin{align}\label{eq:3-loop ansatz}
    G^{(3)}(\chi)&=G^{(3)}_{\log^0}(\chi)+G^{(3)}_{\log^1}(\chi)\log{\chi}+G^{(3)}_{\log^2}(\chi)\log^2{\chi}+G^{(3)}_{\log^3}\log^3{\chi},
\end{align}
where
\begin{align}
    G^{(3)}_{\log^0}(\chi)&=r_1(\chi)+r_2(\chi)\log(1-\chi)+r_3(\chi) \log^2(1-\chi)+r_4(\chi)\polylog{2}(\chi)+r_5(\chi)\log^3(1-\chi)\nonumber\\&\qquad +r_6(\chi)\log(1-\chi)\polylog{2}(\chi)+r_7(\chi)\polylog{3}(\chi)+r_8(\chi)S_{1,2}(\chi)\label{eq:G 3loop no-log ansatz}\\
    G^{(3)}_{\log^1}(\chi)&=r_9(\chi)+r_{10}(\chi)\log(1-\chi)+r_{11}(\chi)\log^2(1-\chi)+r_{12}(\chi)\text{Li}_2(\chi),\label{eq:G 3loop log ansatz}\\
    G^{(3)}_{\log^2}(\chi)&=r_{13}(\chi)+r_{14}(\chi)\log(1-\chi),\label{eq:G 3loop log-squared-ansatz}\\
    G^{(3)}_{\log^3}(\chi)&=r_{15}(\chi).\label{eq:G 3loop log-cubed-ansatz}
\end{align}
We can fix the rational function multiplying the highest log term using \eqref{eq:CB expansion 10}. Expressing the tree-level anomalous dimension in terms of the quadratic Casimir, we find
\begin{equation}\label{eq:G3 log-cubed}
 \begin{aligned}
     G^{(3)}_{\rm log^3}(\chi)&=r_{15}(\chi)=\frac{1}{3!}\sum_n \hat{a}_n^{(0)} (\hat{\gamma}_n^{(1)})^3 G_{\Delta_n}(\chi)=\frac{-1}{3!\times 8^3}(\hat{\mathcal{C}}-4)^3 \left[\chi^4+\frac{\chi^4}{(1-\chi)^4}\right]\\&=-\frac{\chi^4}{96(1-\chi)^7}(450 \chi^{10}-3775 \chi^9+14049 \chi^8-30459 \chi^7+42441 \chi^6\\&\qquad\qquad -39501 \chi^5+24675 \chi^4-10056 \chi^3+2754 \chi^2-160 \chi+32).
 \end{aligned}
 \end{equation}
In fact, as can be seen from \eqref{eq:CB expansion 1}-\eqref{eq:CB expansion 10}, not only the leading logarithm but any term with more than one logarithm only depends on conformal data from lower orders. In practice, we evaluate the infinite sums by assuming an ansatz of the form,
 \begin{equation}
     \frac{P(\chi)}{\chi^n (1-\chi)^m}
 \end{equation}
where $P$ is a polynomial,  for each of the rational coefficients. We increase its degree and $n$ and $m$ until we find a function that empirically matches all OPE coefficients. For the second second highest log term, given by eq.~\eqref{eq:CB expansion 9}, we find:
\begin{align}\label{eq:G3 log-squared}
    G^{(3)}_{\log^2}(\chi)&=r_{13}(\chi)+r_{14}(\chi)\log(1-\chi)\\&=\frac{1}{960 (1 - \chi)^6}\big[120-720 \chi +1810 \chi ^2-2450 \chi ^3-109 \chi ^4+7216 \chi ^5 -61732 \chi ^6\nonumber\\&\qquad +157570 \chi ^7-248460 \chi ^8+243662 \chi ^9-146119 \chi ^{10}+49212 \chi ^{11}-7150 \chi
   ^{12}\big]\nonumber\\&+\frac{1}{32 (1 - \chi)^6 \chi}\big[4-26 \chi +72 \chi ^2-110 \chi ^3+100 \chi ^4-38 \chi ^5-368 \chi ^6+2431 \chi ^7-8470 \chi ^8\nonumber\\& \qquad +17710 \chi ^9-23590 \chi ^{10}+20258 \chi ^{11}-10898 \chi ^{12}+3350
   \chi ^{13}-450 \chi ^{14}\big]\log(1-\chi).\nonumber
\end{align}
Thus, resumming all the terms in the conformal block expansion that involve only the lower order OPE data fixes $r_{13}$, $r_{14}$, and $r_{15}$.

Next, we impose the cyclic and Bose symmetries. We need the following identities for $\polylog{3}(\chi)$ and $S_{1,2}(\chi)$:
\begin{align}
    \polylog{3}(1-\chi)&=\zeta(3)+\zeta(2)\log(1-\chi)-\frac{1}{2}\log^2(1-\chi)\log(\chi)-\log(1-\chi)\polylog{2}(\chi)-S_{1,2}(\chi),\\
    \polylog{3}(\frac{\chi}{\chi-1})&=\frac{1}{6}\log^3(1-\chi)+\log(1-\chi)\polylog{2}(\chi)-\polylog{3}(\chi)+S_{1,2}(\chi),\\
    S_{1,2}(1-\chi)&=\zeta(3)+\frac{1}{2}\log^2{\chi}\log(1-\chi)+\log(\chi)\polylog{2}(\chi)-\polylog{3}(\chi)
    \\
    S_{1,2}(\frac{\chi}{\chi-1})&=-\frac{1}{6}\log^3(1-\chi)+S_{1,2}(\chi).
\end{align}
in addition to the $\polylog{2}(\chi)$ identities \eqref{eq:dilog identity 1} and \eqref{eq:dilog identity 2}. Then, cyclic symmetry gives the following independent conditions:
\begin{align}
\frac{(1-\chi)^4}{\chi^4} r_1(\chi)&=r_1(1-\chi)+\zeta(2) r_4(1-\chi)+\zeta(3)r_7(1-\chi)+\zeta(3)r_8(1-\chi),\\
\frac{(1-\chi)^4}{\chi^4} r_2(\chi)&=\zeta(2)r_7(1-\chi)+r_9(1-\chi)+\zeta(2)r_{12}(1-\chi),\\
\frac{(1-\chi)^4}{\chi^4}r_3(\chi)&=r_{13}(1-\chi),\\
\frac{(1-\chi)^4}{\chi^4}r_4(\chi)&=-r_4(1-\chi),\\
\frac{(1-\chi)^4}{\chi^4}r_5(\chi)&=r_{15}(1-\chi),\\
\frac{(1-\chi)^4}{\chi^4}r_6(\chi)&=-r_7(1-\chi)-r_{12}(1-\chi),\\
\frac{(1-\chi)^4}{\chi^4}r_7(\chi)&=-r_8(1-\chi),\\
\frac{(1-\chi)^4}{\chi^4}r_{10}(\chi)&=r_{10}(1-\chi)-r_4(1-\chi),\\
\frac{(1-\chi)^4}{\chi^4}r_{11}(\chi)&=-\frac{1}{2}r_7(1-\chi)-r_{12}(1-\chi)+r_{14}(1-\chi),
\end{align}
while braiding symmetry gives the following independent conditions:
\begin{align}
r_1(\chi)&=r_1(\frac{\chi}{\chi-1}),\\
 r_2(\chi)&=-r_2(\frac{\chi}{\chi-1})-r_9(\frac{\chi}{\chi-1}),\\
r_3(\chi)&=r_3(\frac{\chi}{\chi-1})-\frac{1}{2}r_4(\frac{\chi}{\chi-1})+r_{10}(\frac{\chi}{\chi-1})+r_{13}(\frac{\chi}{\chi-1}),\\
r_4(\chi)&=-r_4(\frac{\chi}{\chi-1}),\\
r_5(\chi)&=-r_5\left(\frac{\chi}{\chi-1}\right)+\frac{1}{2} r_6\left(\frac{\chi}{\chi-1}\right)+\frac{1}{6} r_7\left(\frac{\chi}{\chi-1}\right)-\frac{1}{6}
   r_8\left(\frac{\chi}{\chi-1}\right)\nonumber\\&\qquad\qquad -r_{11}\left(\frac{\chi}{\chi-1}\right)+\frac{1}{2}
   r_{12}\left(\frac{\chi}{\chi-1}\right)-r_{14}\left(\frac{\chi}{\chi-1}\right)-r_{15}\left(\frac{\chi}{\chi-1}\right),\\
r_6(\chi)&=r_6\left(\frac{\chi}{\chi-1}\right)+r_7(\frac{\chi}{\chi-1})+r_{12}(\frac{\chi}{\chi-1}),\\
r_7(\chi)&=-r_7(\frac{\chi}{\chi-1}),\\
r_8(\chi)&=r_8(\frac{\chi}{\chi-1})+r_7(\frac{\chi}{\chi-1}),\\
r_9(\chi)&=r_9(\frac{\chi}{\chi-1}),\\
r_{10}(\chi)&=-r_{10}(\frac{\chi}{\chi-1})-2r_{13}(\frac{\chi}{\chi-1}),\\
r_{11}(\chi)&=r_{11}\left(\frac{\chi}{\chi-1}\right)-\frac{1}{2} r_{12}\left(\frac{\chi}{\chi-1}\right)+2 r_{14}\left(\frac{\chi}{\chi-1}\right)+3 r_{15}\left(\frac{\chi}{\chi-1}\right),\\
r_{12}(\chi)&=-r_{12}(\frac{\chi}{\chi-1}),\\
r_{13}(\chi)&=r_{13}(\frac{\chi}{\chi-1}),\\
r_{14}(\chi)&=-r_{14}(\frac{\chi}{\chi-1})-3r_{15}(\frac{\chi}{\chi-1}),\\
r_{15}(\chi)&=r_{15}(\frac{\chi}{\chi-1})
\end{align}

We first focus on the rational functions multiplying the remaining highest logarithmic terms, namely all except $r_1$, $r_2$, and $r_9$. We find the following unique solution:
\begin{align}
r_4(\chi) &= r_6(\chi) = 0, \nonumber\\[4pt]
r_3(\chi) &= \frac{\chi^4}{(1-\chi)^4}\, r_{13}(1-\chi), 
\qquad
r_5(\chi) = \frac{\chi^4}{(1-\chi)^4}\, r_{15}(1-\chi), \nonumber\\[6pt]
r_{10}(\chi) &= -r_{13}(\chi)
+ \chi^4 r_{13}(\chi^{-1})
- \frac{\chi^4}{(1-\chi)^4} r_{13}(1-\chi), \nonumber\\[6pt]
r_{11}(\chi) &= -\frac{1}{2} r_8(\chi)
+ \frac{\chi^4}{(1-\chi)^4} r_{14}(1-\chi), \\[6pt]
r_7(\chi) &= -r_{12}(\chi)
= -\frac{\chi^4}{(1-\chi)^4} r_8(1-\chi) \nonumber\\
&= -r_{14}(\chi)
+ r_{14}\!\left(\frac{\chi}{\chi-1}\right)
+ \chi^4 r_{14}\!\left(\frac{1}{1-\chi}\right)
- \frac{\chi^4}{(1-\chi)^4} r_{14}(1-\chi).\nonumber
\end{align}
Combined with eq.~\eqref{eq:G3 log-cubed} and eq.~\eqref{eq:G3 log-squared}, this determines all the rational functions, except $r_1$, $r_2$ and $r_9$.

Next, we analyze the cyclic and braiding relations for $r_1$, $r_2$, and $r_9$. Using the results obtained for the highest logarithmic terms, these relations simplify to:
\begin{align}
&\frac{(1-\chi)^4}{\chi^4}\, r_1(\chi)
=
r_1(1-\chi)
+\zeta(3)\!\left[
r_7(1-\chi)
-\frac{(1-\chi)^4}{\chi^4} r_7(\chi)
\right],
\\[8pt]
&\frac{(1-\chi)^4}{\chi^4}\, r_2(\chi)
=
r_9(1-\chi),
\qquad
r_1(\chi)
=
r_1\!\left(\frac{\chi}{\chi-1}\right),
\\[8pt]
&0
=
r_9\!\left(\frac{\chi}{\chi-1}\right)
+
r_2(\chi)
+
r_2\!\left(\frac{\chi}{\chi-1}\right),
\qquad
r_9(\chi)
=
r_9\!\left(\frac{\chi}{\chi-1}\right).
\end{align}
We can write the solution for $r_1(\chi)$ as
\begin{align}\label{eq:fghjuytfdse}
    r_1(\chi)&=\tilde{r}_1(\chi)+\zeta(3)\bigg[\chi^4 r_{14}\bigg(\frac{1}{\chi}\bigg)+\frac{\chi^4}{(1-\chi)^4}r_{14}\bigg(\frac{\chi-1}{\chi}\bigg)\bigg],
\end{align}
where $\tilde r_1(\chi)$ satisfies the same equations as $r_1(\chi)$, with the source term proportional to $\zeta(3)$ set to zero. It then follows that $\tilde r_1(\chi)$, $r_2(\chi)$, and $r_9(\chi)$ obey the same equations as the three rational functions appearing in the tree-level ansatz. We can therefore express the general solution in terms of the functions $\theta_n$ and $\eta_n$ defined in eqs.~\eqref{eq:theta n}–\eqref{eq:eta n}:
\begin{align}\label{eq:rtyju65red}
    \tilde{r}_1(\chi)&=\sum_{n=0}c_{1,n} \theta_n(\chi),&r_9(\chi)&=\sum_{n=0}c_{9,n}\eta_n(\chi), &r_2(\chi)&=\frac{\chi^4}{(1-\chi)^4}r_9(1-\chi).
\end{align}
Finally, we impose correct scaling of the full correlator in the OPE and Regge limits, given in eqs.~\eqref{eq:OPE constraint} and~\eqref{eq:Regge constraint 2}. Just as explained under equation \eqref{eq:tyedsrt54}, the general solution for $\tilde{r}_1$ and $r_9$ is a sum of a particular solution and a generic linear combination of homogeneous solutions. The particular solution for the coefficients $c_{1,n}$ and $c_{9,n}$ is:
\begin{align}
    c_{1,n}&=\left(-\frac{454555}{27648}+\frac{159 \zeta (3)}{16}\right)\delta_{n,0}+\left(\frac{1105}{384}-\frac{171 \zeta
   (3)}{16}\right)\delta_{n,1}+\frac{25 \zeta (3)}{16}\delta_{n,2},\label{eq:c1n particular solution}\\
    c_{9,n}&=\frac{5383}{1536}\delta_{n,0}-\frac{851}{1920}\delta_{n,2}.\label{eq:c9n particular solution}
\end{align}
There are three homogeneous solutions are, given in eqs.~\eqref{eq:tree-level bootstrap solution 1}, \eqref{eq:tree-level bootstrap solution 2}, and ~\eqref{eq:certy54ewsd}. The 2-loop correlator can thus be written as
\begin{align}
    G^{(3)}(\chi)=G^{(3)}_{\rm particular}(\chi)+A_1\xi_{z^4}(\chi)+A_2\xi_{\rm NG}(\chi)+A_3\xi_{\rm HC}(\chi),
\end{align}
where the $\xi_\bullet(\chi)$ are given in eqs.~\eqref{eq:z-to-the-fourth contact diagram final expression}-\eqref{eq:tree level four point function}, and $G^{(3)}_{\rm particular}(\chi)$ is found by substituting eq.~\eqref{eq:c1n particular solution}-\eqref{eq:c9n particular solution} into eq.~\eqref{eq:rtyju65red}, eq.~\eqref{eq:rtyju65red} into eq.~\eqref{eq:fghjuytfdse}, and then substituting all the results for the rational functions into eqs.~\eqref{eq:3-loop ansatz}-\eqref{eq:G 3loop log-cubed-ansatz}. 

The final step is to fix $A_1$, $A_2$ and $A_3$ by imposing the integral constraint in eq.~\eqref{eq:integral constraint}, the definition of the coupling in eq.~\eqref{eq:anomalous dimension of z2 and definition of coupling}, and the flat-space limit of the correlator in eq.~\eqref{eq:flat space limit of correlator}. These fix $A_1=\frac{14827}{2304}$, $A_2=-\frac{81607}{57600}+\frac{15\pi^2}{128}\gamma_{\rm SM}$, and $A_3=-\frac{15\pi^2}{128}\gamma_{\rm SM}$. Note the residual dependence on the Wilson coefficient $\gamma_{\rm SM}$ from the branon scattering phase in flat space.

For completeness, we present our final result for the two-loop four-point function. For the convenience of the reader, we do so in parts. The four-point function takes the form in \eqref{eq:3-loop ansatz}. In addition to $G^{(3)}_{\log^3}(\chi)$ and $G^{(3)}_{\log^2}(\chi)$, given explicitly in eq.~\eqref{eq:G3 log-cubed} and \eqref{eq:G3 log-squared}, we find:

\begin{equation}
\begin{aligned}
&G^{(3)}_{\log^1}(\chi)=\\&
\bigg[-102120 + 510600 \chi - 1030710 \chi^2 + 1059240 \chi^3 - 836313 \chi^4
+ 946119 \chi^5 \\
&\quad -2915865 \chi^6 + 4469445 \chi^7 -
3844690 \chi^8 + 1744294 \chi^9 - 326428 \chi^{10}
\biggr]\frac{1}{115200(1-\chi)^5} \\[1ex]
&+ \frac{
75\pi^2 \chi^6
(-11 + 44 \chi - 112 \chi^2 + 182 \chi^3 - 196 \chi^4
+ 140 \chi^5 - 64 \chi^6 + 17 \chi^7 - 2 \chi^8)
}{64(1-\chi)^7}\,\gamma_{\rm SM} \\[1ex]
&+ \bigg[
-838 + 4549 \chi - 10175 \chi^2 + 11965 \chi^3 - 7785 \chi^4 + 3695 \chi^5
- 15363 \chi^6 + 57740 \chi^7 \\
&\quad -122030 \chi^8 + 150380 \chi^9 - 108726 \chi^{10}
+ 42900 \chi^{11} - 7150 \chi^{12}\bigg]
\frac{\log(1-\chi)}{480(1-\chi)^5 \chi}\\[1ex]
&+ \bigg[
-75 + 447 \chi - 1121 \chi^2 + 1525 \chi^3 - 1205 \chi^4 + 545 \chi^5
- 183 \chi^6 + 861 \chi^7 - 4805 \chi^8 
 \\
&\quad + 14085 \chi^9 - 24285 \chi^{10} + 25685 \chi^{11}
- 16451 \chi^{12} + 5875 \chi^{13} - 900 \chi^{14}\bigg]
\frac{\log^2(1-\chi)}{64(1-\chi)^5 \chi^2} \\[1ex]
&+ \bigg[
10 - 65 \chi + 180 \chi^2 - 275 \chi^3 + 250 \chi^4 - 135 \chi^5
+ 124 \chi^6 - 299 \chi^7 + 720 \chi^8 \\
&\quad  
- 1065 \chi^9+986 \chi^{10} - 561 \chi^{11} + 180 \chi^{12} - 25 \chi^{13}
\bigg]
\frac{\operatorname{Li}_2(\chi)}{32(1-\chi)^6 \chi}
\end{aligned}
\end{equation}

\begin{equation}
\begin{aligned}
&G^{(3)}_{\log^0}(\chi)=\\&\big[10144-40576 \chi+56661 \chi^2-27967 \chi^3+13620 \chi^4\\&\quad -27967 \chi^5+56661 \chi^6-40576 \chi^7+10144 \chi^8\big]\frac{1}{230400 (1-\chi)^4}\\
&+\big[2-8 \chi+166 \chi^2-470 \chi^3+790 \chi^4-806 \chi^5+496 \chi^6-170 \chi^7+25 \chi^8\big]\frac{\chi^4 \zeta (3)}{32 (1-\chi)^6}\\&-\big[120-840 \chi+2530 \chi^2-4260 \chi^3+4349 \chi^4-2715 \chi^5+999 \chi^6-246 \chi^7+999 \chi^8\\&\quad -2715 \chi^9+4349 \chi^{10}-4260 \chi^{11}+2530 \chi^{12}-840 \chi^{13}+120 \chi^{14}\big]\frac{5 \pi ^2 \gamma_{\rm SM} }{256 (1-\chi)^6 \chi^2}\\&-\big[326428-1519986 \chi+2835304 \chi^2-2664851 \chi^3+1310254 \chi^4-472178 \chi^5\\&\quad +1310254 \chi^6-2664851 \chi^7+2835304 \chi^8-1519986 \chi^9+326428 \chi^{10}\big]\frac{\log (1-\chi)}{115200 (1-\chi)^4 \chi}\\&-\big[2-3 \chi+\chi^2+\chi^8-3 \chi^9+2 \chi^{10}\big]\frac{75 \pi ^2 \gamma_{\rm SM}  \log (1-\chi)}{64 \chi^3}\\&-\big[7150-36588 \chi+76687 \chi^2-83868 \chi^3+50127 \chi^4-15398 \chi^5+3660 \chi^6\\&\quad -15398 \chi^7+50127 \chi^8-83868 \chi^9+76687 \chi^{10}-36588 \chi^{11}+7150 \chi^{12}\big]\frac{\log ^2(1-\chi)}{960 (1-\chi)^4 \chi^2}\\&-\big[450-725 \chi+324 \chi^2-33 \chi^3-33 \chi^7+324 \chi^8-725 \chi^9+450 \chi^{10}\big]\frac{\log ^3(1-\chi)}{96 \chi^3}\\&-\big[10-65 \chi+180 \chi^2-275 \chi^3+250 \chi^4-135 \chi^5+124 \chi^6-299 \chi^7\\&\quad +720 \chi^8-1065 \chi^9+986 \chi^{10}-561 \chi^{11}+180 \chi^{12}-25 \chi^{13}\big]\frac{\operatorname{Li}_3(\chi)}{32 (1-\chi)^6 \chi}\\&+\big[25-170 \chi+496 \chi^2-806 \chi^3+790 \chi^4-470 \chi^5+164 \chi^6-164 \chi^8\\&\quad +470 \chi^9-790 \chi^{10}+806 \chi^{11}-496 \chi^{12}+170 \chi^{13}-25 \chi^{14}\big]\frac{\operatorname{S}_{1,2}(\chi)}{32 (1-\chi)^6 \chi^2}
\end{aligned}
\end{equation}

\paragraph{Extracting OPE data.} Finally, we can extract the anomalous dimensions and OPE coefficients from the correlator using eqs.~\eqref{eq:CB expansion 7} and~\eqref{eq:CB expansion 8}. Because of mixing (see discussion below the referenced equations), we can only extract the combinations of OPE coefficients and anomalous dimensions given in eq.~\eqref{eq:accessible mixed OPE data}. The lowest two dimension operators, $[z^2]$, $[z^2]_2$, are the unique $++$ parity operators with classical dimension $4$ and $6$ and thus are mixing free; the next lowest operators, $[z^2]_4,[z^4]_0$ and $[z^2]_6,[z^4]_2$ are only doubly degenerate and thus exhibit minimal mixing. Taken together, we can extract the following OPE coefficients:
\begin{align}
    \hat{a}^{(3)}_0&=\frac{1912081}{6912000}+\frac{\zeta (3)}{16}-\frac{165 \pi ^2 \gamma_{\rm SM} }{448},\\
    \hat{a}^{(3)}_1&=\frac{438075464651}{92177326080}+\frac{695 \zeta (3)}{144}-\frac{57695 \pi ^2 \gamma_{\rm SM} }{3584},\\
    \langle a^{(3)}\rangle_2&=-\frac{188283300608327}{12929536908288}+\frac{27125 \zeta
   (3)}{2288}-\frac{130551925 \pi ^2 \gamma_{\rm SM} }{5710848},\\
    \langle a^{(3)}\rangle_3&=-\frac{12571458963080090550107}{460861229817660672000}+\frac{36939 \zeta
   (3)}{3536}+\frac{17877027 \pi ^2 \gamma_{\rm SM} }{10579712},
\end{align}
and the following linear combinations of conformal data:
\begin{align}
    \hat{\gamma}^{(3)}_1&=\frac{1994231}{7526400}-\frac{1485 \pi ^2 \gamma_{\rm SM} }{512},\nonumber\\
    \hat{\gamma}^{(3)}_2&+\frac{\braket{a^{(2)}\gamma^{(1)}}_2-\braket{a^{(2)}}_2\hat{\gamma}^{(1)}_2}{\hat{a}^{(0)}_2}=-\frac{641173}{1451520}-\frac{6105 \pi ^2 \gamma_{\rm SM}}{256},\\
    \hat{\gamma}^{(3)}_3&+\frac{\braket{a^{(2)}\gamma^{(1)}}_3-\braket{a^{(2)}}_3\hat{\gamma}^{(1)}_3}{\hat{a}^{(0)}_3}=-\frac{15547313009}{2732083200}-\frac{54795 \pi ^2 \gamma_{\rm SM}}{512},\nonumber
\end{align}
We do not attempt to give closed form expressions for the OPE data for general $n$, as we did at lower orders. 

\subsection{Summary of select  OPE data}\label{sec:summary of select OPE data}

Finally, we summarize some of the OPE data that can be extracted from our results. Specifically, we give the (potentially mixed) conformal dimensions and OPE coefficients of the lowest dimension operators. For the conformal dimensions we have:
\begin{align}
    \Delta_{z}&=2,\qquad \text{(protected)},\\
    \Delta_{z^2}&=4-\varsigma, \qquad \text{(definition of $\varsigma$)},\\
    \Delta_{[z^2]_2}&=6-\frac{13}{4}\varsigma+\frac{9787}{8960}\varsigma^2+\left[\frac{1994231}{7526400}-\frac{1485 \pi ^2 \gamma_{\rm SM} }{512}\right]\varsigma^3+O(\varsigma^3),\\
    \frac{a_{[z^2]_4}\Delta_{[z^2]_4}+a_{[z^4]}\Delta_{[z^4]}}{a_{[z^2]_4}+a_{[z^4]}}&={\Delta_{[z^2]_4}} + O(\varsigma^3)\\
    &=8-\frac{13}{2}\varsigma+\frac{32713}{6720}\varsigma^2-\left[\frac{641173}{1451520}+\frac{6105 \pi ^2 \gamma_{\rm SM}}{256}\right]\varsigma^3+O(\varsigma^4),\\
    \frac{a_{[z^2]_6}\Delta_{[z^2]_6}+a_{[z^4]_2}\Delta_{[z^4]_2}}{a_{[z^2]_6}+a_{[z^4]_2}}&={\Delta_{[z^2]_6}} + O(\varsigma^3)\\
    &=10-\frac{43}{4}\varsigma+\frac{11310049}{887040}\varsigma^2-\left[\frac{15547313009}{2732083200}+\frac{54795 \pi ^2 \gamma_{\rm SM}}{512}\right]\varsigma^3+O(\varsigma^4)
\end{align}
For the squared OPE coefficients, we have:
\begin{align}
    a_{[z^2]}&=2-\frac{251}{60}\varsigma+\frac{11011}{4800}\varsigma^2+\left[\frac{1912081}{6912000}+\frac{\zeta (3)}{16}-\frac{165 \pi ^2 \gamma_{\rm SM} }{448}\right]\varsigma^3+O(\varsigma^4),\\
    a_{[z^2]_2}&=\frac{40}{9}-\frac{11455}{2268}\varsigma-\frac{1219193}{326592}\varsigma^2\nonumber\\&\qquad +\left[\frac{438075464651}{92177326080}+\frac{695 \zeta (3)}{144}-\frac{57695 \pi ^2 \gamma_{\rm SM} }{3584}\right]\varsigma^3+O(\varsigma^4),\\
    a_{[z^2]_4}+a_{[z^4]}&=a_{[z^2]_4}+O(\varsigma^2) = \frac{350}{143}+\frac{330295}{113256}\varsigma-\frac{1624478875}{194347296}\varsigma^2\\&\qquad+\left[-\frac{188283300608327}{12929536908288}+\frac{27125 \zeta
   (3)}{2288}-\frac{130551925 \pi ^2 \gamma_{\rm SM} }{5710848}\right]\varsigma^3+O(\varsigma^4),\nonumber\\
   a_{[z^2]_6}+a_{[z^4]_2}&=a_{[z^2]_6}+O(\varsigma^2)= \frac{168}{221}+\frac{20114059}{5372510}\varsigma+\frac{839102849209}{870416951040}\varsigma^2\\&\qquad +\left[-\frac{12571458963080090550107}{460861229817660672000}+\frac{36939 \zeta
   (3)}{3536}+\frac{17877027 \pi ^2 \gamma_{\rm SM} }{10579712}\right]\varsigma^3+O(\varsigma^4).
\end{align}

\paragraph{The two-point function of the displacement.}
Beyond the homogeneous constraint \eqref{eq:integral constraint}, correlators involving the displacement operator satisfy an additional integrated constraint \cite{Gabai:2025zcs}. It takes the form
\begin{align}\label{eq:integral constraint inhomog}
\int_0^\infty dt \,
\Big[
(t^2-1)\log t
-
2(1+2t)\log\!\left(\frac{t}{1+t}\right)
\Big]
\left[
G\!\left(\frac{t}{1+t}\right)
-
G^{(0)}\!\left(\frac{t}{1+t}\right)
\right]
=
\frac{4}{\Lambda},
\end{align}
where $\Lambda$ denotes the natural normalization of the displacement two-point function. Since the correlator is known up to two loops, this constraint allows us to extract $\Lambda$ to the same order:
\begin{align}\label{eq:displacement two-pt function}
\frac{1}{\Lambda}
&=
\frac{\pi^2}{12}\,\varsigma
+\frac{269\pi^2}{5760}\,\varsigma^2
+\left[
\frac{8731\pi^2}{153600}
-
\frac{\pi^4}{576}
\left(1+\frac{45\,\gamma_{\rm SM}}{8}\right)
\right]
\varsigma^3
+O(\varsigma^4).
\end{align}

\section{Bridging small and large AdS radius with Pad\'e}\label{sec:bridging small and large AdS radius with Pade}

In this section, we combine our results with those in \cite{Gabai:2025hwf} and interpolate various observables between small and large AdS using double sided Pad\'e approximants. Our goal at present is not to draw any rigorous conclusions, but rather to summarize our results graphically and to assess the convergence of the currently accessible perturbative results towards a smooth interpolation.

Given a function $f(x)$ that we want to approximate on an interval $x\in[a,b]$, we define its $[M/N]$ double-sided Pad\'e approximant with $L$ orders of perturbative input at $x=a$ and $R$ orders of perturbative input at $x=b$ to be the rational function
\begin{align}\label{eq:Pade approximant}
    f_{[M/N]}(x)&\equiv\frac{\sum_{m=0}^M p_m x^m}{1+\sum_{n=1}^N q_n x^n},
\end{align}
such that
\begin{align}
    f_{[M/N]}(x)-f(x)&\overset{x\to a}{=}O((x-a)^{L+1}),&\text{and}&&f_{[M/N]}(x)-f(x)&\overset{x\to b}{=}O((x-b)^{R+1}).
\end{align}
These conditions can be rewritten as a system of linear equations for the coefficients $p_m$ and $q_n$. We always pick $M$ and $N$ such that $M+N=L+R+1$, so that a solution for the coefficients generically exists and is unique. Note that the dependence of $f_{[M/N]}$ on the interval $[a,b]$ and on $L$ and $R$ is left implicit. Also, we have set $q_0=1$ in eq.~\eqref{eq:Pade approximant} to fix the freedom of simultaneously rescaling all the $p_m$ and $q_n$ without changing the approximant; this condition needs to be modified if there is a pole at $x=0$. 

We will use Pad\'e approximants to approximate the conformal dimensions and OPE coefficients of the first few two-particle operators, as well as the norm $\Lambda$ of the displacement operators, as functions of $\varsigma\equiv 4-\Delta_{z^2}$. We will use as input our perturbative results at both $\varsigma=0$ from the present work and $\varsigma=1$ from \cite{Gabai:2025hwf}. For these observables, we can typically fix the behavior of the approximant to $O(\varsigma^3)$ at large AdS radius and to $O((1-\varsigma)^0)$ or $O((1-\varsigma)^1)$ at small AdS radius.

Pad\'e approximation has a long history and has proven to be a useful tool in many physics contexts \cite{Baker_Graves-Morris_1996,Basdevant:1972fe}. For example, it can be used to accurately determine critical exponents from the high temperature expansion \cite{PhysRev.124.768}, to study strongly coupled phenomena \cite{Zinn-Justin:1971sac}, and to bridge the $2+\epsilon$ and $4-\epsilon$ expansions of the Ising and $O(N)$ CFTs in general dimensions (see, e.g., \cite{Henriksson:2025vyi}).\footnote{See also \cite{Alday:2025pmg,Dempsey:2025yiv} for some recent work in the context of $\mathcal{N}=4$ SYM that found that Pad\'e approximants for various observables either combine well with or are in good agreement with relatively tight numerical bootstrap bounds.} Moreover, the empirical utility of the Pad\'e method is supported by rigorous mathematical results guaranteeing convergence under relatively general conditions  \cite{nuttall1970convergence}. Nonetheless, convergence theorems are of limited use when one has access to only a few orders of perturbation theory and in such cases interpreting Pad\'e results requires care. To draw any convincing conclusions, one should adopt a sound procedure for distinguishing between well-behaved and poorly-behaved Pad\'es corresponding to different choices of $M$ and $N$, and for handling Pad\'e approximants that have poles in the interval of interest. 

Because our goal at present is to be illustrative rather than precise, we will largely sidestep these issues. We will instead present various Pad\'es for the different observables of interest, and then suggest some tentative lessons without attempting to identify any best Pad\'e approximant or combination thereof. To lend some credence to our observations, we will first examine the Pad\'e approximants for certain observables of the supersymmetric Wilson line in $\mathcal{N}=4$ SYM. It seems plausible that this set-up is sufficiently similar to the flux-tube in AdS$_3$ for the conclusions that we draw to be transferrable; at the same time, the availability of exact results makes it possible to precisely assess the quality of different Pad\'e approximants. 

\subsection{Pad\'e approximants for the supersymmetric Wilson line in $\mathcal{N}=4$ SYM}

There are many analytical and precise numerical results available for defect observables on the supersymmetric Wilson line in planar $\mathcal{N}=4$ SYM, even at intermediate couplings. This is due to the confluence of techniques from AdS/CFT \cite{Giombi:2017cqn}, supersymmetric localization \cite{Correa:2012at, Giombi:2018qox}, analytic bootstrap \cite{Ferrero:2021bsb,Ferrero:2023gnu,Ferrero:2023znz}, and integrability \cite{Grabner:2020nis,Cavaglia:2021bnz,Cavaglia:2022qpg}. 

Let us briefly review the set-up: the supersymmetric Wilson line in $\mathcal{N}=4$ SYM takes the form $\mathcal{W}=\text{P exp}\left(\int [i A_0 + \Phi_{\parallel}]dx^0\right)$. Here, $A_\mu$ is the gauge field, and $\phi^\parallel$ is one of the six scalars of $\mathcal{N}=4$ SYM; we denote the other scalars $\phi_m^\perp$, $m=1,\ldots,5$. As in the set-up we consider in this paper, one can define correlators on the supersymmetric Wilson line by inserting adjoint operators along the straight line contour \cite{Drukker:2006xg}. A superconformal multiplet of primary importance is the short multiplet whose bosonic conformal primaries are the five scalars $\phi_m^\perp$ and the three displacement operators $\mathbb{D}_i$, $i=1,2,3$, which have protected dimensions $\Delta_{\phi^\perp_m}=1$ and $\Delta_{\mathbb{D}_i}=2$. In the dual description of the Wilson line as a string in AdS$_5\times S^5$ with AdS$_2$ geometry in AdS$_5$ and sitting at a point in $S^5$, there are three transverse directions $x_i$ in AdS$_5$ and five transverse directions $y_m$ in $S^5$; $x_i$ and $y_m$ are then the duals of $\mathbb{D}_i$ and $\phi^\perp_m$, respectively. One can study the Wilson line and its defect CFT data at any $g_{\rm YM}$ and gauge group rank $N$. We focus on the planar limit $N\to \infty$ with $g\equiv \frac{\sqrt{g_{\rm YM}^2N}}{4\pi}$ fixed; $g\to 0$ is the weak coupling or gauge theory limit and $g\to \infty$ is the strong coupling or string theory limit.

We will consider the three lowest-dimension superconformal primaries that are singlets under the $SO(3)\times SO(5)_R$ global symmetry of the line defect. We denote the three operators $\Psi_a$, $a=1,2,3$.\footnote{We thank Julius Julius and Nika Sokolova for discussions and for sharing with us a notebook of high precision numerical results for the dimensions of these three operators, as computed using the quantum spectral curve (QSC). We also refer the reader to \cite{Julius:2025dce} for a complementary and comprehensive study of Pad\'e interpolations between weak and strong coupling for a larger set of observables than what we consider here. That work also proposes an interesting dictionary between the gauge theory and string theory alphabets for at least a subset of operators on the BPS Wilson line.} At weak coupling and strong coupling, the first operator takes the form $\Psi_1\sim \phi^\parallel$ and $\Psi_1\sim y_m y_m$. The second and third operators take the following forms at weak coupling, $\Psi_2\sim (\phi^\parallel)^2-\frac{1}{\sqrt{5}}\phi_m^\perp \phi_m^\perp$ and $\Psi_3\sim (\phi^\parallel)^2+\frac{1}{\sqrt{5}}\phi_m^\perp \phi_m^\perp$, and are given, schematically, by two linear combinations of $\partial^2 y_my_m$ and $(y_my_m)(y_ny_n)$ at strong coupling \cite{Julius:2025dce}.

We note the following perturbative data of these operators at weak coupling \cite{Cavaglia:2022qpg}:
\begin{align}
    \Delta_{\Psi_1}&=1+4g^2-16g^4+O(g^6),\\
    \Delta_{\Psi_2}&=2+[5-\sqrt{5}]g^2-2[9-4\sqrt{5}]g^4+O(g^6)\\
    \Delta_{\Psi_3}&=2+[5+\sqrt{5}]g^2-2[9+4\sqrt{5}]g^4+O(g^6).
\end{align}
(Higher orders are available, but these suffice for our purposes.) At strong coupling:
\begin{align}
    \Delta_{\Psi_1}&=2-\frac{5}{4\pi g}+\frac{295}{384\pi^2 g^2}-\frac{305}{1024\pi^3 g^3}+\left(\frac{351845}{3538944 \pi^4}-\frac{75\zeta(3)}{512\pi^4}\right)\frac{1}{g^4}+O\left(\frac{1}{g^5}\right),\\
    \Delta_{\Psi_2}&=4-\frac{7}{2\pi g}+\frac{7231-7\sqrt{62569}}{1920\pi^2g^2}+\frac{-0.0436\pm 0.0005}{g^3}+O\left(\frac{1}{g^4}\right),\\
    \Delta_{\Psi_3}&=4-\frac{7}{2\pi g}+\frac{7231+7\sqrt{62569}}{1920\pi^2g^2}+\frac{-0.167\pm 0.001}{g^3}+O\left(\frac{1}{g^4}\right).
\end{align}
The analytic results come from \cite{Giombi:2017cqn,Ferrero:2019luz}; we estimated the cubic coefficients for $\Delta_{\Psi_2}$ and $\Delta_{\Psi_3}$ by fitting polynomials to the numerical values generated by the quantum spectral curve (QSC), and looking for convergence as both the order of the polynomial fit and the interval on which the fit is performed are varied. More generally, the QSC results determine $\Delta_{\Psi_i}$ for $0\leq g\lesssim 4$ with very high precision.

In analogy with what we do for the flux tube in AdS$_3$, we will use $\sigma\equiv 2-\Delta_{\Psi_1}$ as the interpolator between weak and strong coupling. At weak coupling, the dimensions of $\Psi_2$ and $\Psi_3$ can be expanded in $1-\sigma$ as
\begin{align}
    \Delta_{\Psi_2}&=2+\frac{5-\sqrt{5}}{4}(1-\sigma)+\frac{1+2\sqrt{5}}{8}(1-\sigma)^2+O((1-\sigma)^3),\\
    \Delta_{\Psi_3}&=2+\frac{5+\sqrt{5}}{4}(1-\sigma)+\frac{1-2\sqrt{5}}{8}(1-\sigma)^2+O((1-\sigma)^3)
\end{align}
At strong coupling, they can be expanded in $\sigma$ as
\begin{align}
    \Delta_{\Psi_2}&=4-\frac{14}{5}\sigma+\frac{7(443+\sqrt{62569})}{3000}\sigma^2+(-0.6414\pm 0.005)\sigma^3+O(\sigma^4),\\
    \Delta_{\Psi_3}&=4-\frac{14}{5}\sigma+\frac{7(443-\sqrt{62569})}{3000}\sigma^2+(0.1775\pm 0.0015)\sigma^3+O(\sigma^4).
\end{align}

\begin{figure}[h!] 
  \centering

  % ---------- Row 1 ----------
  \begin{subfigure}[c]{0.49\textwidth}
    \centering
    \includegraphics[width=\textwidth]{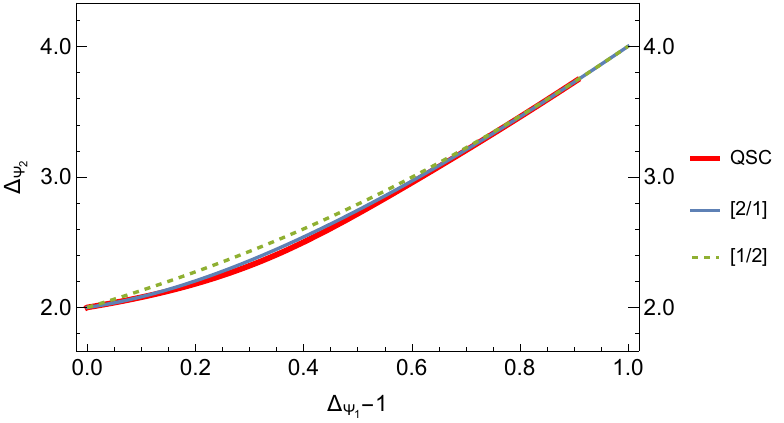}
    %\caption{}
  \end{subfigure}\hfill
  \begin{subfigure}[c]{0.49\textwidth}
    \centering
    \includegraphics[width=\textwidth]{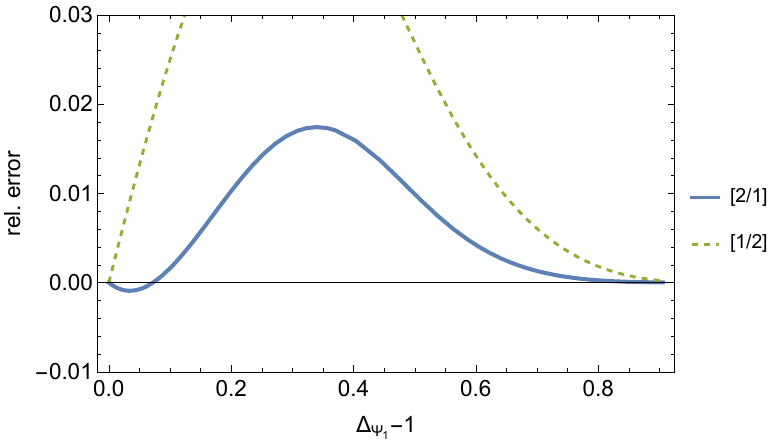}
    %\caption{}
  \end{subfigure}

  \vspace{1em}

  % ---------- Row 2 ----------
  \begin{subfigure}[c]{0.49\textwidth}
    \centering
    \includegraphics[width=\textwidth]{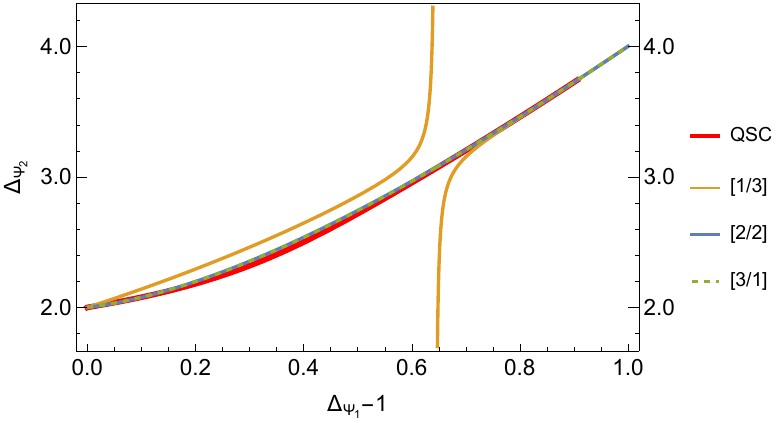}
    %\caption{}
  \end{subfigure}\hfill
  \begin{subfigure}[c]{0.49\textwidth}
    \centering
    \includegraphics[width=\textwidth]{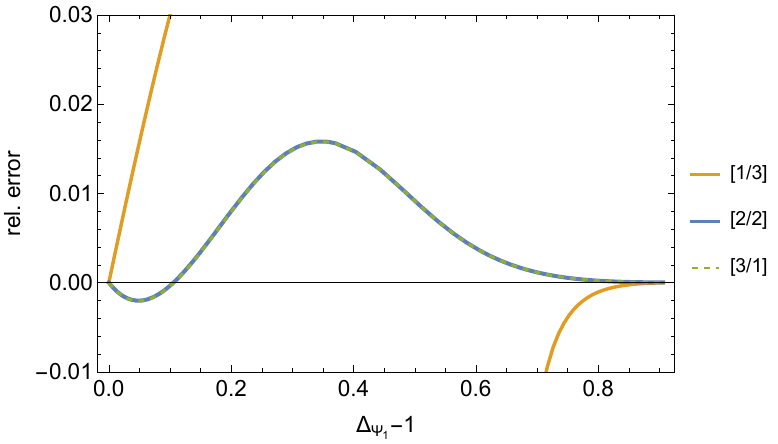}
    %\caption{}
  \end{subfigure}

  \vspace{1em}

    % ---------- Row 3 ----------
  \begin{subfigure}[c]{0.49\textwidth}
    \centering
    \includegraphics[width=\textwidth]{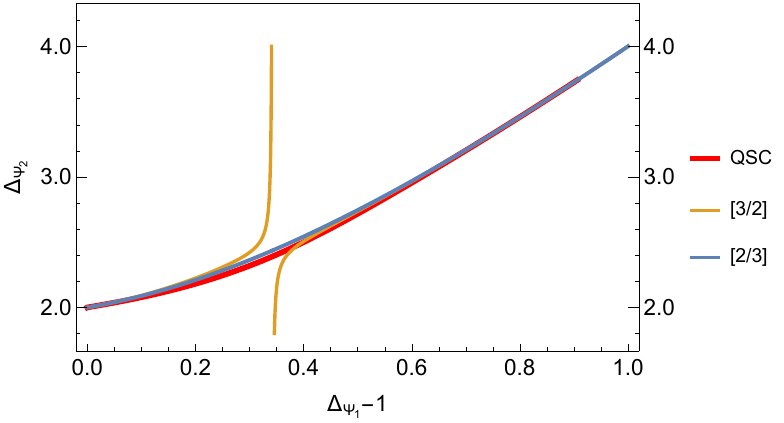}
    %\caption{}
  \end{subfigure}\hfill
  \begin{subfigure}[c]{0.49\textwidth}
    \centering
    \includegraphics[width=\textwidth]{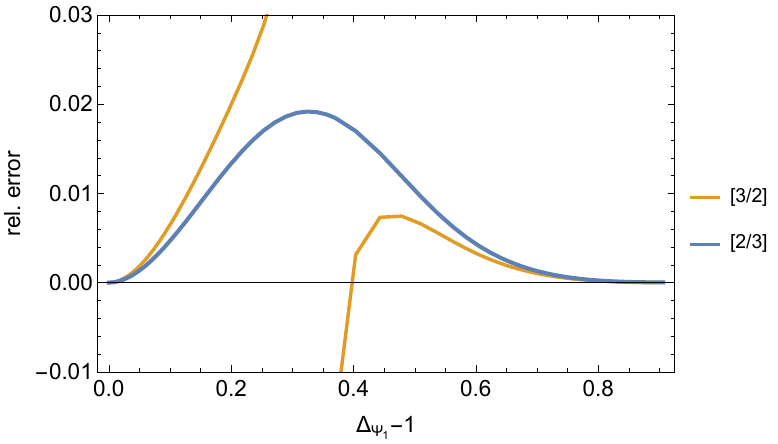}
    %\caption{}
  \end{subfigure}

      % ---------- Row 4 ----------
  \begin{subfigure}[c]{0.49\textwidth}
    \centering
    \includegraphics[width=\textwidth]{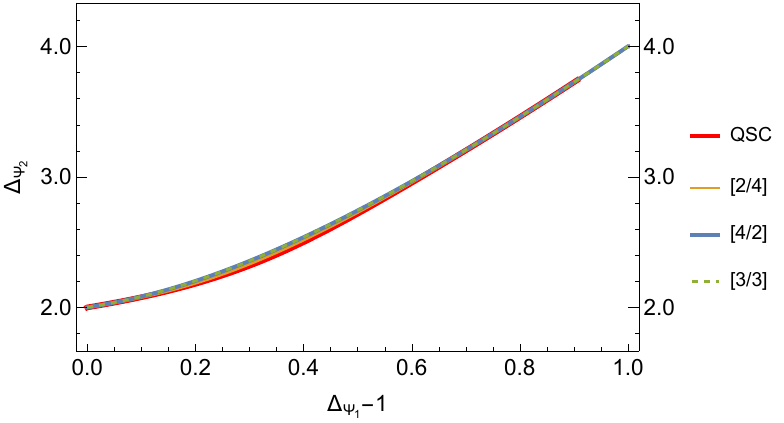}
    %\caption{}
  \end{subfigure}\hfill
  \begin{subfigure}[c]{0.49\textwidth}
    \centering
    \includegraphics[width=\textwidth]{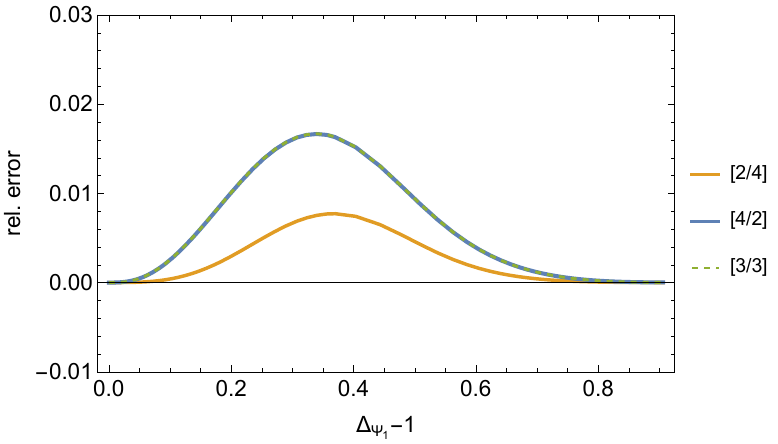}
    %\caption{}
    \end{subfigure}

  \caption{(Supersymmetric Wilson line in $\mathcal{N}=4$ SYM). Various Pad\'e approximants for $\Delta_{\Psi_2}$ (left column) and the relative error compared with exact results from QSC (right column). \textit{First row:} $[2/1]$ and $[1/2]$ Pad\'e approximants with $O(\sigma^2)$ and $O((1-\sigma)^0)$ perturbative input. \textit{Second row:} $[2/2]$, $[1/3]$, $[3/1]$ Pad\'e approximants with $O(\sigma^3)$ and $O((1-\sigma)^0)$ perturbative input, which is analogous to what we currently have for the Yang-Mills flux tube. \textit{Third row:} $[2/3]$ and $[3/2]$ Pad\'e approximants with $O(\sigma^3)$ and $O((1-\sigma)^1)$ perturbative. \textit{Fourth row:} $[4/2]$, $[3/3]$, $[2/4]$ Pad\'e approximants with $O(\sigma^3)$ and $O((1-\sigma)^2)$ perturbative input.}
  \label{fig:N=4 SYM Pades for dimensions 1}
\end{figure}

\begin{figure}[h!] % use figure* for two-column papers
  \centering

  % ---------- Row 1 ----------
    \begin{subfigure}[t]{0.49\textwidth}
    \centering
    \includegraphics[width=\textwidth]{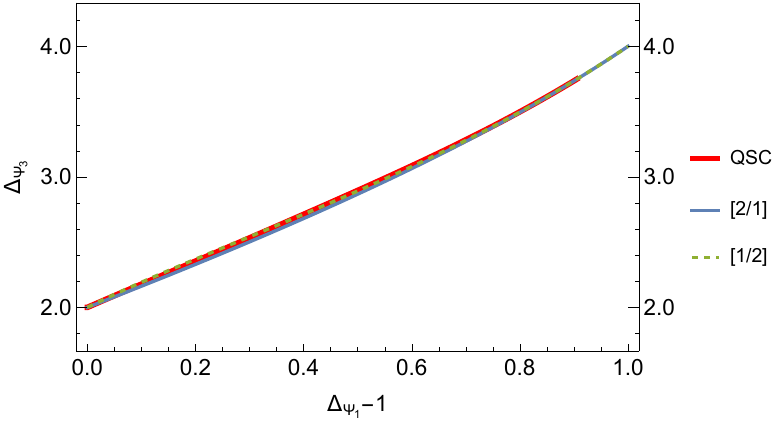}
    %\caption{}
  \end{subfigure}\hfill
  \begin{subfigure}[t]{0.49\textwidth}
    \centering
    \includegraphics[width=\textwidth]{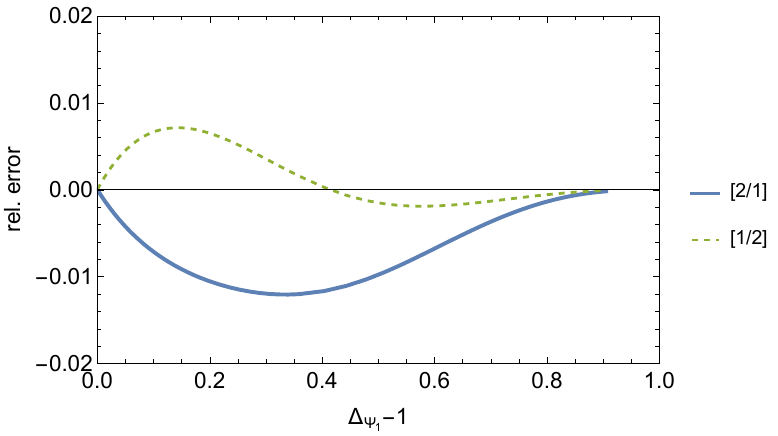}
    %\caption{}
  \end{subfigure}

  \vspace{1em}

  % ---------- Row 2 ----------
  \begin{subfigure}[t]{0.49\textwidth}
    \centering
    \includegraphics[width=\textwidth]{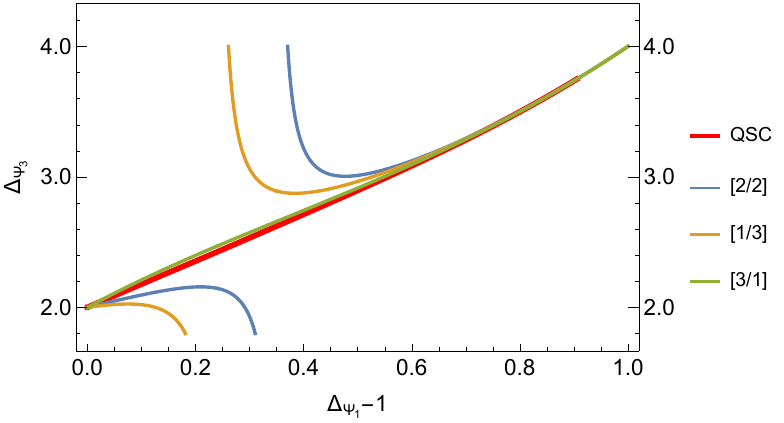}
    %\caption{}
  \end{subfigure}\hfill
  \begin{subfigure}[t]{0.49\textwidth}
    \centering
    \includegraphics[width=\textwidth]{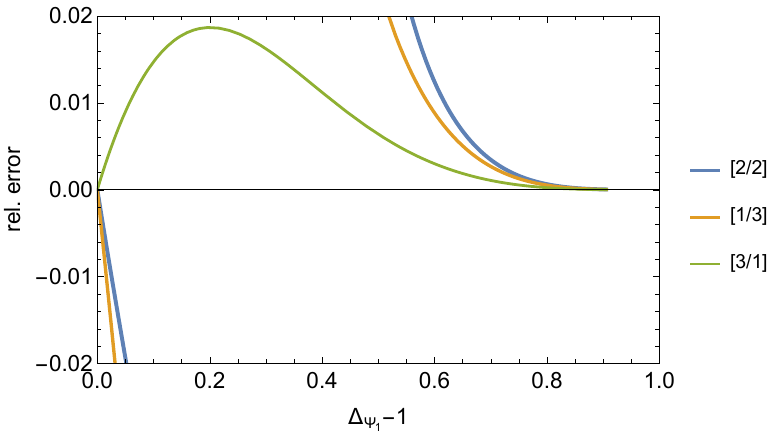}
    %\caption{}
  \end{subfigure}

  \vspace{1em}

    % ---------- Row 3 ----------
  \begin{subfigure}[t]{0.49\textwidth}
    \centering
    \includegraphics[width=\textwidth]{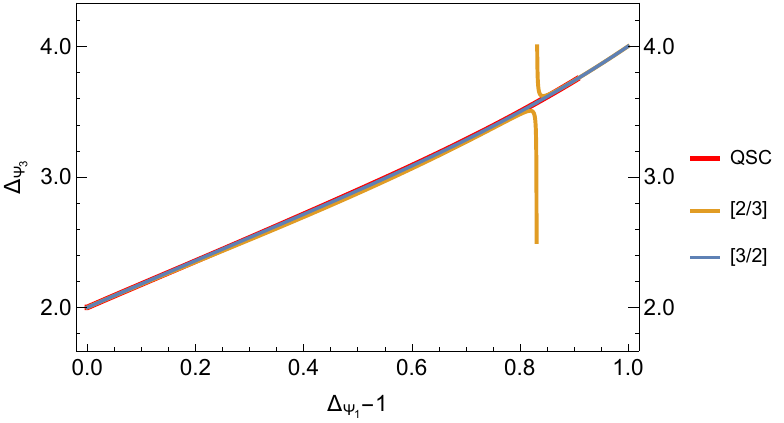}
    %\caption{}
  \end{subfigure}\hfill
  \begin{subfigure}[t]{0.49\textwidth}
    \centering
    \includegraphics[width=\textwidth]{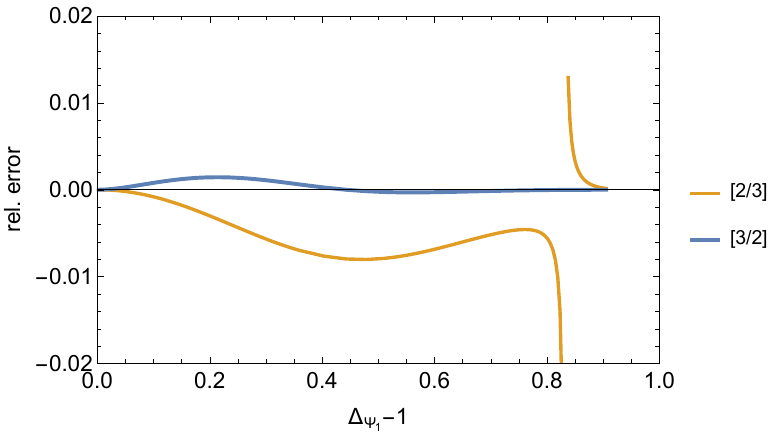}
    %\caption{}
  \end{subfigure}

      % ---------- Row 4 ----------
    \begin{subfigure}[t]{0.49\textwidth}
        \centering
        \includegraphics[width=\textwidth]{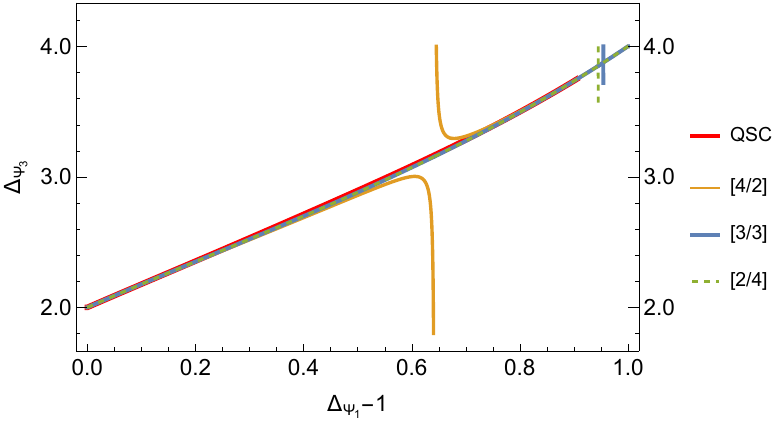}
    %\caption{}
    \end{subfigure}
    \hfill
  \begin{subfigure}[t]{0.49\textwidth}
    \centering
    \includegraphics[width=\textwidth]{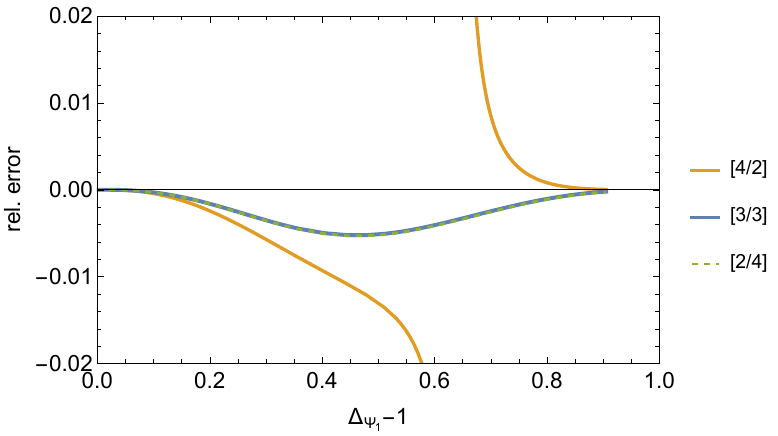}
    %\caption{}
  \end{subfigure}

  \caption{(Supersymmetric Wilson line in $\mathcal{N}=4$ SYM). Various Pad\'e approximants for $\Delta_{\Psi_3}$ (left column) and the relative error compared with exact results from QSC (right column). \textit{First row:} $[2/1]$ and $[1/2]$ Pad\'e approximants with $O(\sigma^2)$ and $O((1-\sigma)^0)$ perturbative input. \textit{Second row:} $[2/2]$, $[1/3]$, $[3/1]$ Pad\'e approximants with $O(\sigma^3)$ and $O((1-\sigma)^0)$ perturbative input, which is analogous to what we currently have for the Yang-Mills flux tube. \textit{Third row:} $[2/3]$ and $[3/2]$ Pad\'e approximants with $O(\sigma^3)$ and $O((1-\sigma)^1)$ perturbative input. \textit{Fourth row:} $[4/2]$, $[3/3]$, $[2/4]$ Pad\'e approximants with $O(\sigma^3)$ and $O((1-\sigma)^2)$ perturbative input.}
  \label{fig:N=4 SYM Pades for dimensions 2}
\end{figure}

We will now determine and assess Pad\'e approximants for $\Delta_{\Psi_2}$ and $\Delta_{\Psi_3}$. We choose $M$ and $N$ such that the resulting Pad\'e approximants are diagonal ($M=N$) or nearly diagonal ($|M-N|=1,2$). We also vary the number of orders of perturbation theory that we input at both weak and strong coupling. The resulting Pad\'es are depicted in Figures~\ref{fig:N=4 SYM Pades for dimensions 1} and~\ref{fig:N=4 SYM Pades for dimensions 2} and compared against the QSC results. The first row in each figure depicts the results with $O((1-\sigma)^0)$ and $O(\sigma^2)$ perturbative input. The second row depicts the results with $O((1-\sigma)^0)$ and $O(\sigma^3)$ perturbative input, which is analogous to the amount of perturbative information we currently have for the lowest two-particle operators on the Yang-Mills flux tube. The third and fourth row depict the results when we add one or two extra orders at weak coupling.

\begin{figure}[t!]
    \centering
    \begin{subfigure}[t]{0.49\textwidth}
        \centering
        \includegraphics[width=\textwidth]{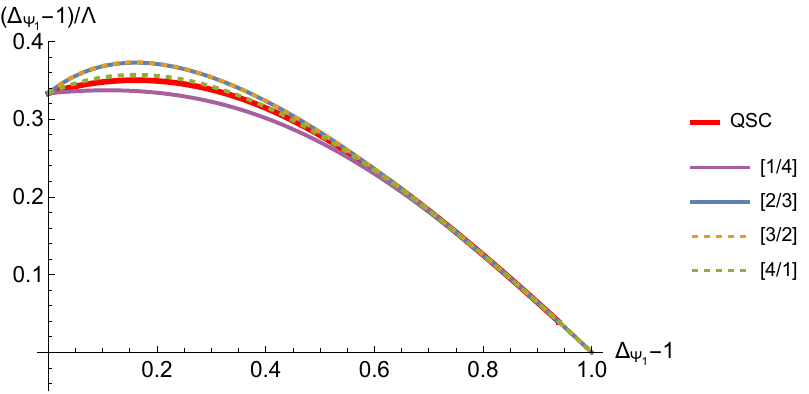}
    \end{subfigure}\hfill
    \begin{subfigure}[t]{0.49\textwidth}
        \centering
        \includegraphics[width=\textwidth]{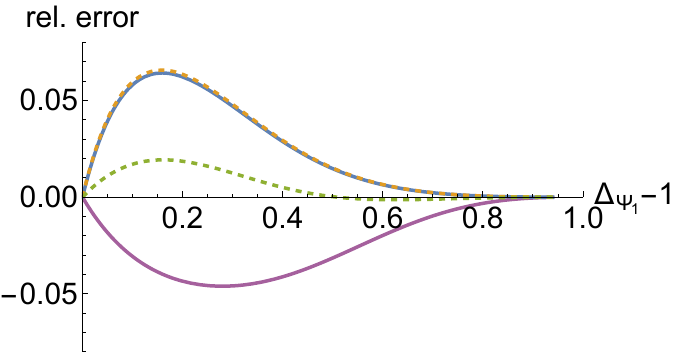}
    \end{subfigure}
    \caption{(Supersymmetric Wilson line in $\mathcal{N}=4$ SYM). Various Pad\'e approximants for the norm of the displacement.}
    \label{fig:N=4SYM Inverse Lambda Pade}
\end{figure}

Although we should be wary of generalizing too much from these results, we make the following observations:
\begin{itemize}
    \item The Pad\'e approximants without poles generally approximate the exact QSC results quite well. All but one have relative errors that are less than $2\%$.  
    \item Pad\'es often develop poles even when approximating smooth functions, and the same is true here. Nonetheless, even some of the Pad\'e approximants \textit{with} poles approximate the exact QSC results quite well, away from the poles. This is especially true for the Pad\'e approximants in the third and fourth rows in Figures~\ref{fig:N=4 SYM Pades for dimensions 1} and ~\ref{fig:N=4 SYM Pades for dimensions 2}, where we input additional orders at weak coupling. The relative errors of those Pad\'es with poles are less than $2\%$ on more than $85\%$ of the interval.

    Furthermore, some poles, such as those in the [3/3] and [2/4] Pad\'es in the bottom left plot of Figure~\ref{fig:N=4 SYM Pades for dimensions 2}, are completely harmless.
    \item The Pad\'es generally improve as we add more orders at weak coupling. In particular, the Pad\'e approximants with $O((1-\sigma)^1)$ input look significantly better than those with $O((1-\sigma)^0)$ input. This suggests that performing perturbative computations to one or two more orders at weak coupling in the study of the Yang-Mills flux tube is likely to be worthwhile. 
    However, this is not an iron rule: for example, the Pad\'e approximants in Figures~\ref{fig:N=4 SYM Pades for dimensions 1} and~\ref{fig:N=4 SYM Pades for dimensions 2} with only $O(\sigma^2)$ and $O((1-\sigma)^0)$ input are surprisingly good.
\end{itemize}

Finally, we can also use Pad\'e approximants to interpolate between the weak and strong coupling expansions of the displacement norm, $\Lambda_{\rm SYM}$. In $\mathcal{N}=4$ SYM, $\Lambda_{\rm SYM}$ is known exactly for any $g_{\rm YM}$ and $N$ \cite{Correa:2012at}. In the planar limit, $\Lambda_{\rm SYM}^{-1}=\frac{\pi}{12g}\frac{I_1(4\pi g)}{I_2(4\pi g)}$ and the perturbative expansions at weak coupling and strong coupling are $\Lambda_{\rm SYM}^{-1}=\frac{1}{12g^2}+O(g^0)$ and $\Lambda_{\rm SYM}^{-1}=\frac{\pi}{12g}+\frac{1}{32g^2}+\frac{5}{512\pi g^3}+O(1/g^4)$. In terms of $1-\sigma$, the weak coupling expansion becomes:
\begin{align}
    \Lambda_{\rm SYM}^{-1}&=\frac{1}{3(1-\sigma)}+O((1-\sigma)^0).
\end{align}
In terms of $\sigma$, the strong coupling expansion becomes
\begin{align}
    \Lambda_{\rm SYM}^{-1}&=\frac{\pi^2}{15}\sigma+\frac{19\pi^2}{360}\sigma^2+\frac{5047\pi^2}{108000}\sigma^3+O(\sigma^4).
\end{align}
To accommodate the pole at $\sigma=1$, we now consider Pad\'e approximants of the form $\Lambda^{-1}=\big[\sum_{m=0}^M p_m \sigma^m\big]/\big[(1-\sigma)(1+\sum_{n=0}^{N-1} q_n \sigma^n)\big]$. In Figure~\ref{fig:N=4SYM Inverse Lambda Pade}, we depict the $[1/4]$, $[2/3]$, $[3/2]$, and $[4/1]$ Pad\'e approximants. To make the curve bounded, it is convenient to plot $(1-\sigma)/\Lambda_{\rm SYM}$. The Pad\'e approximants all agree with the QSC exact result with $<6.5\%$ error on the entire interval. In addition, the error is $<3.5\%$ for $\Delta_{\Psi_1}\in [1.5,2]$, due to the many orders of input at strong coupling. Notably, none of these Pad\'es develop poles.

\FloatBarrier

\subsection{Pad\'e approximants for the flux tube in AdS$_3$}

We now turn to the Pad\'e approximants for observables on the Yang-Mills flux tube in AdS$_3$. We focus on the conformal dimensions of the two-particle operators $[z^2]_2$, $[z^2]_4$ (mixed with $z^4$), and $[z^2]_6$ (mixed with $[z^4]_2$), the squared OPE coefficient of the first two-particle operator, $a_{z^2}=c_{zzz^2}^2$, and the norm of the displacement operator. These observables have the most perturbative results and are least affected by mixing.

We take as input the perturbative results valid at large AdS radius that we have extracted in the present work from the analytic bootstrap applied to the effective string description of the flux tube, as well as the perturbative results valid at small AdS radius extracted from the Wilson line description of the flux tube in \cite{Gabai:2025hwf}.  The former results are summarized in Section~\ref{sec:summary of select OPE data}. From among the latter, we need:
\begin{align}
    \Delta_{[z^2]_2}&\overset{\varsigma\to 1}=4+O(1-\varsigma),\label{eq:dimension at weak coupling 1}\\
    \frac{a_{[z^2]_4}\Delta_{[z^2]_4}+a_{[z^4]}\Delta_{[z^4]}}{a_{[z^2]_4}+a_{[z^4]}}&\overset{\varsigma\to 1}=5+O(1-\varsigma),\label{eq:dimension at weak coupling 2}\\
    \frac{a_{[z^2]_6}\Delta_{[z^2]_6}+a_{[z^4]_2}\Delta_{[z^4]_2}}{a_{[z^2]_6}+a_{[z^4]_2}}&\overset{\varsigma\to 1}=6+O(1-\varsigma),\label{eq:dimension at weak coupling 3}
\end{align}
for the conformal dimensions,
\begin{align}
    a_{[z^2]}\overset{\varsigma\to 1}{=}\frac{5}{12}(1-\varsigma)+O((1-\varsigma)^2)
\end{align}
for the OPE coefficient, and
\begin{align}
    \Lambda^{-1}&=\frac{5}{12}\frac{1}{1-\varsigma}+O((1-\varsigma)^0),
\end{align}
for the norm of the displacement. Note that even though the pairs of operators $[z^2]_4,[z^4]$ and $[z^2]_6,[z^4]_2$ mix, our spectrum matching arguments in section~\ref{sec:counting operators} tell us that the pair of operators with dimension $\Delta=8$ at large AdS radius both have dimension $\Delta=5$ at small AdS radius and that the pair of operators with dimension $\Delta=10$ at large AdS radius both have dimension $\Delta=6$ at small AdS radius. This can be seen from Figures~\ref{table:counts of primaries in different representations} and~\ref{fig:Delta EST to Delta WL and square root fit}. Thus, the predictions in eqs.~\eqref{eq:dimension at weak coupling 2} and \eqref{eq:dimension at weak coupling 3} are unambiguous.

We recall that our perturbative results at strong coupling are parameterized by the coefficient $\gamma_{\rm SM}$ that appears in the low energy expansion of the flat space phase shift in eq.~\eqref{eq:ujmhgfs}. We will present the Pad\'es with $\gamma_{\rm SM}=0.3$ and $\gamma_{\rm SM}=0$. The first value of the phase shift, $\gamma_{\rm SM}=0.3$, comes from extrapolating the phase shifts for the flux tube branons in 3d $SU(N)$ Yang-Mills, as extracted from the lattice, from $N=2,3,6$ to $N=\infty$ \cite{Guerrieri:2024ckc}. To the extent that the lattice estimates and the extrapolation to infinite $N$ can be trusted, this value should characterize the flat space limit of the flux tube in Yang-Mills in AdS$_3$. The second value of the phase shift, $\gamma_{\rm SM}=0$, corresponds to the integrable theory of an infinitely long free bosonic string in flat space \cite{Dubovsky:2012wk}. We include it to illustrate the sensitivity of the Pad\'es on $\gamma_{\rm SM}$, and also because it is interesting to note that the Pad\'es behave especially well for this value.

\begin{figure}[t!]
    \centering
    \begin{subfigure}[c]{0.9\textwidth}
    \centering
        \includegraphics[width=\textwidth]{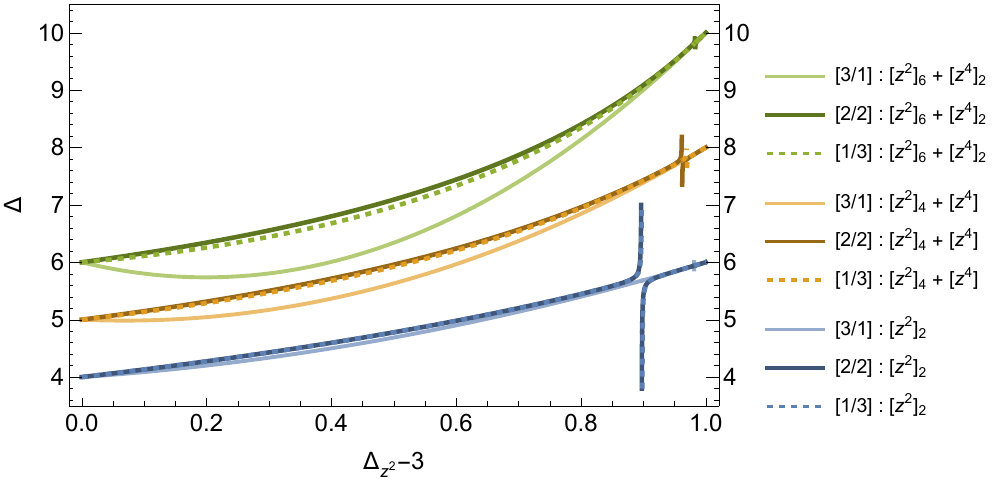}
        \caption{$\gamma_{\rm SM}=0.3$}
    \end{subfigure}

    \vspace{1em}

        \begin{subfigure}[c]{0.9\textwidth}
    \centering
        \includegraphics[width=\textwidth]{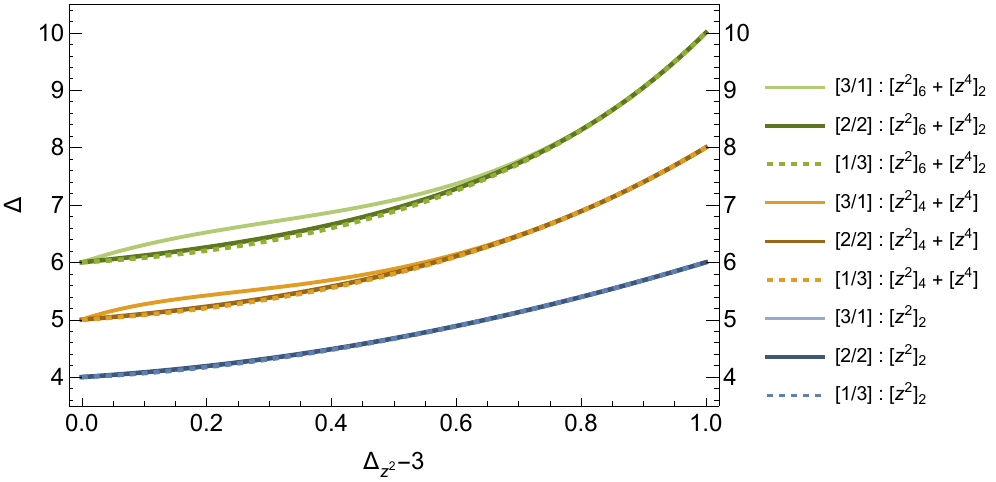}
        \caption{$\gamma_{\rm SM}=0$}
    \end{subfigure}
    
    \caption{(Yang-Mills flux tube in AdS$_3$). Various Pad\'e approximants for the (mixing-averaged) conformal dimensions of the first three two-particle operators after $[z^2]$. We present the approximants for both $\gamma_{\rm SM}=0.3$ (the estimated value for 3d Yang-Mills with gauge group $SU(\infty)$) and $\gamma_{\rm SM}=0$ (corresponding to the integrable worldsheet theory in flat-space).}
    \label{fig:Yang-Mills Flux Tube Dimensions Pade}
\end{figure}

\begin{figure}[h!] 
  \centering
  
    % ---------- Row 1 ----------
  \begin{subfigure}[t]{\textwidth}
    \centering
    \includegraphics[width=0.8\textwidth]{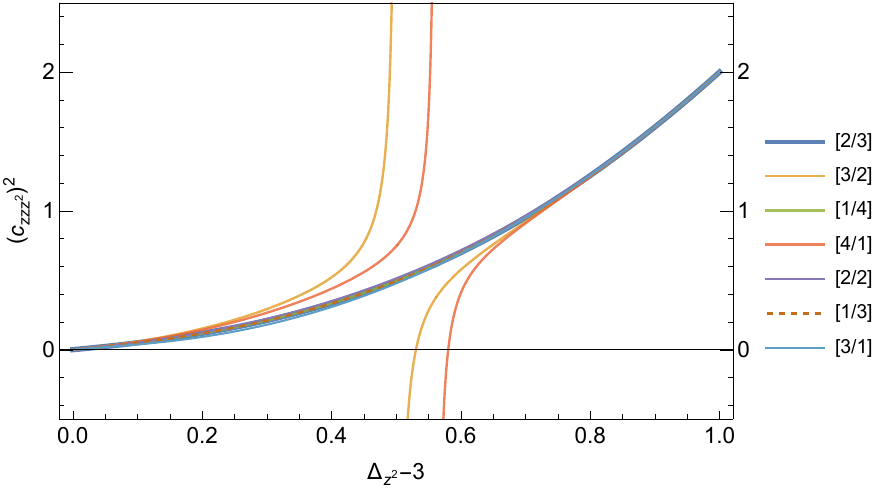}
    \caption{$\gamma_{\rm SM}=0.3$}
  \end{subfigure}

    \vspace{1em}
  
  \begin{subfigure}[t]{\textwidth}
    \centering
    \includegraphics[width=0.8\textwidth]{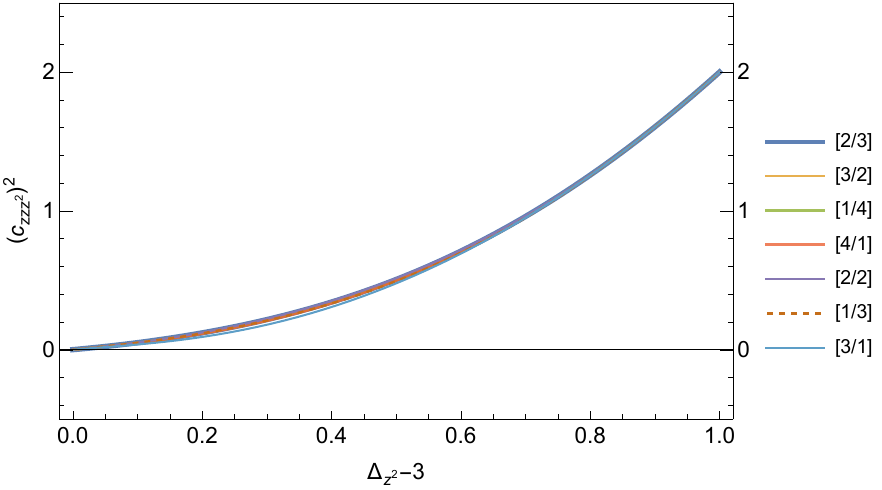}
    \caption{$\gamma_{\rm SM}=0$}
  \end{subfigure}
  
  \vspace{1em}

  \caption{(Yang-Mills flux tube in AdS$_3$). Various Pad\'e approximants for the squared OPE coefficient of $[z^2]$, for both $\gamma_{\rm SM}=0.3$ and $\gamma_{\rm SM}=0$.}
  \label{fig:YM FT Pade for OPE coefficient}
\end{figure}

\begin{figure}
    \centering 
        % ---------- Row 2 ----------
  \begin{subfigure}[t]{\textwidth}
    \centering
    \includegraphics[width=0.9\textwidth]{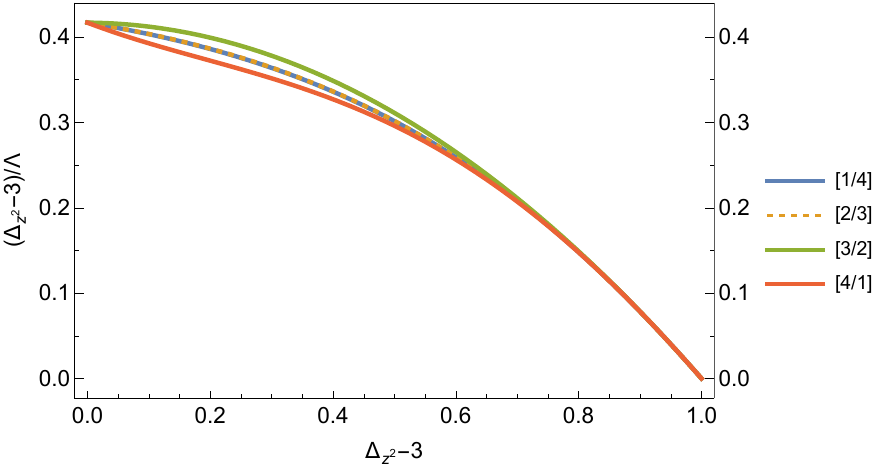}
    \caption{$\gamma_{\rm SM}=0.3$}
  \end{subfigure}

  \vspace{1em}

    \begin{subfigure}[t]{\textwidth}
    \centering
    \includegraphics[width=0.9\textwidth]{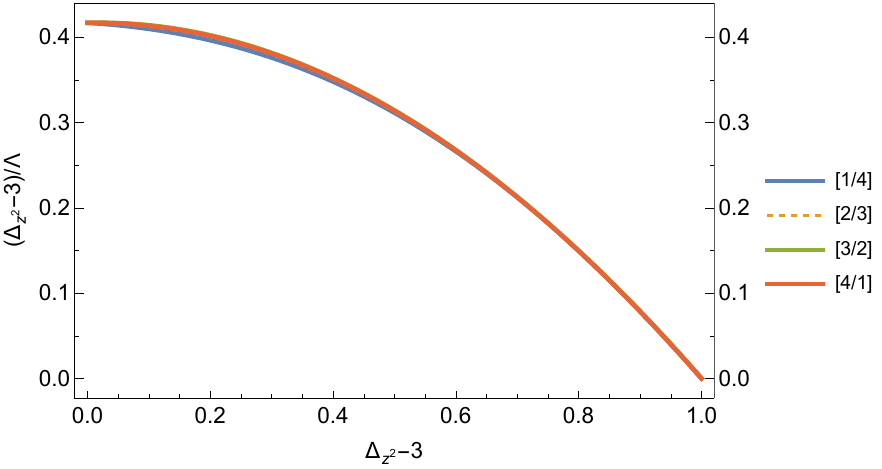}
    \caption{$\gamma_{\rm SM}=0$}
  \end{subfigure}
  \caption{(Yang-Mills flux tube in AdS$_3$). Various Pad\'e approximants for $\Lambda^{-1}(\Delta_{z^2}-3)$, the normalization of the displacement two-point function, for both $\gamma_{\rm SM}=0.3$ and $\gamma_{\rm SM}=0$. (We multiply $\Lambda^{-1}$ by $(\Delta_{z^2}-3)$ to remove the pole at $\Delta_{z^2}=3$.}
  \label{fig:YM FT Pade for disp norm}
\end{figure}

We present the Pad\'e approximants for the conformal dimensions, using perturbative results up to order $\varsigma^3$ at large AdS radius and at order $(1-\varsigma)^0$ at small AdS radius, in Figure~\ref{fig:Yang-Mills Flux Tube Dimensions Pade}. We present the Pad\'es for the OPE coefficient of $[z^2]$ using perturbative data up to order $\varsigma^3$ and $(1-\varsigma)$ in Figure~\ref{fig:YM FT Pade for OPE coefficient}. We present the Pad\'es for the normalization of the displacement in Figure~\ref{fig:YM FT Pade for disp norm}.

We make the following tentative observations:
\begin{itemize}
    \item Focusing first on the Pad\'es with $\gamma_{\rm SM}=0.3$, we generally see pretty good agreement between the different Pad\'es. This is especially true for the OPE coefficient and the displacement norm. 
    \item We also see pretty good agreement between the $[2/2]$ and $[1/3]$ Pad\'es for the conformal dimensions. They develop poles, but they seem relatively harmless. On the other hand, the $[3/1]$ Pad\'es differ significantly from the other two.
    \item The fact that the $[3/1]$ Pad\'es differ significantly from the $[2/2]$ and $[1/3]$ Pad\'es especially near $\Delta_{z^2}=3$ (the different Pad\'es for the third operator even differ in the sign of the first derivative) suggests that having one extra perturbative order at small AdS could significantly improve the precision of the Pad\'e. This aligns with the trends we observed in the Pad\'es for the $\mathcal{N}=4$ SYM observables.
    \item We want to emphasize that the conformal dimensions for operators on the flux tube in AdS behave qualitatively similarly to those on the supersymmetric Wilson line in $\mathcal{N}=4$ SYM. To the extent that we can trust the Pad\'es, we see that the conformal dimensions vary smoothly and are generally monotonic, with a change in dimension that is of order 1. Even the Pad\'es are generally similar for the two set-ups. In particular, poles arise generically and should not be taken as evidence that there is an error in the perturbative results or in the mapping between the Wilson line and effective string operators.
    \item Finally, it is interesting to note the behavior of the Pad\'es for the various observables when we set $\gamma_{\rm SM}=0$. Although this sets $\gamma_{\rm SM}$ to a value other than what we expect from the lattice, it does not seem to ``break’’ the Pad\'es. In fact, they ``improve,’’ in the sense that there are no poles and there is much better agreement between the different Pad\'es with different choices of $M$ and $N$. The first observation indicates that the perturbative data that we currently have access to cannot be used to provide an independent, analytic estimate of $\gamma_{\rm SM}$. Such an estimate should in principle be possible, given enough orders of perturbative input at both small and large AdS radius, but our present results are not there yet. The second observation --- that the Pad\'es for all observables are much ``better behaved'' when $\gamma_{\rm SM}=0$--- is puzzling; it could be a coincidence (e.g., the Pad\'es for the $\mathcal{N}=4$ SYM observables with the same amount of perturbative input were not so nicely behaved) or it could be an indication of some connection between the Yang-Mills flux tube and integrable worldsheet toy models. We leave the resolution of this puzzle for future work.
\end{itemize}

\FloatBarrier

\section{Conclusions}

In this work, continuing the program initiated in \cite{Gabai:2025hwf}, we have studied the Yang-Mills flux tube in AdS$_3$. We first derived the effective string action up to $O(\ell_s^4)$ and matched the symmetry-resolved spectra of operators at small AdS radius in the Wilson line description and at large AdS radius in the effective string description. We then computed the branon/displacement four-point function in the effective string description up to two-loops using the ansatz bootstrap. The correlator is fixed uniquely by imposing crossing, compatibility with the conformal block expansion, integrated correlator identities, and the flat space limit. Knowing the four-point function allowed us to extract conformal dimensions and OPE coefficients of the two-particle defect operators, modulo mixing, to three non-trivial orders in perturbation theory. Finally, we presented preliminary attempts to interpolate between small and large AdS radius using Pad\'e approximants, with the perturbative CFT data we computed here and in \cite{Gabai:2025hwf} as inputs. The results, while tentative, look promising.

There are many natural next steps that we plan to pursue. Firstly, it would be nice to push the ansatz bootstrap to three-loops, as was done in \cite{Ferrero:2021bsb,Ferrero:2023gnu,Ferrero:2023znz} for the supersymmetric Wilson line in $\mathcal{N}=4$ SYM. To achieve this, we must first solve the problem of mixing between the two-particle and higher particle operators, which may be possible by simultaneously bootstrapping the four-point functions of a suitable family of two-particle and multi-particle operators in addition to the branon four-point function. Extending the ansatz bootstrap analysis by one order would represent technical progress, and would also potentially yield Pad\'e interpolations that are more sensitive to the flat space phase shift parameter $\gamma_{\rm SM}$. We could conceivably use this sensitivity to put rough bounds on $\gamma_{\rm SM}$. While not fully rigorous, such bounds would be entirely analytic and could be compared with results from the lattice and rigorous bounds from the S-matrix bootstrap.

We would also like to push the perturbative analysis in the gauge theory description of the flux tube, valid in the regime of small AdS radius, to higher orders. This would put our perturbative results at small and large AdS radius on more equal footing. Moreover, the analysis in section~\ref{sec:bridging small and large AdS radius with Pade} and in \cite{Julius:2025dce} of the Pad\'e approximants for the $\mathcal{N}=4$ Wilson line, for which very precise results are available from integrability \cite{Grabner:2020nis,Cavaglia:2021bnz,Cavaglia:2022qpg}, suggests that, if we can indeed perform these next order calculations, the Pad\'e approximants for the flux tube observables may converge with a few percent precision at any radius. Given this motivation, it seems eminently achievable to compute the Witten diagrams contributing to the displacement four-point function at sub-leading order using the tools introduced in \cite{Gabai:2025hwf}. It may also be possible to shortcut the computation of diagrams using the integrated correlator bootstrap approach that was successfully implemented at leading order in \cite{Gabai:2025hwf}. Furthermore, this approach could potentially be combined with the ansatz bootstrap methods that we used in the effective string description in the present work, although the increased density of the spectrum of operators at small AdS radius, as compared with the spectrum at large AdS radius, may make the ansatz bootstrap more challenging. Finally, yet another way to study the weakly coupled gauge theory is to generalize and apply the formalism developed in \cite{Loparco:2026fki}, which recasts renormalization group flow for QFTs in AdS as an infinite system of coupled ODEs for the boundary CFT data, to Yang-Mills in AdS$_3$. This formalism could potentially also be used to further study the effective string theory description of the flux tube, which can be viewed as an (effective) quantum field theory in AdS$_2$, the specific case considered in \cite{Loparco:2026fki}. 

Another interesting direction is to employ the one dimensional conformal bootstrap to constrain line defect observables. The multi-correlator bootstrap, combined with the integral equations imposing non-linearly realized symmetry and conservative gap assumptions, is expected to give strong bounds on low lying conformal data. The comparison of these bounds with our perturbative results would provide a check of our set-up and, in particular, of the conjecture of smooth interpolation between small and large AdS radius.\footnote{We are grateful to Philine van Vliet for discussions and ongoing work, including promising early results, in this direction.}

We would also like to extend our analysis to the Yang-Mills flux tube in AdS$_4$. This would bring us one step closer to drawing conclusions that are relevant for real world QCD. On a technical level, we believe many of the tools we have used can be transferred without significant complications. On a qualitative level, we are hopeful that putting the flux tube in AdS$_4$ will give us an analytic handle on the worldsheet axion, which has been discovered in the lattice studies of flat-space flux tubes \cite{Dubovsky:2013gi,Athenodorou:2024loq}. Another interesting generalization is the inclusion of  matter fields and finite-$N$ corrections. These modifications can lead to additional states in the defect spectrum or to string breaking. In an optimistic scenario, sufficiently precise input from the small-radius regime could either exclude or necessitate the presence of such extra states within the effective string theory description.

%We would also like to extend our analysis to the Yang-Mills flux tube in AdS$_4$. This would bring us one step closer to drawing conclusions that are relevant for real world QCD. On a technical level, we believe many of the tools we have used can be transferred without significant complications. On a qualitative level, we are hopeful that putting the flux tube in AdS$_4$ will give us an analytic handle on the worldsheet axion, which has been discovered in the lattice studies of flat-space flux tubes \cite{Dubovsky:2013gi,Athenodorou:2024loq}. Another interesting generalization is the inclusion of  matter fields and finite-$N$ corrections. From the perspective of the defect spectrum, these modifications manifest themselves by the appearance of additional states, thereby altering the gaps between energy levels,  appearance of massive states on the worldsheet, or leading to string breaking. In an optimistic scenario, sufficiently precise input from the small-radius regime could either exclude or necessitate the presence of such extra states within the effective string theory description.

While a straight flux tube in a pure AdS background with Neumann boundary conditions in the most symmetric setting in which to study a confining string and interpolate to a perturbative regime, it is also worthwhile to consider less symmetric configurations that may better capture certain aspects of confinement. Examples include rotating flux tubes in AdS or flux tubes in backgrounds of the form $\mathrm{AdS}_d \times S^n$. Such less symmetric setups appear to be related to the study of Wilson line correlators in \cite{Albert:2026fqj}, where the short-distance behavior is likewise governed by perturbative Yang--Mills theory.

\section*{Acknowledgements}
We thank Sergei Dubovsky, Simone Giombi, Johan Henriksson, Julius Julius, Ami Katz, Shota Komatsu, Marco Meineri, Miguel Paulos, Joao Penedones,  Amit Sever, Philine van Vliet, and Sasha Zhiboedov for useful discussions. We especially thank Jiaxin Qiao for collaboration on related topics, for his contributions at the early stages of this work, and for many discussions. 
The work of BG and VG is supported by Simons Foundation grant 994310 (Simons Collaboration on Confinement and QCD Strings).

\appendix

\section{Details of Witten diagrams and D-functions in AdS$_2$}\label{app:Witten diagrams and D functions}

In this appendix, we provide the missing details needed to evaluate the contact diagrams in eqs.~\eqref{eq:0 deriv contact diagram}-\eqref{eq:8 deriv contact diagram S channel}. The relevant tools were introduced in the context of strings with AdS$_2$ geometry in \cite{Giombi:2017cqn}. 

The basic object for computing four-point contact diagrams are the $D$-functions:
\begin{align}\label{eq:D function def}
    D_{\Delta_1\Delta_2\Delta_3\Delta_4}(\tau_1,\tau_2,\tau_3,\tau_4)=\int \frac{dzdx}{z^2} \prod_{i=1}^4 \tilde{K}_{\Delta_i}(z,x;\tau_i).
\end{align}
They can be put in the form
\begin{align}\label{eq:D and reduced D functions}
    D_{\Delta_1\Delta_2\Delta_3\Delta_4}=\frac{\sqrt{\pi}\Gamma(\Sigma-\frac{1}{2})}{2\prod_{i=1}^4 \Gamma(\Delta_i)} \frac{\tau_{14}^{2(\Sigma-\Delta_1-\Delta_4)}\tau_{34}^{2(\Sigma-\Delta_3-\Delta_4)}}{\tau_{13}^{2(\Sigma-\Delta_4)}\tau_{24}^{2\Delta_2}}\bar{D}_{\Delta_1\Delta_2\Delta_3\Delta_4}(\chi),
\end{align}
where $\Sigma\equiv \frac{1}{2}\sum_i \Delta_i$. The $\bar{D}$ functions can be explicitly evaluated in terms of rational functions and logs; we provide explicit expressions for relevant ones below.

The zero-derivative contact diagram in eq.~\eqref{eq:0 deriv contact diagram} is proportional to $D_{2222}(\vec{\tau})$. By contrast, the four-derivative and eight-derivative contact diagrams in eq.~\eqref{eq:4 deriv contact diagram S channel} and eq.~\eqref{eq:8 deriv contact diagram S channel} involve boundary-to-bulk propagators dressed by derivatives. The four-derivative contact diagram can be reduced to sums of $D$-functions using the identity:
\begin{equation}
\begin{aligned}\label{eq:dKdK identity}
    &g^{\mu\nu}\partial_\mu \tilde{K}_{\Delta_1}(z,x;\tau_1)\partial_\nu \tilde{K}_{\Delta_2}(z,x;\tau_2)\\&\qquad\qquad=\Delta_1\Delta_2[\tilde{K}_{\Delta_1}(z,x;\tau_1)\tilde{K}_{\Delta_2}(z,x;\tau_2)-2\tau_{12}^2\tilde{K}_{\Delta_1+1}(z,x;\tau_1)\tilde{K}_{\Delta_2+1}(z,x;\tau_2)],
\end{aligned}
\end{equation}
which lets us trade derivatives acting on boundary-to-bulk propagators for higher conformal dimensions.

An analogous identity exists when two derivatives act on boundary-to-bulk propagators:
\begin{equation}
\begin{aligned}\label{eq:higher derivative bdy-to-bulk identity}
    &h^{\mu\nu}h^{\rho\sigma}\nabla_\mu \partial_\rho \tilde{K}_{\Delta_1}(z,x,\tau_1) \nabla_{\nu}\partial_{\sigma} \tilde{K}_{\Delta_2}(z,x,\tau_2)\\&\qquad\qquad=\Delta_1\Delta_2(1+\Delta_1\Delta_2)\tilde{K}_{\Delta_1}(z,x,\tau_1)\tilde{K}_{\Delta_1}(z,x,\tau_2)\\&\qquad\qquad\qquad -4\Delta_1\Delta_2(1+\Delta_1)(1+\Delta_2)\tau_{12}^2 \tilde{K}_{\Delta_1+1}(z,x,\tau_1)\tilde{K}_{\Delta_2+1}(z,x,\tau_2)\\&\qquad\qquad\qquad +4\Delta_1\Delta_2(1+\Delta_1)(1+\Delta_2)\tau_{12}^4 \tilde{K}_{\Delta_1+2}(z,x,\tau_1)\tilde{K}_{\Delta_2+2}(z,x,\tau_2)
\end{aligned}
\end{equation}
Here, $\nabla_\mu \partial_\nu f = \partial_\mu \partial_\nu f -\hat{\Gamma}^\rho_{\mu\nu}\partial_\rho f$, with the non-zero components of the connection on AdS$_2$ in Poincar\'e coordinates given by $\hat{\Gamma}^x_{xz}=\hat{\Gamma}^x_{zx}=-\frac{1}{z},
    \hat{\Gamma}^z_{xx}=\frac{1}{z},
    \hat{\Gamma}^z_{zz}=-\frac{1}{z}.$

Substituting these identities into eq.~\eqref{eq:4 deriv contact diagram S channel} and eq.~\eqref{eq:8 deriv contact diagram S channel}, it follows then that the contact diagrams are:
\begin{align}
     X_{0-\text{der}}(\vec{\tau})=&24\mathcal{C}_2^4 D_{2222}(\vec{x}),\\
     X_{4-\text{der}}^{\rm S}(\vec{\tau})=&128\mathcal{C}_2^4 \big[D_{2222}(\vec{x})-2\tau_{12}^2 D_{3322}(\vec{x})-2\tau_{34}^2 D_{2233}(\vec{x})+4\tau_{12}^2\tau_{34}^2D_{3333}(\vec{x})\big],\\
     X_{8-\text{der}}^{\rm S}(\vec{\tau})=&128\mathcal{C}_2^4\big[25D_{2222}(\vec{x})+1296\tau_{12}^2\tau_{34}^2D_{3333}(\vec{x})+1296\tau_{12}^4\tau_{34}^4D_{4444}(\vec{x})\nonumber\\&-180\tau_{12}^2 D_{3322}(\vec{x})-180\tau_{34}^2D_{2233}(\vec{x})+180\tau_{12}^4 D_{4422}(\vec{x})\\&+180\tau_{34}^4 D_{2244}(\vec{x})-1296\tau_{12}^4\tau_{34}^2 D_{4433}(\vec{x})-1296\tau_{12}^2\tau_{34}^4D_{3344}(\vec{x})\big]\nonumber
\end{align}
Replacing the $D$-functions by their reduced forms via eq.~\eqref{eq:D and reduced D functions} yields
\begin{align}
    X_{0-\text{der}}(\chi)=&\frac{\mathcal{C}^{2}_2}{ \tau_{12}^4 \tau_{34}^4}\frac{10 \chi^4}{\pi}\bar{D}_{2222}(\chi),\\
    X^{\rm S}_{4-\text{der}}(\chi)=&\frac{\mathcal{C}^{2}_2}{ \tau_{12}^4 \tau_{34}^4}\bigg[\frac{160\chi^4}{3\pi}\bar{D}_{2222}(\chi)-\frac{280\chi^6}{3\pi}\bar{D}_{3322}(\chi)\\&-\frac{280\chi^4}{3\pi}\bar{D}_{2233}(\chi)+\frac{210\chi^6}{\pi}\bar{D}_{3333}(\chi)\bigg],\\
    X^{\rm S}_{8-\text{der}}(\chi)=&\frac{\mathcal{C}^{2}_2}{ \tau_{12}^4 \tau_{34}^4}\bigg[\frac{4000}{3\pi}\chi^4\bar{D}_{2222}(\chi)+\frac{68040\chi^6}{\pi}\bar{D}_{3333}(\chi)+\frac{30030\chi^8}{\pi}\bar{D}_{4444}(\chi)\nonumber\\&-\frac{8400\chi^4}{\pi}\bar{D}_{2233}-\frac{8400\chi^6}{\pi}\bar{D}_{3322}+\frac{4200\chi^8}{\pi}\bar{D}_{4422}(\chi)\nonumber\\&+\frac{4200\chi^4}{\pi}\bar{D}_{2244}(\chi)-\frac{41580\chi^8}{\pi}\bar{D}_{4433}-\frac{41580\chi^6}{\pi}\bar{D}_{3344}\bigg].
\end{align}
(Here, $\mathcal{C}_2\equiv \frac{2\pi}{3}$.) Summing over the different channels for the four-derivative and eight-derivative contact diagrams in accordance with eqs.~\eqref{eq:4 deriv contact diagram}-\eqref{eq:8 deriv contact diagram} and using the explicit expressions for the reduced $D$ functions, which we provide in the next sub-section, yields the explicit expressions for the $n$-derivative contact diagrams. Finally, taking the linear combinations prescribed in eq.~\eqref{eq:z-to-the-fourth contact diagram def}-\eqref{eq:higher curvature contact diagram def}, this leads to the final expressions for the $z^4$, Nambu-Goto, and higher-curvature contact diagrams given in eqs.~\eqref{eq:z-to-the-fourth contact diagram final expression}-\eqref{eq:higher curvature contact diagram final expression}.

\subsection{Explicit expressions for reduced $D$ functions}

\begin{align}
    \bar{D}_{2222}(\chi)&=\frac{\left(2 \chi ^2-5 \chi +5\right) \log | \chi |}{15 (\chi -1)^3}-\frac{\left(2 \chi ^2+\chi +2\right) \log | 1-\chi |}{15 \chi ^3}-\frac{2 ((\chi -1) \chi
   +1)}{15 (\chi -1)^2 \chi ^2},\\
    \bar{D}_{3322}(\chi)&=\chi^{-2}\bar{D}_{2233}(\chi)=\frac{\left(9 \chi ^2-28 \chi +28\right) \log | \chi |}{105 (\chi -1)^4}+\frac{\left(-9 \chi ^3-8 \chi ^2-6 \chi -12\right) \log | 1-\chi |}{105 \chi
   ^5}\nonumber\\&+\frac{-18 \chi ^4+29 \chi ^3-5 \chi ^2-48 \chi +24}{210 (\chi -1)^3 \chi ^4},\\
    \bar{D}_{3333}(\chi)&=\frac{\left(8 \chi ^4-36 \chi ^3+64 \chi ^2-56 \chi +28\right) \log | \chi |}{105 (\chi -1)^5}\nonumber\\&+\frac{\left(-8 \chi ^4-4 \chi ^3-4 \chi ^2-4 \chi -8\right) \log |
   1-\chi |}{105 \chi ^5}\nonumber\\&+\frac{-24 \chi ^6+72 \chi ^5-74 \chi ^4+28 \chi ^3-74 \chi ^2+72 \chi -24}{315 (\chi -1)^4 \chi ^4},\\
    \bar{D}_{4422}(\chi)&=\chi^{-4}\bar{D}_{2244}(\chi)=\frac{\left(8 \chi ^2-27 \chi +27\right) \log | \chi |}{105 (\chi -1)^5}\nonumber\\&+\frac{\left(-8 \chi ^4-13 \chi ^3-12 \chi ^2-10 \chi -20\right) \log | 1-\chi |}{105 \chi
   ^7}\nonumber\\&+\frac{-48 \chi ^6+90 \chi ^5-7 \chi ^4-46 \chi ^3-277 \chi ^2+360 \chi -120}{630 (\chi -1)^4 \chi ^6},\\
    \bar{D}_{4433}(\chi)&=\chi^{-2}\bar{D}_{3344}(\chi)=\frac{\left(100 \chi ^4-528 \chi ^3+1122 \chi ^2-1188 \chi +594\right) \log | \chi |}{1155 (\chi -1)^6}\nonumber\\&+\frac{\left(-100 \chi ^5-72 \chi ^4-54 \chi ^3-56 \chi ^2-60
   \chi -120\right) \log | 1-\chi |}{1155 \chi ^7}\nonumber\\&+\frac{-300 \chi ^8+1134 \chi ^7-1540 \chi ^6+794 \chi ^5+8 \chi ^4-1494 \chi ^3+2178 \chi ^2-1440 \chi +360}{3465
   (\chi -1)^5 \chi ^6},\\
    \bar{D}_{4444}(\chi)&=\frac{\left(600 \chi ^6-3900 \chi ^5+10764 \chi ^4-16302 \chi ^3+14586 \chi ^2-7722 \chi +2574\right) \log | \chi |}{5005 (\chi -1)^7}\nonumber\\&+\frac{\left(-600 \chi ^6-300
   \chi ^5-264 \chi ^4-246 \chi ^3-264 \chi ^2-300 \chi -600\right) \log | 1-\chi |}{5005 \chi ^7}\nonumber\\&+\frac{1}{5005 (\chi -1)^6 \chi ^6}\big[-600 \chi ^{10}+3000 \chi ^9-6014 \chi ^8+6056 \chi ^7-3112
   \chi ^6\nonumber\\&\qquad\qquad\qquad +740 \chi ^5-3112 \chi ^4+6056 \chi ^3-6014 \chi ^2+3000 \chi -600\big],
\end{align}

\section{Operator mixing and anomalous dimensions at tree level}\label{app:unmixing at tree level}

In this appendix, we extract free and tree-level OPE data of some of the lowest dimension primaries from the four-point function computed in section~\ref{sec:tree-level four point function via Witten diagrammatics}. Because the extraction of OPE data via the conformal block expansion can suffer from mixing issues, in this appendix we proceed via more direct but laborious methods to unmix, at least partially, the operators and their OPE data.\footnote{The mixing problem in the context of the half-BPS Wilson line in $\mathcal{N}=4$ SYM is discussed in detail in \cite{Ferrero:2023gnu}.}

Let us first briefly define what we mean by mixing and why it presents challenges when extracting OPE data via the conformal block expansion. The term ``mixing'' in this context can refer to two related but distinct concepts and we might distinguish between them by calling the first ``operator mixing'' and the second ``OPE data mixing.'' This is best illustrated with a concrete example. As we can see from Table~\ref{tab:strong coupling}, there are two independent primaries with dimension $\Delta=8$ and $++$ symmetry in the free-limit of the EST; this is the lowest dimension where the mixing issue arises in the analysis of the branon four point function. Let us call these two primaries $O_1$ and $O_2$. They satisfy $\braket{O_i(\tau_1)O_j(\tau_2)}=\delta_{ij}/|\tau_{12}|^{16}$. Of course, this basis is not unique in the free theory because we can always define rotated operators $\tilde{O}_i=R_{ij}O_j$, such that $R^TR=I_{2\times 2}$, which are also primaries and have orthogonal two point functions. However, once the interaction coupling $\varsigma$ is turned on, the primaries will pick up anomalous dimensions, the degeneracy will typically be lifted, and we can pick out the ``good basis'' of degenerate primaries in the free theory by tracing the two non-degenerate primaries at finite $\varsigma$ back to $\varsigma=0$. Thus, ``operator mixing'' is the CFT version of the state mixing that occurs in degenerate perturbation theory in quantum mechanics, and ``unmixing operators'' means identifying the good basis of primaries order by order in perturbation theory.

To understand the related concept of OPE data mixing, let us now consider the contribution of the operators $O_1$ and $O_2$ to the branon four point function:
\begin{align}
    G(\chi)\ni a_{O_1} \mathfrak{f}_{\Delta_{O_1}}(\chi)+a_{O_2} \mathfrak{f}_{\Delta_{O_2}}(\chi),
\end{align}
where $a_{O_i}\equiv c_{zzO_i}^2$ is the squared OPE coefficient and $\mathfrak{f}_\Delta(\chi)\equiv \chi^\Delta {_2F_1}(\Delta,\Delta,2\Delta,\chi)$ is the conformal block in 1d. Perturbatively, $a_{O_i}\equiv a_{O_i}^{(0)}+\varsigma a_{O_i}^{(1)}+O(\varsigma^2)$ and $\Delta_{O_i}\equiv 8+\varsigma \gamma_{O_i}^{(1)}+O(\varsigma^2)$, so that the contributions of $O_1$ and $O_2$ to the four point function are:
\begin{align}
    &G^{(0)}(\chi)+\varsigma G^{(1)}(\chi)\ni (a_{O_1}^{(0)}+a_{O_2}^{(0)})\mathfrak{f}_{8}(\chi)\\&+\left[(a_{O_1}^{(0)}\gamma_{O_1}^{(1)}+a_{O_2}^{(0)}\gamma_{O_1}^{(1)})\left(\log(\chi)\mathfrak{f}_8(\chi)+\chi^x \partial_x(\chi^{-x}\mathfrak{f}_x(\chi))\big\rvert_{x=8}\right)+(a_{O_1}^{(1)}+a_{O_2}^{(1)})\mathfrak{f}_8(\chi)\right]\varsigma.\nonumber
\end{align}
Thus, knowing $G^{(0)}$ and $G^{(1)}$ and expanding them in conformal blocks allows one to, at most, determine the following combinations of OPE coefficents and anomalous dimensions:
\begin{align}
    a_{O_1}^{(0)}+a_{O_2}^{(0)},\qquad a_{O_1}^{(0)}\gamma_{O_1}^{(1)}+a_{O_2}^{(0)}\gamma_{O_2}^{(1)}, \qquad a_{O_1}^{(1)}+a_{O_2}^{(1)},
\end{align}
because they always appear together as coefficients of the same conformal blocks and their derivatives. In particular, it is not possible using just the branon four point function to determine the OPE coefficients and anomalous dimensions of $O_1$ and $O_2$ separately. This phenomenon is general: decomposing a perturbative four-point function in the conformal block expansion will typically yield ``mixed OPE data'' whenever there are degenerate operators in the free theory, and this problem persists at all orders in perturbation theory. As we discuss in the section~\ref{sec:ansatz bootstrap}, OPE data mixing is a major obstacle to the analytic bootstrap.

In the remainder of this appendix, we will try to better understand mixing in the branon four-point function by directly computing some of the OPE data and anomalous dimensions up to order $\varsigma^1$ of various composite primaries. This is possible because we can construct the primaries explicitly as linear combinations of products of derivatives acting on the branon field $z$, and because at order $\varsigma^1$ the only interacting diagram that contributes is the four-point contact diagram that we computed in eq.~\eqref{eq:tree level four point function}.

\subsection{Primaries in the free theory}

We will now construct the first few primary operators in the generalized free theory with a single fundamental field $z$ of dimension $\Delta_z=2$. We counted the operators in Fig.~\ref{tab:strong coupling}; now we give their explicit expressions.

\begin{figure}[t]
    \centering
    \begin{subfigure}[t]{0.32\textwidth}
        \centering
        \includegraphics[width=\textwidth]{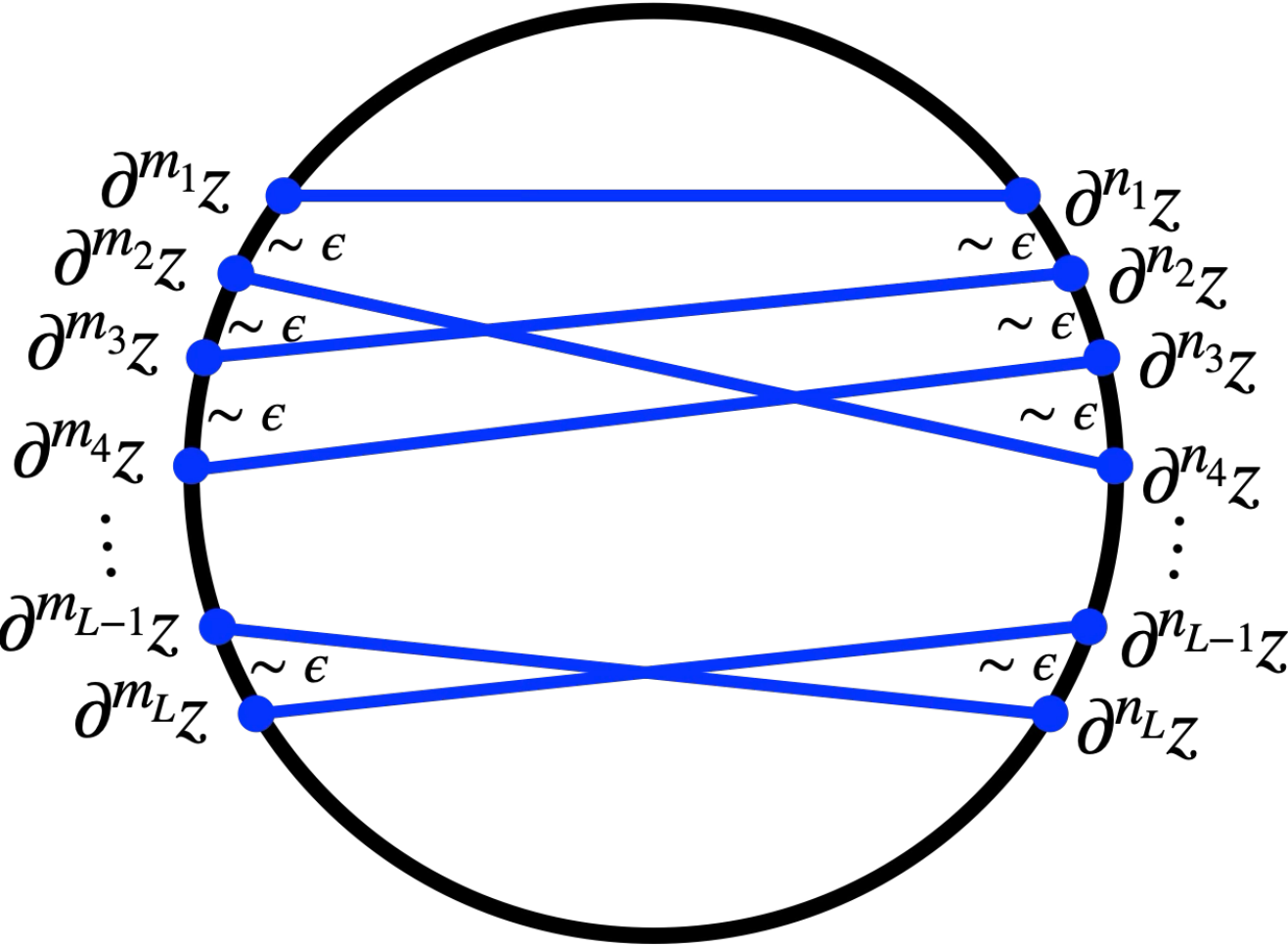}
        \caption{}
        \label{fig:all wick contractions}
    \end{subfigure}\hfill
    \begin{subfigure}[t]{0.32\textwidth}
        \centering
        \includegraphics[width=\textwidth]{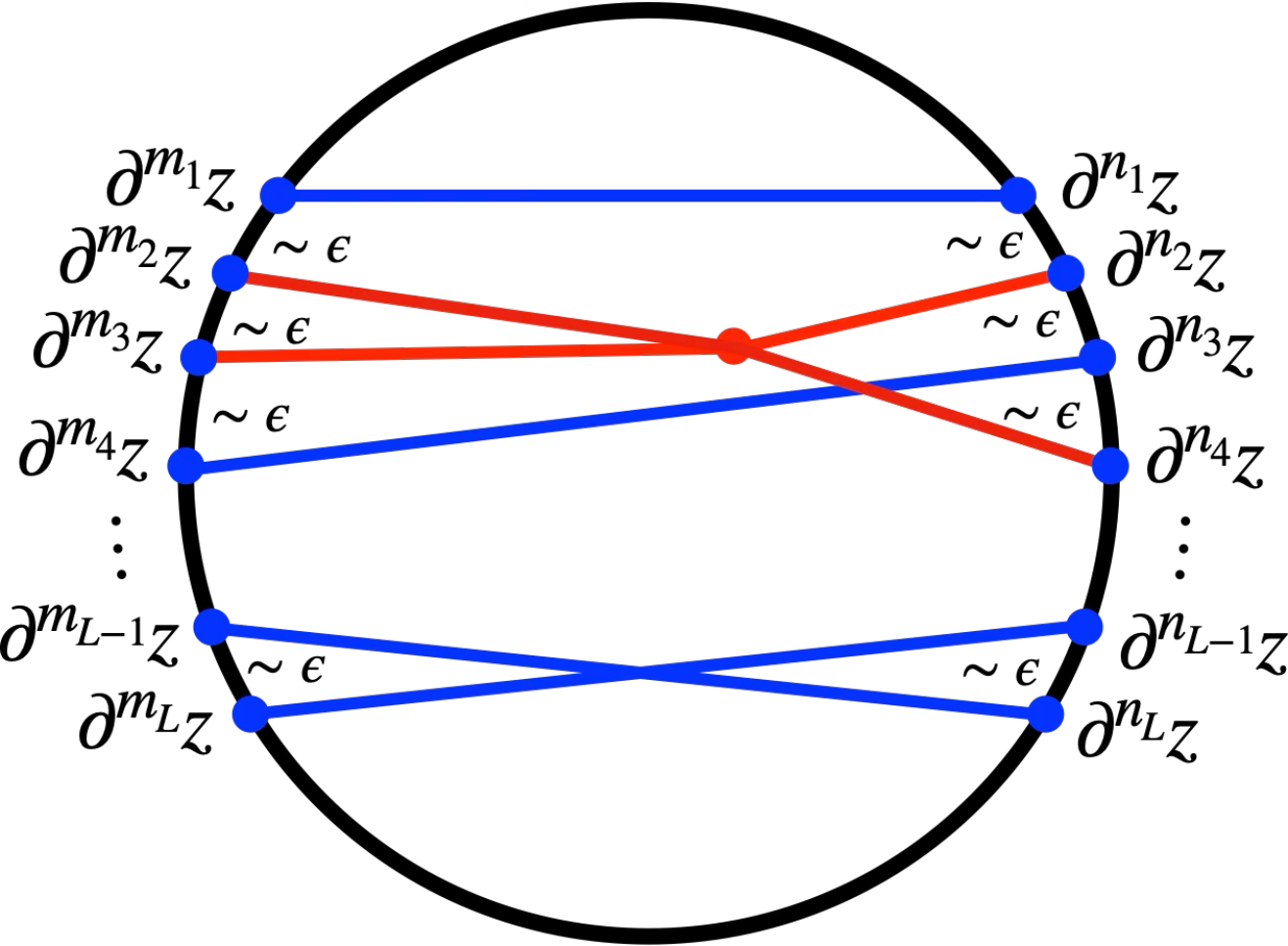}
        \caption{}
    \end{subfigure}\hfill
    \begin{subfigure}[t]{0.32\textwidth}
        \centering
        \includegraphics[width=\textwidth]{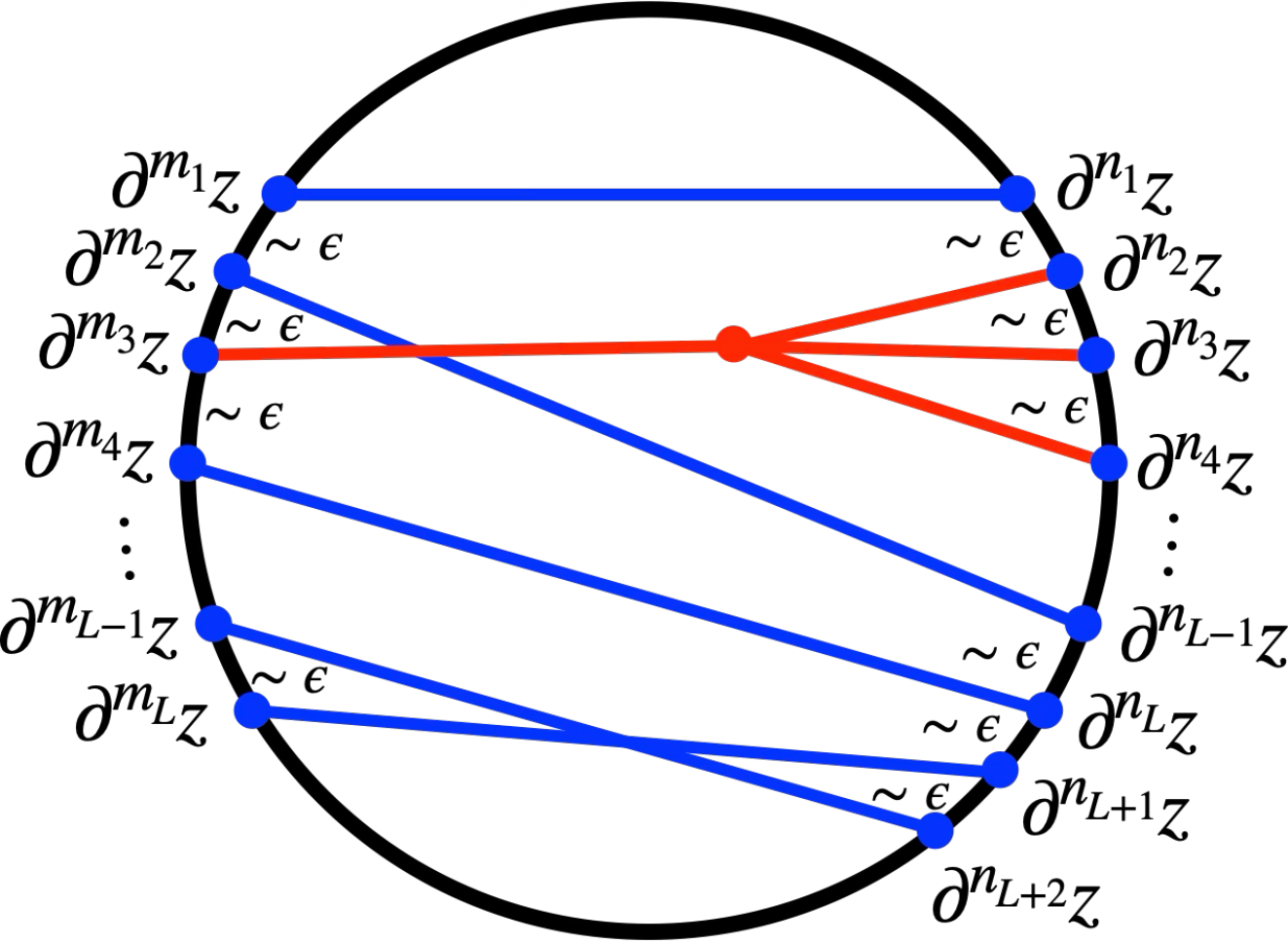}
        \caption{}
    \end{subfigure}
    \caption{There are three classes of diagrams that can contribute to the two-point function of point-split-regularized operators up to order $\varsigma$. \textbf{(a)} One contribution at order $\varsigma^0$ to the two-point function of two $\epsilon$-regularized $L-$particle operators $O_1=\prod_{i=1}^L \partial^{m_i}z$ and $O_2=\prod_{i=1}^L \partial^{n_i}z$ in which all fields are Wick contracted. \textbf{(b)} One contribution to the same two-point function at order $\varsigma^1$ in which two fields each are contracted by the four-point contact diagram, and the remaining fields are Wick contracted. \textbf{(c)} One contribution to the two-point function of $\epsilon$-regularized $L-$ and $(L+2)-$ particle operators  $O_1=\prod_{i=1}^L \partial^{m_i}z$ and $O_2=\prod_{i=1}^{L+2} \partial^{n_i}z$, in which one field from $O_1$ and three fields from $O_2$ are contracted by the contact diagram, and the remaining fields are Wick contracted. As discussed in the text, only the middle class of diagrams give rise to $\log$s in the limit $\epsilon\to0$. This implies only primaries with equal numbers of fields mix at order $\varsigma^0$.}
    \label{fig: two-pt functions wick contractions}
\end{figure}

To construct the primary $O$ of dimension $\Delta^{(0)}$, we consider a general linear combination of all normal-ordered composite operators of the form $\partial^{n_1}z \partial^{n_2}z \ldots \partial^{n_k} z$, $n_1\leq n_2\leq \ldots \leq n_k$, with $k\geq 0$ and $n_i\geq 0$ satisfying $2k+\sum_{i=1}^k n_i=\Delta^{(0)}$, and then fix the coefficients so that the resulting operator is primary. There are two ways to fix the coefficients. First, we can demand that the linear combination be annihilated by the special conformal generator, $K$. Recall that $Kz(0)=0$ (because $z$ is primary), $K$ obeys the Leibniz rule (i.e., $K(O_1O_2)=(KO_1)O_2+O_1(KO_2)$), and $K(\partial^n z(0))=[n(n-1)+2n\Delta_z]\partial^{n-1}z(0)$.\footnote{This follows from the conformal algebra, $[D,P]=P$, $[D,K]=-K$, $[K,P]=2D$, together with $[P,z]=\partial z$ and $[D,z]=\Delta_z z$.}
Alternatively, one can proceed recursively and fix the coefficients of the primary at dimension $\Delta$ by demanding that its two point function with primaries of lower dimensions be zero. (Two-point functions are computed by performing Wick contractions between pairs of fields in the two composite operators using the branon two-point function, $\braket{z(\tau_1)z(\tau_2)}=\frac{1}{\tau_{12}^4}$; see Figure~\ref{fig:all wick contractions} for an example. All contractions between copies of $z$ within the same composite operator are excluded, one consequence of which is that two-point functions between operators with different numbers of $z$ are zero.) The last step is to normalize the operator. These two methods of constructing the primaries are equivalent, and the explicit expressions for all the primaries up to $\Delta^{(0)}=12$ are given in Table~\ref{tab:table of GFF primaries}.

Note that there are many degenerate operators with the same \textbf{R} and \textbf{CT} parities, so the primaries given in Table~\ref{tab:table of GFF primaries} merely reflect one possible choice of basis. To partially resolve this ambiguity, we always define the GFF primaries so that they involve a linear combination of terms with a fixed number of copies of $z$. This makes it natural to use the notation 
\begin{align}
    [z^k]_n
\end{align}
to denote a primary with $k$ copies of $z$ and $n$ derivatives. We call this a $k$-particle primary. When this does not fully resolve the degeneracy (e.g., because there are multiple independent primaries with $k$ fields and dimension $2k+n$), then we add an extra subscript $A=1,2,\ldots$ to distinguish the different orthogonal primaries (that we define arbitrarily). For example, there are two three-particle primaries, $[z^3]_{6,1}$ and $[z^3]_{6,2}$, and two four-particle primaries, $[z^4]_{4,1}$ and $[z^4]_{4,2}$, of dimension $\Delta=12$.

\begin{table}[h!]
\centering
\renewcommand{\arraystretch}{1.65}
\begin{tabular}{c|c|c|c|c}
\textbf{Operator} & $\Delta$ & \textbf{Particle no.} & \textbf{R} & \textbf{CT} \\ \hline
$1$ & 0 & $0$ & $+$ & $+$ \\
\hline
$z$ & 2 & $1$ & $-$ & $+$ \\
\hline
$[z^2]_0\equiv \frac{z^2}{\sqrt{2}}$ & 4 & $2$ & $+$ & $+$ \\
\hline
$[z^2]_2\equiv \frac{z\partial^2 z-\frac{5}{4}(\partial z)^2}{3\sqrt{10}}$ & 6 & $2$ & $+$ & $+$ \\
$[z^3]_0\equiv \frac{z^3}{\sqrt{6}}$ & 6 & $3$ & $-$ & $+$ \\
\hline
 $[z^2]_4\equiv \frac{z\partial^4 z-7 \partial z\partial^3 z +\frac{63}{10} (\partial^2 z)^2}{12\sqrt{2002}}$ & 8 & $2$ &  $+$& $+$ \\
 $[z^3]_2\equiv \frac{z^2 \partial^2 z - \frac{5}{4}z (\partial z)^2}{\sqrt{130}}$ & 8 & $3$ & $-$ & $+$ \\
$[z^4]_0\equiv \frac{z^4}{2\sqrt{6}}$ & 8 & $4$ & $+$ & $+$ \\
\hline
$[z^3]_3\equiv i\frac{z^2 \partial^3 z -\frac{9}{2} z \partial z\partial^2 z +\frac{15}{4}(\partial z)^3}{12\sqrt{70}}$ & 9 & $3$ &  $-$ & $-$ \\
\hline
 $[z^2]_6\equiv \frac{z\partial^6 z-\frac{27}{2} \partial z \partial^5 z +54 \partial^2z\partial^4 z -42 (\partial^3z)^2 }{720\sqrt{9282}}$ & 10 & $2$ & $+$ & $+$ \\
 $[z^3]_4\equiv \frac{z^2 \partial^4 z-7 z \partial z \partial^3 z+\frac{63}{10} z (\partial^2 z )^2}{36\sqrt{238}}$ & 10 & $3$ & $-$ & $+$ \\
$[z^4]_2\equiv \frac{z^3 \partial^2 z -\frac{5}{4} z^2 (\partial z)^2}{2\sqrt{85}}$ & 10 & $4$ & $+$ & $+$ \\
 $[z^5]_0\equiv \frac{z^5}{2\sqrt{30}}$ & 10 & $5$ & $-$ & $+$ \\
 \hline
 $[z^3]_5\equiv i\frac{z^2 \partial^5 z-10 z \partial z \partial^4 z +\frac{28}{5} z \partial^2 z \partial^3 z+28 (\partial z)^2\partial^3 z-\frac{126}{5}\partial z(\partial^2 z)^2}{96\sqrt{3990}}$& 11 & $3$ & $-$ & $-$ \\
 $[z^4]_3\equiv i\frac{z^3 \partial^3 z-\frac{9}{2} z^2 \partial z\partial^2 z+\frac{15}{4} z (\partial z)^3}{90\sqrt{2}}$ & 11 & $4$ & $+$ & $-$ \\
 \hline
 $[z^2]_8\equiv \frac{z\partial^8 z -22\partial z \partial^7z +154 \partial^2 z \partial^6 z -462 \partial^3 z \partial^5 z+330 (\partial^4 z)^2}{120960\sqrt{17765}}$ & 12 & $2$ & $+$ & $+$ \\ $[z^3]_{6A}\equiv \frac{z^2 \partial^6z-\frac{27}{2}z \partial z\partial^5 z +54 z \partial^2 z\partial^4 z-42 z (\partial^3 z)^2}{720\sqrt{9366}}$
 & 12 & $3$ & $-$ & $+$ \\
 $[z^3]_{6B}\equiv${\scriptsize$\frac{13}{504\sqrt{7923190}}\bigg[{\tiny z^2 \partial^6z -\frac{27}{2} z \partial z \partial^5 z +\frac{33}{13} z \partial^2 z \partial^4 z +\frac{469}{26}z(\partial^3 z)^2}$} & 12 & $3$ & $-$ & $+$ \\
 \qquad\qquad\qquad\qquad {\scriptsize $ +\frac{3345}{52}(\partial z)^2 \partial^4 z-\frac{4683}{26}\partial z \partial^2 z \partial^3 z+\frac{14049}{130}(\partial^2 z)^3\bigg]$} & & & & 
 \\
 
$[z^4]_{4A}\equiv \frac{z^3 \partial^4 z-7 z^2 \partial z\partial^3 z+\frac{63}{10} z^2 (\partial^2 z)^2}{24\sqrt{1141}}$ & 12 & $4$ & $+$ & $+$ \\ $[z^4]_{4B}\equiv \frac{z^3 \partial^4 z -7 z^2 \partial z\partial^3 z-10 z^2 (\partial^2 z)^2+\frac{163}{4} z (\partial z)^2 \partial^2 z -\frac{815}{32} (\partial z)^4}{\sqrt{7154070}}$
 & 12 & $4$ & $+$ & $+$ \\
 $[z^5]_2\equiv \frac{z^4 \partial^2 z -\frac{5}{4} z^3 (\partial z)^2}{6\sqrt{35}}$ & 12 & $5$ & $-$ & $+$ \\
 $[z^6]_0\equiv \frac{z^6}{12\sqrt{5}}$ & 12 & $6$ & $+$ & $+$
\end{tabular}
\caption{All primaries in the generalized free theory with dimension $\Delta\leq 12$. }
\label{tab:table of GFF primaries}
\end{table}

\subsection{Tree-level anomalous dimensions and free-OPE coefficients}\label{eq:tree-level dimensions and free OPE coefficients}

Next, we turn to the two-point functions of the primaries at tree-level. We will only sketch the construction of the primaries up to order $\varsigma^1$, as this is sufficient to determine the tree-level anomalous dimensions and free-level OPE coefficients of all primaries. 

When computing correlation functions of composite operators up to order $\varsigma$, in addition to contracting pairs of $z$'s with the two-point function $\braket{z(\tau_1)z(\tau_2)}=\frac{1}{\tau_{12}^4}$, we can also contract four copies of $z$ with the four-point function diagram given in eq.~\eqref{eq:tree level four point function}. This leads to divergences when two or more of the legs of the contact diagram contract copies of $z$ within the same composite operator. One way to handle these divergences is using point-splitting, in which different copies of $z$ within the same composite operator are separated by distances of order $\epsilon$, and then we eventually take $\epsilon\to 0$. This introduces a length scale $\epsilon$ that breaks the conformal symmetry, as well as a lot of scheme dependence depending on how precisely the composite operators are split. However, both conformal symmetry and scheme-independence are restored when $\epsilon\to 0$. 

We can define the primaries in the point-splitting scheme recursively, as follows. We write the general expression for the primary of dimension $\Delta=\Delta^{(0)}+O(\varsigma)$ up to order $\varsigma$ as a sum of all point-split products of $k$ copies of $z$, dressed with $n$ derivatives, for all values of $k$ and $n$ that satisfy $2k+n\leq \Delta^{(0)}$ and are consistent with the parity of the primary. By dimensional analysis, each term in the linear combination must be multiplied by a factor of $\epsilon^{2k+n-\Delta^{(0)}}$. (A concrete manifestation of the fact that point splitting breaks conformal symmetry is that it is necessary to include composite operators of lower classical dimension multiplied by negative powers of $\epsilon$.) The coefficient of each term in the linear combination is of the form $\#+\# \varsigma$. Requiring that the operator reduce to a primary in the generalized free theory as $\varsigma\to 0$ fixes the zeroth order coefficients (up to possible rotations between multiple primaries with the same classical dimension; we'll say more about this below). Meanwhile, the order $\varsigma$ terms in the coefficients are fixed by ensuring that the primary has zero two-point function with all lower primaries and other primaries with the same classical dimension, and is unit-normalized. When computing the two-point functions, we include diagrams of all three types in Figure~\ref{fig: two-pt functions wick contractions}, and expand to linear order in $\varsigma$ and zeroth order in $\epsilon$.

To make the construction of primaries in the previous paragraph more concrete, consider an example. The lowest dimension primary with parity $++$ in the point splitting representation takes the form:
\begin{align}\label{eq:zSq primary pointsplit at order varsigma}
    [z^2]_0^\epsilon(0)\equiv \left(\frac{1}{\sqrt{2}}+a_{01} \varsigma \right)z(\epsilon)z(-\epsilon).
\end{align}
The next lowest takes the form:
\begin{equation}
\begin{aligned}\label{eq:zdSqz primary pointsplit at order varsigma}
    [z^2]_2^\epsilon(0)&\equiv \left(\frac{1}{3\sqrt{10}}+a_{21} \varsigma\right) [z(\epsilon) \partial^2 z(-\epsilon)+z(-\epsilon)\partial^2 z(\epsilon)]\\&+\left(-\frac{5}{12\sqrt{10}}+a_{22}\varsigma\right)\partial z(\epsilon)\partial z(-\epsilon)+\frac{a_{23} \varsigma}{\epsilon^2}z(\epsilon)z(-\epsilon).
\end{aligned}
\end{equation}
We added the superscript $\epsilon$ to distinguish the primaries in the interacting theory from those in the free theory. We read off the order $\varsigma^0$ coefficients from Table~\ref{tab:table of GFF primaries}; the order $\varsigma^1$ coefficients $a_{01}$, $a_{21},a_{22},a_{23}$ need to be determined. Requiring that $[z^2]_0^\epsilon$ have a unit-normalized two-point function of the form $\braket{O(\tau)O(0)}=\frac{(2\epsilon)^{2\gamma_O}}{\tau^{2\Delta_O}}$ fixes $a_{01}=\frac{251}{240\sqrt{2}}$. Likewise, requiring that $[z^2]^\epsilon_2$ have zero two-point function with $[z^2]^\epsilon_0$ and that it be unit normalized fixes $a_{21}=\frac{24859}{4032}$, $a_{22}=\frac{116231}{20160}$, $a_{23}=\frac{3}{16\sqrt{10}}$. From the calculation of the two point function we also directly read off $\gamma_{[z^2]}=-1$ and $\gamma_{[z^2]_2}=-\frac{13}{4}$.

This procedure quickly becomes computationally costly and it is challenging to construct the primaries beyond classical dimension $\Delta^{(0)}=8$ or $10$. This is due to the proliferation of terms appearing in the primaries, to the combinatorial increase in the number of diagrams contributing to the two-point functions of composite operators, and because it is costly to take many derivatives of the four-point contact diagram with respect to the external points. Fortunately, if we restrict our goal to determining only the tree-level anomalous dimensions and free-level OPE coefficients of all the primaries (but not the explicit expressions for the primaries or their tree-level OPE coefficients), then we can simplify the computation significantly.

In particular, in pursuit of our more modest goal, it is sufficient to focus on $\log$ terms in the two-point functions, which are insensitive to many of the details of the four-point contact diagram and the point-splitting procedure. Let us review why these are so. First, recall that the two-point function of a primary $O$ with dimension $\Delta=\Delta^{(0)}+\gamma^{(1)} \varsigma+O(\varsigma^2)$ takes the form:
\begin{align}
    \braket{O(\tau_1)O(\tau_2)}&=\frac{\epsilon^{2(\Delta-\Delta^{(0)})}}{\tau_{12}^{2\Delta}}=\frac{1}{\tau_{12}^{2\Delta^{(0)}}}\left(1-\varsigma \gamma^{(1)} \log{\frac{\tau_{12}^2}{\epsilon^2}}+O(\varsigma^2)\right).
\end{align}
Thus, the anomalous dimension is the coefficient of the log term. More generally, the two-point functions of classically-degenerate primaries $O_i$ with dimensions $\Delta_i=\Delta^{(0)}+\gamma_i^{(1)} \varsigma+O(\varsigma^2)$ are
\begin{align}\label{eq:45yu654e}
    \braket{O_i(\tau_1)O_j(\tau_2)}&=\frac{\epsilon^{2(\Delta_i-\Delta^{(0)})}}{\tau_{12}^{2\Delta_i}}\delta_{ij}=\frac{1}{\tau_{12}^{2\Delta^{(0)}}}\left(1-\varsigma \gamma_i^{(1)} \log{\frac{\tau_{12}^2}{\epsilon^2}}+O(\varsigma^2)\right)\delta_{ij}.
\end{align}
However, before unmixing the primaries up to order $\varsigma$, the more readily accessible observables are the two point functions of some linear combinations of the primaries, $W_i=R_{ij}O_j$, where $R_{ij}^{(0)}+\varsigma R_{ij}^{(1)}+O(\varsigma^2)$. Their two point function is:
\begin{align}
    \braket{W_i(\tau_1)W_j(\tau_2)}&=\frac{1}{\tau_{12}^{2\Delta^{(0)}}}\left(R_{ik}^{(0)}R_{jk}^{(0)}+\varsigma \bigg[-R_{ik}^{(0)}\Gamma_{kl}^{(1)}R_{jl}^{(0)}\log{\frac{\tau_{12}^2}{\epsilon^2}}+R_{ik}^{(0)}R_{jk}^{(1)}+R_{ik}^{(1)}R_{jk}^{(0)}\bigg]+O(\varsigma^2)\right),
\end{align}
where $\Gamma_{ij}^{(1)}\equiv\gamma_i^{(1)} \delta_{ij}$. Thus, by extracting the part of the matrix of the two point functions of the $W_i$ that is proportional to the logs,
\begin{align}
    -\braket{W_i(\tau_1)W_j(\tau_2)}\bigg\rvert_{\varsigma \log{\frac{\tau_{12}^2}{\epsilon^2}}}=(R^{(0)}\Gamma^{(1)} R^{(0)T})_{ij},
\end{align}
and diagonalizing it, we can simultaneously determine the anomalous dimensions $\gamma_i^{(1)}$ and the good basis of primaries at zeroth order.

Restricting to the order $\varsigma$ log terms in the two-point functions simplifies the computation in multiple ways. The log terms can only arise from the contact diagram. This means that we can ignore the order $\varsigma$ terms in the point-splitting definition of the primaries (e.g., in eq.~\eqref{eq:zSq primary pointsplit at order varsigma} and eq.~\eqref{eq:zdSqz primary pointsplit at order varsigma}), and that we only need to consider the two point functions between operators where the number of particles are the same or differ by two; these two cases are illustrated by the second and third diagrams in Figure~\ref{fig: two-pt functions wick contractions}. Furthermore, we find that only the $\log(\chi)$ term is important in the contact diagram contracting two fields at $\tau_1,\tau_2$ in one composite operator and two fields at $\tau_3,\tau_4$ in the other composite primary, since
\begin{align}\label{eq:behavior of logs in point splitting}
    \log(\chi^2)\bigg\rvert_{\substack{\tau_1=y_1+a_1\epsilon,\;\tau_2=y_1+a_2\epsilon\\ \tau_3=y_2+b_1\epsilon,\;\tau_4=y_2+b_2\epsilon }}&\overset{\epsilon\to 0^+}{\sim}2\log{\frac{\epsilon^2}{y_{12}^2}}+\text{finite, no logs}+O(\epsilon),\\
    \log((1-\chi)^2)\bigg\rvert_{\substack{\tau_1=y_1+a_1\epsilon,\;\tau_2=y_1+a_2\epsilon\\ \tau_3=y_2+b_1\epsilon,\;\tau_4=y_2+b_2\epsilon }}&\overset{\epsilon\to 0^+}{\sim}O(\epsilon^2).
\end{align}
Note that the dependence on the detailed point splitting parameterized by the coefficients $a_1,a_2,b_1,b_2$ also drops out, which means we also do not need to be precise about our point-splitting scheme. Meanwhile, no $\log{\frac{\epsilon^2}{y_{12}^2}}$ terms arise from either the $\log(\chi)$ or the $\log((1-\chi))$ terms in the contact diagram contracting one field at $\tau_1$ in one composite operator and three fields at $\tau_2,\tau_3,\tau_4$ in the other composite operator, since:
\begin{align}\label{eq:behavior of logs in point splitting 2}
    \log(\chi^2),\;\log((1-\chi)^2)\bigg\rvert_{\substack{\tau_1=y_1+a_1\epsilon,\;\tau_2=y_2+b_1\epsilon\\ \tau_3=y_2+b_2\epsilon,\;\tau_4=y_2+b_3\epsilon }}&\overset{\epsilon\to 0^+}{\sim}\text{finite, no logs}+O(\epsilon).
\end{align}
In particular, this implies that there are in fact no log terms appearing in the two point functions of any two operators with different numbers of $z$. Finally, it is clear that we only need to keep track of derivatives acting on the prefactor of the $\log(\chi^2)$ term.

Given all of our preparatory work, we can start computing the two-point functions of the lowest order primaries. Let us illustrate the results for the $++$ operators for simplicity. The two-point function of the primary of classical dimension $4$ is:
\begin{align}
    \braket{[z^2]_0^\epsilon(y_1)\;[z^2]_0^\epsilon(y_2)}&=\frac{1}{y_{12}^8}\left(1+\varsigma\big[\log{\frac{y_{12}^2}{\epsilon^2}}+\ldots\big]\right)
\end{align}
Here, $\ldots$ represents finite, non-logarithmic terms. We read off the anomalous dimension $\gamma^{(1)}_{z^2}=-1$. Likewise, the two point function of the primary of classical dimension $6$ is:
\begin{align}
    \braket{[z^2]_2^\epsilon(y_1)\;[z^2]_2^\epsilon(y_2)}&=\frac{1}{y_{12}^{12}}\left(1+\varsigma\big[\frac{13}{4}\log{\frac{y_{12}^2}{\epsilon^2}}+\ldots\big]\right)
\end{align}
We see $\gamma^{(1)}_{[z^2]_2}=-\frac{13}{4}$. Meanwhile, there are two primaries of classical dimension $8$, and their matrix of two point functions takes the form:
\begin{align}
    \left(\begin{array}{cc}\braket{[z^2]_4^\epsilon(y_1)\; [z^2]_4^\epsilon(y_2)}& \braket{[z^2]_4^\epsilon(y_1)\; [z^4]_0^\epsilon(y_2)}\\ \braket{[z^4]_0^\epsilon(y_1)\;[z^2]_4^\epsilon(y_2) }& \braket{[z^4]_0^\epsilon(y_1)[z^4]_0^\epsilon(y_2)}\end{array}\right)&=\left(\begin{array}{cc} 1& 0 \\ 0 & 1\end{array}\right)+\varsigma\left(\begin{array} {cc}\frac{13}{2}& 0 \\  0 &6 \end{array}\right)\log{\frac{y_{12}^2}{\epsilon^2}}+\ldots.
\end{align}
Note in particular that, in accordance with eq.~\eqref{eq:behavior of logs in point splitting 2}, the off diagonal terms of the matrix multiplying $\log{\frac{y_{12}^2}{\epsilon^2}}$ are zero. This means that the ``good basis'' of operators at zeroth order is indeed given by the operators with a fixed number of particles, which is the basis we chose in Table~\ref{tab:table of GFF primaries}. We read off their dimensions as $\gamma_{[z^2]_4}^{(1)}=-\frac{13}{2}$ and $\gamma_{z^4}^{(1)}=-6$. 

We can repeat this procedure for various other primaries. The only other remaining case that is novel is when we consider the two-point functions of degenerate primaries with the same number of particles. This first occurs at $\Delta^{(0)}=12$, where there are two three-particle primaries and two four-particle primaries. For the latter, we find:
\begin{align}
    \left(\begin{array}{cc}\braket{[z^4]_{4,1}^\epsilon(y_1)\; [z^4]_{4,1}^\epsilon(y_2)}& \braket{[z^4]_{4,1}^\epsilon(y_1)\; [z^4]_{4,2}^\epsilon(y_2)}\\ \braket{[z^4]_{4,2}^\epsilon(y_1)\;[z^4]_{4,1}^\epsilon(y_2) }& \braket{[z^4]_{4,2}^\epsilon(y_1)[z^4]_{4,2}^\epsilon(y_2)}\end{array}\right)&=\left(\begin{array}{cc} 1& 0 \\ 0 & 1\end{array}\right)+\varsigma\left(\begin{array} {cc}\frac{31}{2}& 0 \\  0 &\frac{31}{2} \end{array}\right)\log{\frac{y_{12}^2}{\epsilon^2}}+\ldots.
\end{align}
In particular, the degeneracy of these operators persists at order $\varsigma^1$, with $\gamma_{[z^4]_{4,1}}^{(1)}=\gamma_{[z^4]_{4,2}}^{(1)}=-\frac{31}{2}$. The degeneracy likewise persists for the three-particle primaries. 

Finally, we can summarize the tree-level anomalous dimensions of the twenty primaries of classical dimension $\Delta^{(0)}\leq 12$ that we computed in a remarkably compact formula. We find that the tree-level anomalous dimension of a primary with $k-$fields and classical dimension $\Delta^{(0)}$ is:
\begin{align}\label{eq:general formula for tree-level anomalous dimensions}
    \gamma^{(1)}(\Delta^{(0)},k)=-\frac{c_{\Delta^{(0)}}}{8}+\frac{k}{4},
\end{align}
where $c_\Delta\equiv \Delta(\Delta-1)$ is the conformal Casimir. This formula also applies to the identity and the branon. We conjecture that it holds for all operators in the theory. This result can be compared with the analogous result on the half-BPS Wilson line in $\mathcal{N}=4$ SYM, where the leading anomalous dimension for all operators is simply proportional to the superconformal Casimir \cite{Ferrero:2023gnu}.

We should note that the fact that operators with different numbers of particles do not mix at zeroth order holds for any single-scalar theory in AdS$_2$ with quartic interaction, but this particular spectrum --- and the fact the degeneracy between primaries with the same numbers of particles is not lifted at leading order--- appears to be special to the Nambu-Goto interaction. It does not hold for the $m^2=2$ theory with $\phi^4$ interaction, for example.

Our results for the tree-level anomalous dimensions of the primaries also tells us about the OPE coefficients in the GFF limit. In particular, the fact that multi-particle primaries include corrections from two-particle terms at most at order $\varsigma$, and that there are no Wick contractions between two copies of $z$ and a composite operator with more than two $z$'s, implies that the multi-particle OPE coefficients are at most of order $\varsigma$:
\begin{align}
    c_{zz[z^k]_n}\sim O(\varsigma),\qquad \text{for }k>2.
\end{align}
We can also determine the zeroth order OPE coefficients of the two-particle operators simply by performing Wick contractions. Our results are consistent with the general formula:
\begin{align}\label{eq:OPE coefficients zeroth order}
    c_{zz[z^2]_{2n}}^2&=\frac{\Gamma(2n+4)^2\Gamma(2n+7)}{18\Gamma(2n+1)\Gamma(4n+7)}+O(\varsigma).
\end{align}
This result is derived more systematically from the conformal block expansion of the four point function in section~\ref{sec:bootstrap tree-level}. 

\subsection{Comment on tree-level OPE coefficients}
Finally, we can also attempt to compute the tree-level corrections to the OPE coefficients by carefully constructing the primaries following the discussion at the beginning of the previous section (which means computing the full two-point functions between primaries, not just the log terms), and then computing the three-point functions with two insertions of $z$. The results for the first three operators are:
\begin{align}\label{eq:OPE coefficients direct method}
    c_{zz[z^2]}^2&=2-\frac{251}{60}\varsigma+O(\varsigma^2),&
    c_{zz[z^2]_2}^2&=\frac{40}{9}-\frac{11455}{2268}\varsigma+O(\varsigma^2),&
    c_{zz[z^2]_4}^2&=\frac{350}{143}+\frac{330295}{113256}\varsigma+O(\varsigma^2).
\end{align}

Unfortunately, it becomes computationally very expensive to proceed to operators with higher dimensions. More importantly, the methods we have used in this section do not allow us to determine the OPE coefficients of the multi-particle operators, \textit{even in principle}. This is again best illustrated using a concrete example. Let us return to the two primaries that have classical dimension $\Delta^{(0)}=8$, which we now know are at leading order of the form $O_1=[z^2]_4+O(\varsigma)$ and $O_2=[z^4]_0+O(\varsigma)$. These obey the two-point function of the form in eq.~\eqref{eq:45yu654e}. But now it is clear that we can rotate $O_1\to O_1+\alpha \varsigma O_2$ and $O_2\to O_2-\alpha \varsigma O_1$ without changing the two-point function up to order $\varsigma$. Thus, there is again some ambiguity in the basis of primaries, now at order $\varsigma$, which cannot be resolved unless we consider the two-point functions at higher order in perturbation theory. Because the three-point function $\braket{zzO_1}$ is non-zero at zeroth order, this rotation will change the value of the OPE coefficient $c_{zzO_2}$ of the multi-particle operator, which is therefore ambiguous at this order. By contrast, $\braket{zzO_2}$ is zero at zeroth order, which means the $c_{zzO_1}$ OPE coefficient of the two-particle operator is unambiguous up to order $\varsigma$. In particular, the OPE coefficients in eq.~\eqref{eq:OPE coefficients direct method} are unambiguous.

\FloatBarrier

\section{Four-point function via the inversion formula}
\label{app:dispMethod}

In this appendix, we briefly summarize an alternative approach to the analytic bootstrap of 1d CFT four-point functions using dispersion relations derived from the Lorentzian inversion formula \cite{Carmi:2019cub,Mazac:2018qmi,Bonomi:2024lky,Carmi:2024tmp}. Specifically, we focus on the version aimed at perturbative applications given in \cite{Bonomi:2024lky}. We use it to a posteriori justify the transcendentality ansatz used in section \ref{sec:ansatz bootstrap} to compute the four-point function up to two loops.

{We start by giving a self-contained summary of the technical aspects needed to apply the dispersion relations, skipping all derivations. Then, we explain how to specialize the application to perturbative expansions around the generalized free boson. Finally, we explain the strategy implemented to verify the results of the transcendentality ansatz \ref{sec:ansatz bootstrap}, and perform one of the steps explicitly -- extracting the double discontinuity from the correlators.}

\subsection{Summary of the dispersion relations}
The dispersion relation reproduces the $4$-point function from its double discontinuity. For one dimensional theories, the bosonic double discontinuity is,
\begin{equation}\label{eq:dDisc def}
	\text{dDisc}[G(\chi)] = G^{0<\chi<1}(\chi) - \frac{G^{\chi>1}(\chi + i \epsilon) + G^{\chi>1}(\chi - i \epsilon)}{2}\ ,
\end{equation}
{where $G^{0<\chi<1}$ and $G^{\chi>1}$ are defined in eq.~\eqref{eq:G decomposed into three} and in the second term on the RHS we analytically continue $G^{\chi>1}$ from $\chi\in(1,\infty)$ to $\chi\in(0,1)$ both above and below $\chi=1$ and take the average.}
Then, the bosonic dispersion relation for identical external operators of integer dimensions takes the form,
\begin{equation}
\label{eq:dispRelation}
    G^{\mathrm{reg}}(\chi) = \int_0^1 dw\, w^{-2}\, \mathrm{dDisc}\!\left[G^{\mathrm{reg}}(w)\right]\,K^\ell_{\Delta}(\chi,w).
\end{equation}
where $\Delta$ is the dimensions of the external operators. In our case, $\Delta=2$. Let us unpack the various notations introduced in this formula. 

\paragraph{The kernel $K_\Delta^\ell(\chi, w)$.} The label $\ell$ is determined by the behavior of the correlator in the Regge limit. {It is the smallest integer for which}
\begin{equation}
	\left(\frac{1}{2}+it\right)^{-2\Delta_\phi - \ell} G^{\mathrm{reg}}\!\left(\frac{1}{2}+it\right)
< \infty
\qquad \text{for } t \to \infty .
\end{equation}
The kernel itself is given by,\footnote{We present the form of the kernel needed in our analysis of the branon four-point function. For more general expressions, see \cite{Bonomi:2024lky,Carmi:2024tmp}.}
\begin{align}\label{eq:general ell Kernel}
K_{\Delta}^{\ell}(\chi,w)
&\equiv 
K^{-1}_\Delta(\chi,w)
-
\sum_{m,n}
A_{m,n}\,\widehat H^{B}_{m,2}(w)\, \mathcal{C}^n
\left[
\frac{2}{\pi^2}
\left(
\frac{\chi^2 \log \chi}{1-\chi}
+ \chi \log(1-\chi)
\right)
\right]
\nonumber\\
&\quad
-
\sum_{m,n}
\tilde A_{m,n}\,\widehat H^{B}_{m,2}(w)\, \CB_{2+2n}(\chi)\,,
\end{align}
{where the sums over $n$ and $m$ run from zero to infinity}. Here, $\CB$ is the conformal block \eqref{eq:ConfBlocks},
\begin{equation}
    \mathcal{C} = \chi^2 (1-\chi) \partial^2 - \chi^2 \partial\ ,
\end{equation}
and the seed kernel is defined as
\begin{align}
\label{eq:seedKernel}
K_{\Delta}^{-1}(\chi,w)
&=
\frac{w \chi^2 (w-2)\log(1-w) - \chi w^2 (\chi-2)\log(1-\chi)}
{\pi^2 (w-\chi)(w+\chi-w\chi)}
\nonumber\\
&
+ \frac{\chi^2}{\pi^2}
\left[
\log(1-w)\,
\frac{(1-2w)w^{2-2\Delta}}{(w-1)w \chi^2+\chi-1}
+
\frac{\log(1-\chi)}{\chi}\,
\frac{w^{2-2\Delta}}{w\chi-1}
\right.
\\
&\left.
+
\log(\chi)\,
\frac{(1-2w)w^{2-2\Delta}}{(w-1)w \chi^2+\chi-1}
+
\left(
w \to \frac{w}{w-1}
\right)
\right]
\nonumber\\
&
+
\sum_{m=0}^{2\Delta-2}
\sum_{n=0}^{\Delta-4}
\left(
\alpha_{m,n}
+
\beta_{m,n}\log(1-w)
\right)
w^{m+2-2\Delta}
\,\mathcal{C}^n
\left[
\frac{2}{\pi^2}
\left(
\frac{\chi^2\log(\chi)}{1-\chi}
+
\chi\log(1-\chi)
\right)
\right],\nonumber
\end{align}
where the numerical coefficients $\alpha_{m,n}$ and $\beta_{m,n}$ are determined from the condition,
\begin{equation}
    K_{\Delta}^{-1}(\chi,w) = \frac{\chi^{2\Delta}}{(1-\chi)^{2\Delta}}K_{\Delta}^{-1}(1-\chi,w)\ .
\end{equation}
We define $\widehat H^{B}_{m,2}(w)$ separately below. Given it, the coefficients $A$ and $\tilde{A}$ {in \eqref{eq:general ell Kernel}} are determined such that
\begin{equation}
    K^{\ell}_\Delta \sim w^{2+2\ell} 
\end{equation}
at small $w$.

\paragraph{The kernel subtractions $H^{B}_{m,2}(w)$.} 

These are functions of $w$ that {can be added to the seed kernel $K_\Delta^{-1}(\chi,w)$ in order to render the dispersion integral in Eq.~\eqref{eq:dispRelation} convergent in the small-$w$ regime}, without modifying the other desired properties of the kernel \cite{Bonomi:2024lky}. To define them, we first introduce
\begin{align}
p_{\Delta'}(w)
&\equiv
{}_2F_1\!\left(\Delta',1-\Delta';1;w\right),\nonumber
\\[4pt]
q_{\Delta,\Delta'}(w)
&\equiv
a_{\Delta,\Delta'}(w)
+
b_{\Delta,\Delta'}(w)\,\log(1-w),
\\[6pt]
H_{\Delta,\Delta'}(w)
&\equiv
\frac{2\pi}{\sin(\pi\Delta')}
\Bigg[
w^{2-2\Delta}\,p_{\Delta'}(w)
+
\left(\frac{w}{w-1}\right)^{2-2\Delta}
p_{\Delta'}\!\left(\frac{w}{w-1}\right)
+
q_{\Delta,\Delta'}(w)
\Bigg].\nonumber
\end{align}
Here, the $a$'s and $b$'s are linear combinations of the $\alpha$'s and $\beta$'s that we solved for in \eqref{eq:seedKernel},\footnote{Alternatively, they can be determined by demanding that $H_{\Delta,\Delta'}(w)$ is regular at $w=0$.}
\begin{align}
a_{\Delta'}^{\Delta}(w)
&=
\sum_{m=0}^{2\Delta-4}
\sum_{n=0}^{2\Delta-4}
\alpha_{m,n}\,
w^{m+2-2\Delta}\,
{\Delta'}^{\,n}(\Delta'-1)^{n},
\\[6pt]
b_{\Delta'}^{\Delta}(w)
&=
\sum_{m=0}^{2\Delta-4}
\sum_{n=0}^{2\Delta-4}
\beta_{m,n}\,
w^{m+2-2\Delta}\,
{\Delta'}^{\,n}(\Delta'-1)^{n}.
\end{align}
Now we are ready to define 
\begin{equation}
    {\widehat H^{B}_{m,2}(w)} = \pi ^2 \binom{4 m+4 \Delta -2}{2 m+2 \Delta -1} \text{Res}_{\Delta' \to 2\Delta + 2 m}\left[ H_{\Delta,\Delta'}(w) \right].
\end{equation}
The functions $\widehat H^{B}_{m,k}(w)$ for other values of $k$ were defined in \cite{Bonomi:2024lky}, but we will not need them in this work.

\paragraph{The subtracted correlator $G^{\mathrm{reg}}(\chi)$.} The subtracted four point function is tuned to make \eqref{eq:dispRelation} convergent near $w=1$, while still satisfying crossing. For example, we may use the ansatz,
\begin{equation}
\label{eq:regG}
	G^{\mathrm{reg}}(\chi) \equiv G(\chi) - \sum_{m = -\ell}^{\infty} \sum_{n = 0}^{\infty} S_{m,n} \left(\frac{\log^n(1-\chi)}{\chi^m}+\frac{\chi^{2\Delta} \log^n(\chi)}{(1-\chi)^{m+2\Delta}}\right),
\end{equation}
with the coefficients tuned such that the RHS vanishes like $(1-\chi)^{2\ell}$ as $\chi\to 1$.

\subsection{Perturbative expansion}
{The double discontinuity of the correlator has the t-channel conformal block expansion
\begin{align}
    \mathrm{dDisc}[G(\chi)]=\sum_O a_O \frac{\chi^4}{(1-\chi)^4}\mathrm{dDisc}[\mathfrak{f}_{\Delta_O}(1-\chi)],
\end{align}
}
where double discontinuity of the conformal blocks \eqref{eq:ConfBlocks} is explicitly
\begin{align}
\label{eq:dDiscConfBlock}
	\text{dDisc}[\CB_{\Delta_O}(1-\chi)]&=2 \sin^2\left[\frac{\pi}{2}\left(\Delta_O-2\Delta\right)\right]\left(\frac{\chi}{1-\chi}\right)^{2\Delta}\CB_{\Delta_O}(1-\chi).
\end{align}
Since only operators with dimensions of the form $\Delta_n = 2\Delta + 2n + \dots$ are exchanged, the sine-squared term implies that the double discontinuity of the four-point function at order $L$ in the perturbative expansion depends only on conformal data up to order $L-1$.

Combining this observation with Eq.~\eqref{eq:dispRelation}, we conclude that knowledge of all conformal data up to order $L-1$, together with the expected Regge behavior at order $L$, is sufficient to reconstruct the subtracted four-point function $G^{\mathrm{reg}}$ exactly, up to conformal data explicitly appearing in the subtraction. The final step is to invert Eq.~\eqref{eq:regG} to recover the unregularized correlator $G$. The only subtlety is that this inversion depends on a finite set of low-lying conformal data at order $L$. Consequently, the procedure determines the correlator at order $L$ up to a small number of undetermined constants.

\subsection{Application}\label{ref:dispersion bootstrap application}

For our purposes, this method serves only to verify the results of the ansatz bootstrap in section~\ref{sec:ansatz bootstrap}. By construction, the space of correlators allowed by the ansatz bootstrap is a subspace of the space obtained from the procedure described above. Therefore, it suffices to check that the two spaces are in fact identical. We do so by following the steps below, order by order in $\varsigma$:
\begin{enumerate}
	\item Extract the double discontinuity from the ansatz bootstrap result. By Eq.~\eqref{eq:dDiscConfBlock}, this is guaranteed to yield the correct expression.
	\item Construct a suitable subtracted correlator $G^{\text{reg}}$ as described around Eq.~\eqref{eq:regG}.
    \item Write the integral form of the space of correlators $G^{(L)}(\chi)$ whose subtracted form \eqref{eq:regG} solves Eq.~\eqref{eq:dispRelation}, and restrict to the subspace satisfying conditions~4 and~5 of Section~\ref{sec:ansatzBootEnum}. Then verify, analytically or numerically, that this space coincides with the space of solutions to the transcendentality ansatz found in Section~\ref{sec:ansatz bootstrap}.
\end{enumerate}
As long as the resulting solution space has the same dimensionality as that obtained from the transcendentality ansatz, the latter is justified a posteriori. {Using this method, we verified all of the results of section \ref{sec:ansatz bootstrap} -- i.e., the four point function up to two loops.}

\subsection{Computing the double discontinuity}
While we will not present all the details of the verification procedure described in \ref{ref:dispersion bootstrap application}, in the present section we compute the double discontinuity of the branon four-point function up to two loops, which is needed for the first step.

Although the dDisc is defined in eq.~\eqref{eq:dDisc def} in terms of $G^{\chi>1}$ (and $G^{0<\chi<1}$ ), it is convenient for us to write it in terms of $G^{\chi<0}$ using the crossing relation $G^{\chi>1}(\chi)=\frac{\chi^4}{(1-\chi)^4}G^{\chi<0}(1-\chi)$. This is because, first, in the ansatz bootstrap we naturally express $G^{0<\chi<1}$ in the form in \eqref{eq:G(1) expanded in log terms}-\eqref{eq:G(3) expanded in log terms}; second, as mentioned in the discussion of cyclic and braiding symmetry in section~\ref{sec:ansatzBootEnum}, $G^{\chi<0}$ is given by the same expression as $G^{0<\chi<1}$ with $\log^k(\chi)\to \log^k|\chi|$; and, third, when we analytically continue $G^{\chi<0}$, only the $\log^k$ factors behave nontrivially because the $G_{\log^k}^{(L)}$ factors are analytic around $\chi=0$.

Concretely, we can write the double discontinuity as:
\begin{align}\label{eq:dDisc in terms of G0 and Gminus}
    \text{dDisc}\;G(\chi)&=G^{0<\chi<1}(\chi)-\frac{\chi^4}{2(1-\chi)^4}\left[G^{\chi<0}(1-\chi+i\epsilon)+G^{\chi<0}(1-\chi-i\epsilon)\right],
\end{align}
where the arguments of $G^{\chi<0}$ are continued from $(-\infty,0)$ to $(0,1)$ both above and below the branch cut. The analytic continuation is non-trivial only for the $\log(-\chi)$ terms in $G^{\chi<0}(\chi)$, which satisfy
\begin{align}
    \log(-(1-\chi\pm i \epsilon))=\log(1-\chi)\mp i\pi.
\end{align}
Thus, the average of the analytic continuations above and below is trivial for $\log^0(\chi)$ and $\log^1(\chi)$ terms in $G^{\chi<0}(\chi)$, and is non-trivial only for the $\log^k(\chi)$ terms with $k\geq 2$. 

If we write the order $L$ correlators as
\begin{align}
    G^{0<\chi<1,(L)}(\chi)&=\sum_{k=0}^L \log^k(\chi)G_{\log^k}^{(L)}(\chi),&
    G^{\chi<0,(L)}(\chi)&=\sum_{k=0}^L \log^k(-\chi)G_{\log^k}^{(L)}(\chi),
\end{align}
then the analytic continuation from the negative interval to the unit interval gives
\begin{align}
    G^{\chi<0,(L)}(\chi)&=\sum_{k=0}^L (\log(\chi)\pm i \pi)^k G_{\log^k}^{(L)}(\chi).
\end{align}
With this,
\begin{equation}
\begin{aligned}
    &\frac{1}{2}[G^{\chi<0,(L)}(\chi+ i\epsilon)+G^{\chi<0,(L)}(\chi- i\epsilon)]\\&\qquad \qquad =G^{0<\chi<1,(L)}(\chi)-\pi^2 G^{(L)}_{\log^2}(\chi)-3\pi^2 \log{\chi}G^{(L)}_{\log^3}(\chi)+\ldots.
\end{aligned}
\end{equation}
Additional terms indicated by the $\ldots$ are not relevant for $L\leq 3$. Combining this with the crossing relation for $G^{0<\chi<1}$ in eq.~\eqref{eq:cyclic symmetry}, we find that the expression for the dDisc in eq.~\eqref{eq:dDisc in terms of G0 and Gminus} becomes:
\begin{align}
    \text{dDisc}\;G^{(0)}(\chi)&=\text{dDisc}\;G^{(1)}(\chi)=0,\\
    \text{dDisc}\;G^{(2)}(\chi)&=\frac{\pi^2\chi^4}{(1-\chi)^4} G^{(2)}_{\log^2}(1-\chi),\\
    \text{dDisc}\;G^{(3)}(\chi)&=\frac{\pi^2\chi^4}{(1-\chi)^4}\left(G^{(3)}_{\log^2}(1-\chi)+3\log(1-\chi)G^{(3)}_{\log^3}(1-\chi)\right).
\end{align}
Given our results for the one-loop correlator in \eqref{eq:G2 final answer} and two-loop correlator in \eqref{eq:G3 log-cubed} and \eqref{eq:G3 log-squared}, we find:
\begin{align}
    \mathrm{dDisc}\;G^{(2)}(\chi)&=\frac{\pi^2 (25 - 26 \chi + 5 \chi^2 + 5 \chi^6 - 26 \chi^7 + 25 \chi^8)}{8 \chi^2},\\
    \mathrm{dDisc}\;G^{(3)}(\chi)&=-\frac{\pi^2}{960(1-\chi)^4 \chi^2}\bigg(7150 - 36588 \chi + 76687 \chi^2 - 83868 \chi^3 + 50127 \chi^4 - 15398 \chi^5 \nonumber\\& + 
 3660 \chi^6-15398 \chi^7 + 50127 \chi^8 - 83868 \chi^9 + 76687 \chi^{10} - 36588 \chi^{11} + 
 7150 \chi^{12}\bigg)\nonumber\\&-\frac{\pi^2}{32(1-\chi)^5\chi^2}\big(25 - 151 \chi + 385 \chi^2 - 535 \chi^3 + 435 \chi^4 - 205 \chi^5 + 77 \chi^6 - 
 445 \chi^7  \nonumber\\&+2470 \chi^8- 7210 \chi^9 + 12370 \chi^{10} - 13018 \chi^{11} + 
 8298 \chi^{12} - 2950 \chi^{13} + 450 \chi^{14}\big)\log{\chi}\nonumber\\&-\frac{\pi^2 (450 - 725 \chi + 324 \chi^2 - 33 \chi^3 - 33 \chi^7 + 324 \chi^8 - 
   725 \chi^9 + 450 \chi^{10}) \log(1 - \chi)}{32 \chi^3}.
\end{align}

\small
	
\bibliography{string_ads}
\bibliographystyle{utphys}
	
\end{document}